\documentclass{aa}  

\usepackage{graphicx}
\usepackage{txfonts}
\usepackage{comment}

\newcommand{\kms}{$\mathrm{km\,s^{-1}}$}
\newcommand{\kmskpc}{$\mathrm{km\,s^{-1}\,kpc^{-1}}$}
\newcommand{\masyr}{$\mathrm{mas\,yr^{-1}}$}
\newcommand{\Msun}{$\rm M_{\odot}$}
\newcommand{\Lsun}{$\rm L_{\odot}$}
\newcommand{\MLsun}{$\rm M_{\odot}/L_{\odot}$}

\begin{document}

   \title{Equilibrium dynamical models in the inner region of the Large Magellanic Cloud based on Gaia DR3 kinematics}
   \titlerunning{Equilibrium dynamical models for the LMC}
   \authorrunning{Kacharov et al.}


   \author{Nikolay Kacharov\inst{1},
               Behzad Tahmasebzadeh\inst{2},
               Maria-Rosa L. Cioni\inst{1},
               Glenn van de Ven\inst{3},
               Ling Zhu\inst{4},
               Sergey Khoperskov\inst{1}
          }

   \institute{Leibniz Institute for Astrophysics (AIP), An der Sternwarte 16, 14471 Potsdam, Germany\\
              \email{kacharov@aip.de}
         \and
             Department of Astronomy, University of Michigan, Ann Arbor, MI, 48109, USA
          \and
             Institute for Astronomy (IfA), University of Vienna, T\"urkenschanzstrasse 17, A-1180 Vienna, Austria
          \and
             Shanghai Astronomical Observatory, Chinese Academy of Sciences, 80 Nandan Road, Shanghai 200030, PR China
             }

   \date{Received July 15, 2024; accepted September 15, 2024}

 
  \abstract
  {The Large Magellanic Cloud (LMC) has a complex dynamics driven by both internal and external processes. The external forces are due to tidal interactions with the Small Magellanic Cloud and the Milky Way, while internally its dynamics mainly depends on the stellar, gas, and dark matter mass distributions. Despite this complexity, simple physical models often provide valuable insights into the primary driving factors.}
  {We use Gaia Data Release 3 (DR3) data to explore how well equilibrium dynamical models based on the Jeans equations and the Schwarzschild orbit superposition method are able to describe LMC's 5-dimensional phase-space distribution and line-of-sight (LOS) velocity distribution, respectively. In the latter model we incorporate a triaxial bar component for the first time and derive LMC's bar pattern speed.}
   {We fit comprehensive Jeans dynamical models to all Gaia DR3 stars with proper motion and LOS velocity measurements found in the VMC VISTA survey of the LMC using a discrete maximum likelihood approach. These models are very efficient at discriminating genuine LMC member stars from Milky Way foreground stars and background galaxies. They constrain the shape, orientation, and enclosed mass of the galaxy under the assumption for axisymmetry. We use the Jeans model results as a stepping stone to more complex 2-component Schwarzschild models, which include an axisymmetric disc and a co-centric triaxial bar, which we fit to the LMC Gaia DR3 LOS velocity field, using a $\chi^2$ minimisation approach.}
   {The Jeans models describe well the rotation and velocity dispersion of the LMC disc and we find an inclination angle $\theta=25.5^{\circ}\pm0.2^{\circ}$, line of nodes orientation $\psi=124^{\circ}\pm0.4^{\circ}$, and an intrinsic thickness of the disc $q_0^d = \frac{b}{a} = 0.23\pm0.01$ (minor to major axis ratio). However, bound to axisymmetry, these models fail to properly describe the kinematics in the central region of the galaxy, dominated by the LMC bar. We use the derived disc orientation and the Gaia DR3 density image of the LMC to obtain the intrinsic shape of the bar. Using these two components as an input to our Schwarzschild models, we perform orbit integration and weighting in a rotating reference frame fixed to the bar, deriving an independent measurement of the LMC bar pattern speed $\Omega = 11\pm4$\,\kmskpc. Both the Jeans and Schwarzschild models predict the same enclosed mass distribution within a radius of $6.2$\,kpc of $\sim1.4\times10^{10}$\,\Msun.}
   {}


   \maketitle
%

\section{Introduction}

The Large Magellanic Cloud (LMC) is the most massive Milky Way satellite galaxy \citep{mcconnachie2012}.
Situated at a distance of about $50$\,kpc, it lies well within the Milky Way halo alongside its lesser counterpart the Small Magellanic Cloud (SMC $\sim62$\,kpc).
Their relative proximity makes them prominent features in the southern sky, easily visible with the naked eye, known to mankind since the dawn of the southern civilisations \citep{dennefeld2020}.

The past orbit of the LMC and exact epoch of arrival of the Clouds in such a proximity to the Milky Way is difficult to discern, but they are likely on their first approach \citep{besla+2007}, as evidenced by LMC's high tangential velocities, approaching the Milky Way escape velocity \citep{kallivayalil+2006}.
This statement is further supported by the fact that the LMC is still associated with a group of its own satellites that have not been tidally stripped \citep{jethwa+2016, sales+2017, kallivayalil+2018}.
Although \citet{vasiliev2024} proposed a model, where the Clouds are on their second approach, the first being $5-10$\,Gyr ago with a pericentric distance larger than $100$\,kpc, far enough to retain its current system of satellites.

The total mass of the LMC has also been under debate for many years.
The close interactions with the SMC and the Milky Way have led to significant perturbations in the kinematics in the LMC outer regions, where reliable tracers are also scarce, making direct mass measurements almost impossible. 
But a plethora of recent studies using different independent methods (perturbations of stellar streams \citealt{erkal+2019, vasiliev+2021, shipp+2021, koposov+2023}; momentum balance of the Local group \citealt{penarrubia+2016}; census of the LMC satellites \citealt{erkal+belokurov2020}) firmly put it in the range $1-2\times10^{11}$\,\Msun, only $5-10$ times lower than the mass of the Milky Way \citep[see the review by][]{vasiliev2023}.
Most recently, the virial mass of the LMC was estimated by \citet{watkins+2024} to be $1.8\times10^{11}$\,\Msun, using globular clusters as tracers going out to $13.2$\,kpc and assuming a NFW dark matter (DM) halo density.

Still star forming and thus classified as a dwarf irregular (dIrr), owing to its high mass and recent arrival in the vicinity of the Milky Way, the LMC possesses surprisingly regular features.
To a first approximation, it can be described morphologically as a barred spiral disc galaxy, showing a typical rotation pattern, as demonstrated by \citet{luks+rohlfs1992, kim+1998, olsen+massey2007} for the gaseous component and by \citet{vandermarel+2002, olsen+2011, gaiaedr3} for the stellar disc.
\citet{vandermarel+kallivayalil2014} modelled the LMC as a flat rotating disc in three dimensions and \citet{vasiliev2018} used axisymmetric Jeans dynamical models to simultaneously fit its rotation and velocity dispersion using Gaia Data Release 2 \citep[DR2][]{gaiadr2} proper motions (PM).
\citet{niederhofer+2022} showed that the stars in the bar region follow elongated orbits.

A more detailed overview of the morphological and kinematic structure of the LMC, however, reveals the effects of the close interactions with the SMC and the Milky Way.
The LMC - SMC interactions likely formed the Magellanic Stream $\sim2$\,Gyr ago \citep{besla+2012, diaz+bekki2012} - a large body of stripped ionised and neutral gas that follows the orbits of the Clouds and spans more than $200^{\circ}$ on the sky.
In addition the two galaxies are connected through the Magellanic Bridge - a younger gaseous and stellar feature, that probably originated from a direct collision between the two galaxies a few hundred Myr ago \citep{diaz+bekki2012, besla+2013, wang+2019, zivick+2019}.
Multiple morphological and kinematic substructures discovered in the periphery of the LMC attest to its perturbed nature \citep{mackey+2016, mackey+2018, belokurov+erkal2019, elyoussoufi+2021, cullinane+2022a, cullinane+2022b}.

The LMC disc appears elongated \citep{vandermarel2001} and also shows deviations from a simple planar structure, as it is warped and truncated in the west direction towards the SMC \citep{mackey+2018, choi+2018}.
The bar appears off-centre \citep{devaucouleurs+freeman1972} and might be a transient unvirialised structure, formed through the interactions and close encounters between the Clouds \citep{zhao+evans2000, besla+2012}.

The dynamical centre of the LMC has been equally difficult to establish with discrepant results from photometry \citep{vandermarel2001}, stellar \citep{vandermarel+kallivayalil2014, wan+2020, gaiaedr3, niederhofer+2022}, and gas \citep{luks+rohlfs1992, kim+1998} kinematics.
The determination of the dynamical centre from stellar kinematics also depends on the type of data - line of sight (LOS) velocities vs. PMs, and the type of stellar populations used as tracers, e.\,g. young vs. old stars.
It is worth mentioning that the kinematic centre is found very close to the centre of the bar, based on Gaia DR3 PMs - the most complete and precise stellar PM database in the LMC today \citep{gaiaedr3}.

The literature has also not been conclusive about the LMC figure rotation and its bar pattern speed.
\citet{dottori+1996} studied the spatial distributions of two young massive LMC clusters and concluded that their differences can be attributed to an off-centred bar induced perturbation that propagates with $\Omega = 13.7\pm2$\,\kmskpc.
\citet{gardiner+1998} used N-body simulations to study the effects from perturbations due to an off-centred bar to the global distribution of the gas and star formation activity in the LMC.
They found that a spiral structure compatible with the LMC emerges for bar pattern speeds in the range of $40-50$\,\kmskpc, noting that their findings are in tension with \citet{dottori+1996}.
\citet{shimizu+yoshii2012} derived $\Omega = 21\pm3$\,\kmskpc~under the assumption that the Shapley Constellation~III star forming region \citep{shapley1951} is found in the Lagrange point $L_4$ of a rotating non-axisymmetric bar potential.
Most recently, \citet{jimenez-arranz+2024a} used several independent methods to estimate the LMC bar pattern speed using Gaia DR3 PM and LOS velocity data.
They conclude that the \citet{tremaine+weinberg1984} method is not applicable to the LMC owing to its strong dependency on the orientation of the galaxy frame and the viewing angle of the bar perturbation.
Using the \citet{dehnen+2023} method, they find $\Omega=-1\pm0.5$\,\kmskpc~- a bar that barely rotates or even shows a marginal counter rotation, but acknowledge that such a configuration is incompatible with the spiral arms structure of the stellar disc.
Finally, \citet{jimenez-arranz+2024a} find that fitting a bi-symmetric velocity model \citep{gaia2023} to the tangential velocity field yields a co-rotation radius $R_c = 4.20\pm0.25$\,kpc and $\Omega = 18.5\pm1$\,\kmskpc.
With co-rotation to bar radius ratio $R_c/R_b = 1.8\pm0.1$, they argue for a slow bar in the LMC.
In a separate work, \citet{jimenez-arranz+2024b} used the KRATOS suite of N-body simulations to analyse the bar pattern speed of the LMC, finding $\Omega=10-20$\,\kmskpc.
They conclude that recent close interactions between the LMC and SMC do not necessary alter the frame rotation rate and bar size, and that long bars are typically slow.

\citet{tahmasebzadeh+2022} introduced a new method to derive the bar pattern speed using Schwarzschild orbit super-position dynamical models \citep{schwarzschild1979} to fit the stellar motions in a triaxial bar potential with figure rotation.
Initially demonstrated on a simulated galaxy, \citet{tahmasebzadeh+2023} used the method to measure the bar pattern speed in NGC\,4371.
Motivated by the non-axisymmetric kinematics in the inner region of the LMC, we test the performance of the orbit super-position dynamical models in the inner region of the LMC utilising Gaia DR3 LOS velocities, thus deriving the LMC bar pattern speed, independently.

In this study we present a series of equilibrium dynamical models of the LMC, taking advantage of the most recent 3D kinematic data in the LMC from Gaia DR3 \citep{gaiadr3}.
Although the LMC is in a complex dynamical environment and likely significantly out of equilibrium, we concur that our approach allows us to learn about and highlight kinematic and morphological features, which are inherently not accounted for in equilibrium models.
The article is organised as follows:
in Sect.~\ref{sec:data} we describe our dataset;
in Sect.~\ref{sec:jeans} we present 3-dimensional (3D) axisymmetric Jeans dynamical models of the LMC, which results establish the basis for our Schwarzschild models;
in Sect~\ref{sec:phot} we show a bar - disc decomposition of the LMC and derive the galaxy's viewing angles and de-projected morphology, assuming co-centric axisymmetric disc and triaxial bar;
in Sect~\ref{sec:schwarzschild} we present our Schwarzschild models and the results for the LMC bar pattern speed.
Sect~\ref{sec:conclusion} summarises our work and provides an outlook of our plans for the future of LMC dynamical studies.


\section{Data}\label{sec:data}

For this study we utilise Gaia Data Release 3 \citep[DR3\footnote{\url{https://www.cosmos.esa.int/web/gaia/dr3}}][]{gaiadr3}.
We select Gaia DR3 sources with full 6D phase-space information: sky position, parallax, PMs, and LOS velocities from the Gaia radial velocity spectrometer (RVS), crossmatched with the VISTA Magellanic Clouds (VMC\footnote{\url{https://star.herts.ac.uk/~mcioni/vmc/}}) survey DR6 \citep{cioni+2011}, i.e. all our sources have both Gaia and VMC entries.
There are $\sim72\,000$ sources in our selection, which include foreground and background contaminants.
We use this complete catalogue without removing the contaminants for our Jeans modelling (see Sect. \ref{sec:jeans}), where we evaluate the probability of each source of being a genuine LMC member.
However, in this section we apply the LMC membership criteria from \citet{gaiaedr3}, based on parallax and PMs to illustrate some properties of the selected Gaia data ($\sim28\,000$ LMC sources).
Throughout this article we adopt the widely used PM convention $\rm \mu_{RA} \equiv \mu_{RA}\cos\delta$.

   \begin{figure*}
   \resizebox{\hsize}{!}
            {
            \includegraphics[scale=.15]{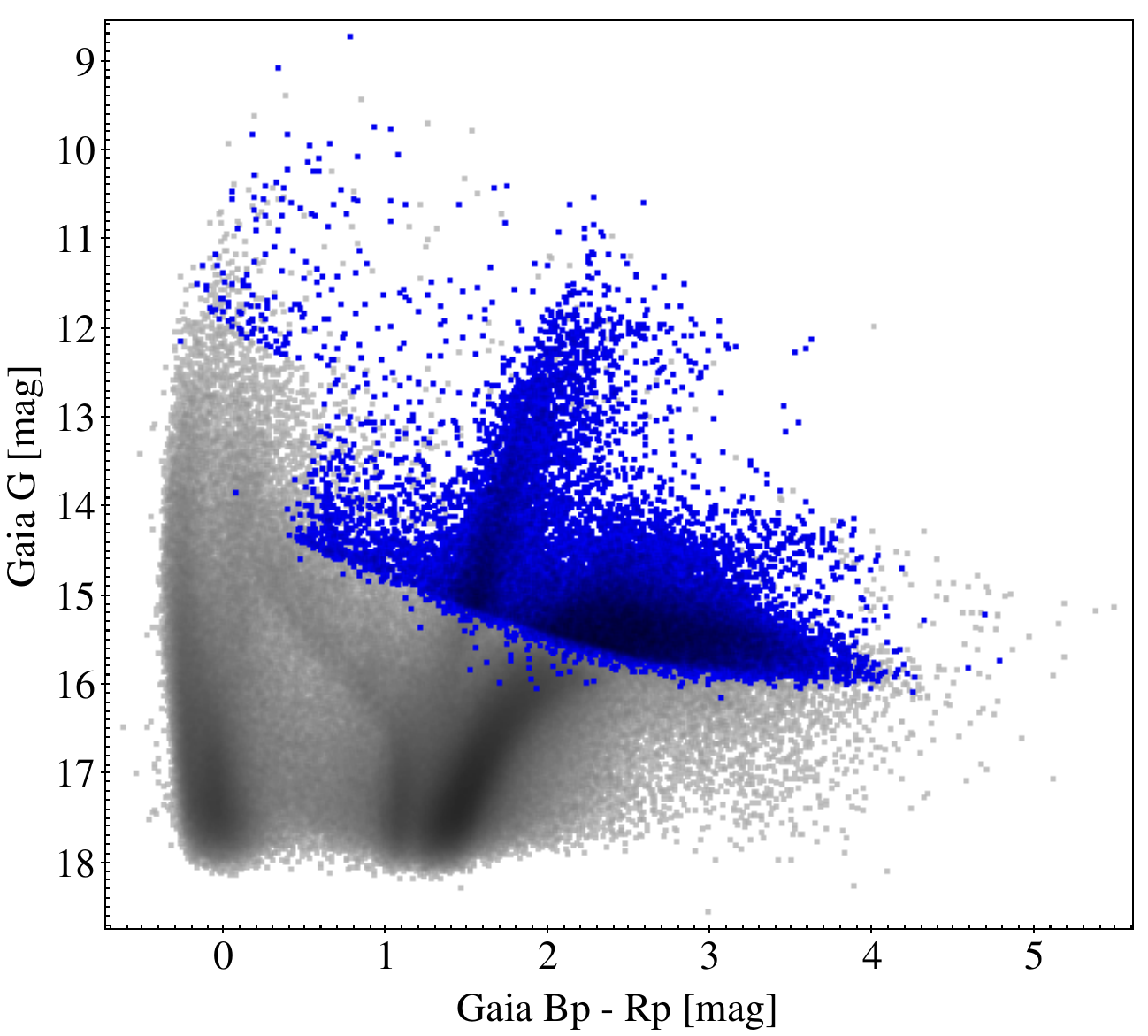}
            \includegraphics[scale=.15]{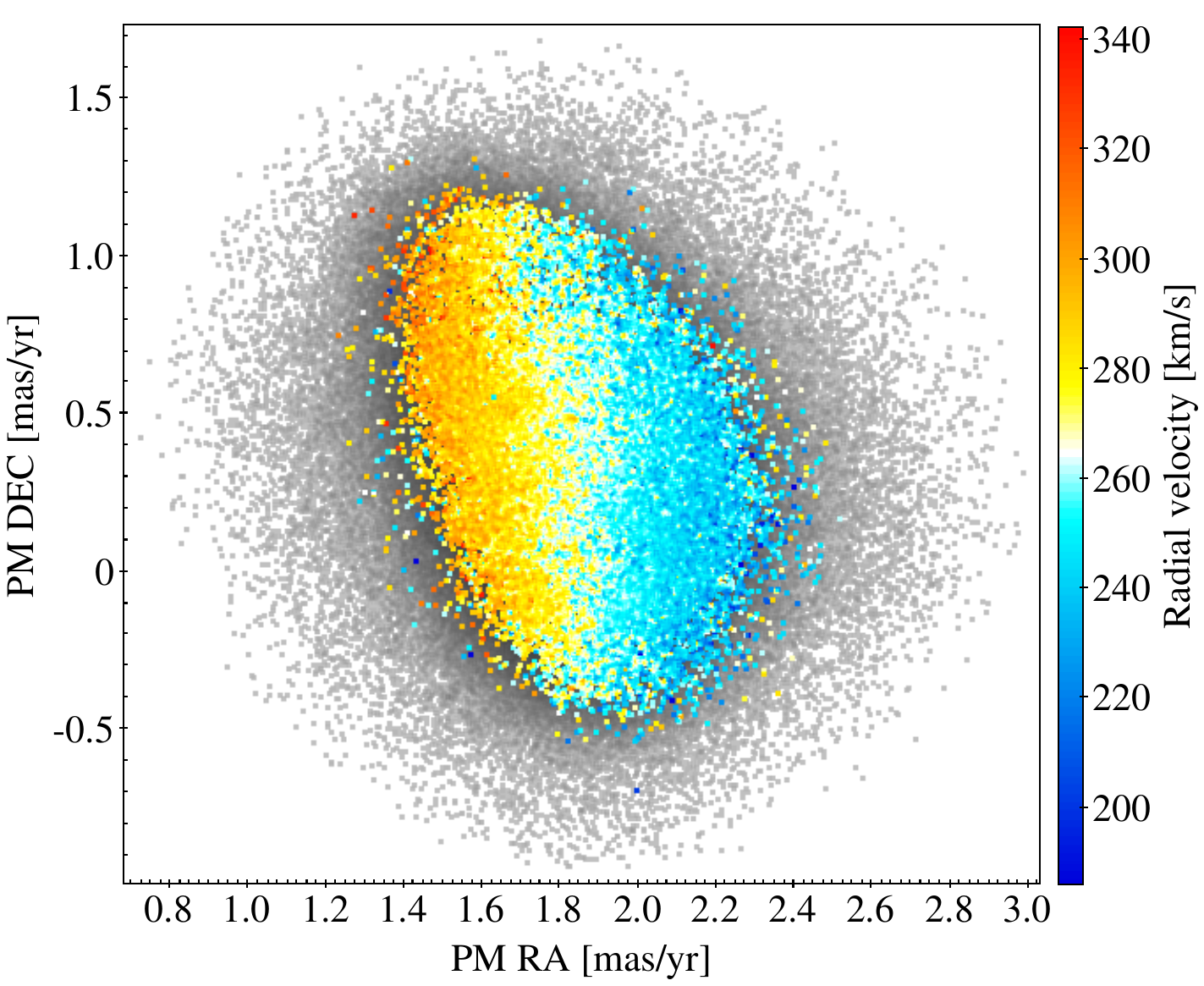}
            \includegraphics[scale=.15]{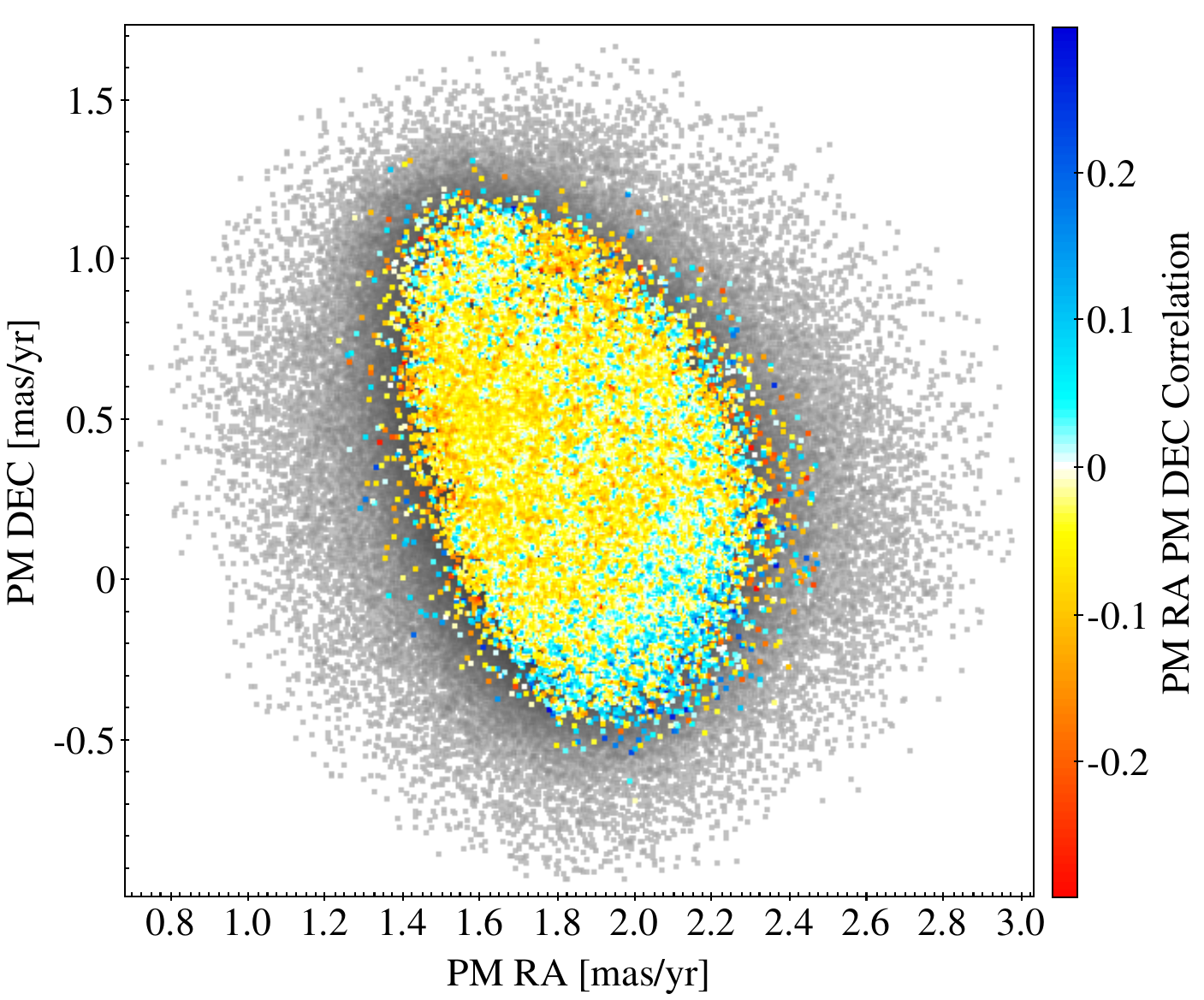}
            }
      \caption{{\it Left panel:} A Gaia CMD of the selected LMC targets with 3D kinematics (blue). {\it Middle panel:} A PM vector diagram of the LMC sources with 3D kinematics, colour coded by the LOS velocity. {\it Right panel:} Same as the middle panel, but colour coded by the $\rm \mu_{RA}\,\mu_{DEC}$ correlation coefficient. The grey points in all panels show additional LMC sources with good Gaia PM measurements ($\rm \mu_{err} < 0.1\,mas\,yr^{-1}$), but no LOS velocity entries. These are the sources used in our 2D Jeans model.}
         \label{fig:data1}
   \end{figure*}

   \begin{figure*}
   \resizebox{\hsize}{!}
            {
            \includegraphics[scale=.15]{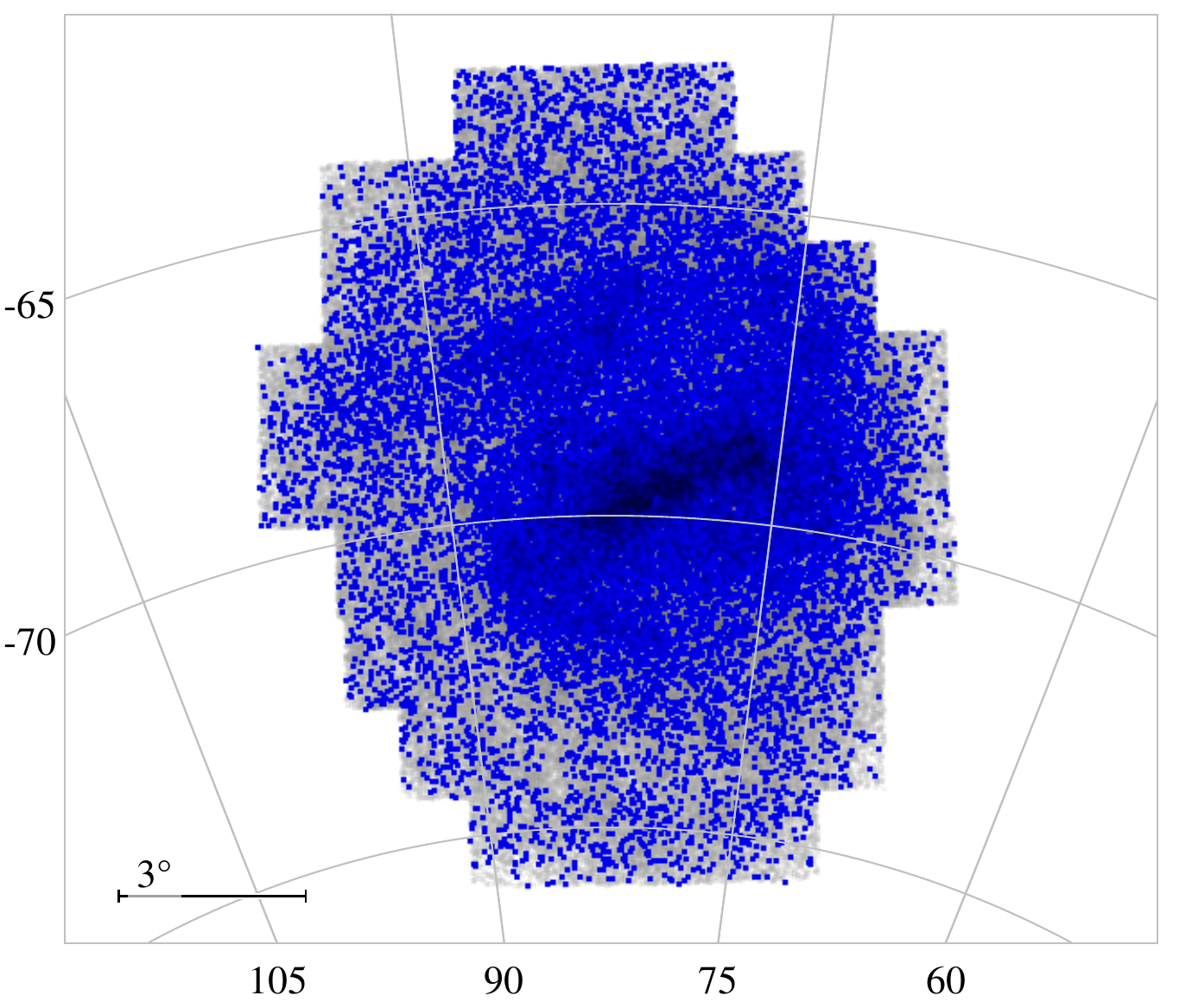}
            \includegraphics[scale=.15]{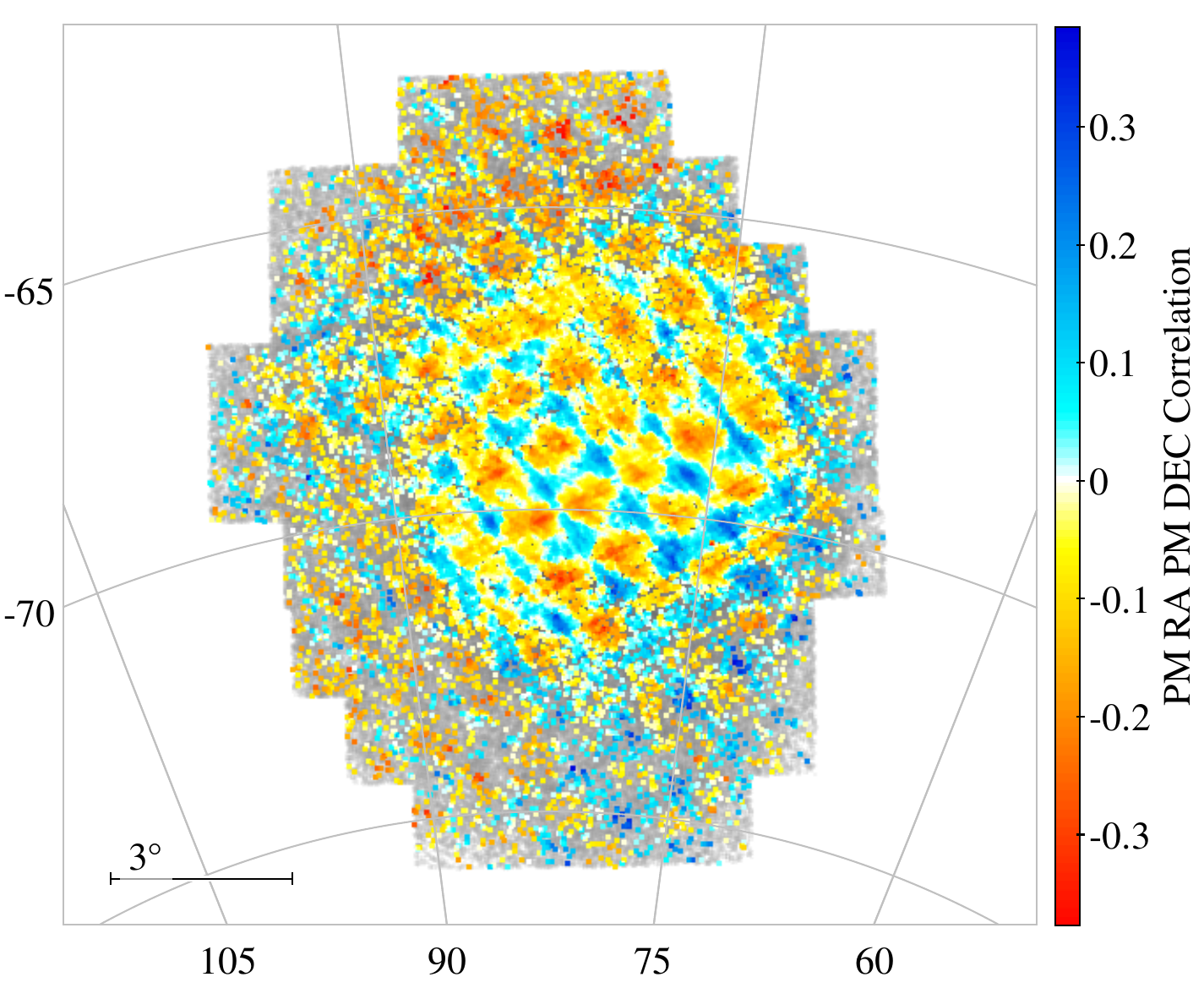}
            \includegraphics[scale=.15]{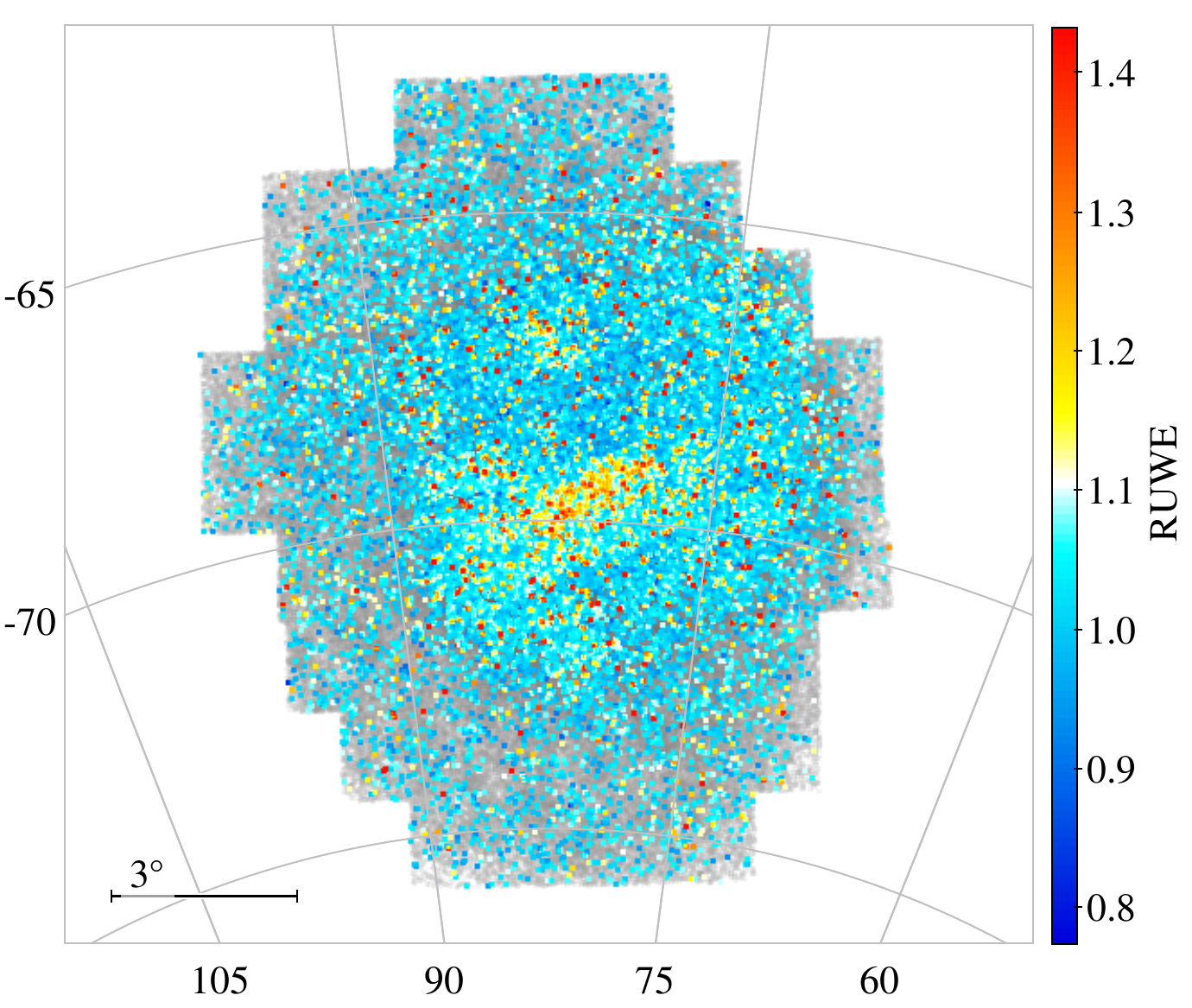}
            }
      \caption{{\em Left panel:} A spatial density distribution of the selected LMC targets with 3D kinematics (blue) with the larger sample of stars with good PM measurements plotted under with light grey dots. {\em Middle panel:} Similar to the left panel, but the stars with 3D kinematics are colour-coded according to the $\rm \mu_{RA}\,\mu_{DEC}$ correlation coefficient. {\em Right panel:} the stars with 3D kinematics are colour-coded according to the RUWE index.}
         \label{fig:data2}
   \end{figure*}

In Fig.~\ref{fig:data1} we show the Gaia CMD of the LMC stars with 3D kinematics, compared to a larger sample of $\sim760\,000$ sources with high precision PM measurements ($\rm \mu_{err} < 0.1\,mas\,yr^{-1}$), but no LOS velocity information, that we analysed in \citep{kacharov+cioni2023} applying 2D Jeans models.
It is evident that the stars, which have Gaia DR3 LOS velocities include only the brightest populations - intermediate age asymptotic giant branch (AGB) stars and young red supergiants.
There are very few main sequence stars.
In the middle and right panels of Fig.~\ref{fig:data1} we show the PM vector diagram of the selected sources, colour coded by the LOS velocity measurements and the $\rm \mu_{RA}\,\mu_{DEC}$ correlation coefficient.
The gradient of the LOS velocity field shows that the galaxy rotation is well resolved.
On the other hand, we do not observe any noticeable gradient in the $\rm \mu_{RA}\,\mu_{DEC}$ correlation coefficient in the PM vector plot.
It is derived from the full 5-dimensional astrometric solution for the source and defines the proper motion error ellipse. It depends on the Gaia scanning pattern. Highly correlated solutions could contribute additional systematic uncertainties in the analysis. The mean correlation in our dataset is $-0.03$ with a standard deviation of $0.11$, which we deem reasonably small. We treat the $\rm \mu_{RA}, \mu_{DEC}, and V_{LOS}$ as independent in our 3D Jeans modelling.

In the left panel of Fig.~\ref{fig:data2} we show the density distribution of the selected LMC sources, where notable features of the galaxy morphology are recognisable like its dense stellar bar and the spiral structure of the disc.
We also plot the spatial distribution of the $\rm \mu_{RA}\,\mu_{DEC}$ correlation in the middle panel that can be directly compared to the stellar density distribution.
The PM correlation map can be directly traced to the Gaia scanning pattern, but there is no dependence with the stellar crowdedness or any other LMC morphological features.
Finally we show the Renormalised Unit Weight Error (RUWE) spatial distribution of the LMC sources in the right panel of Fig.~\ref{fig:data2}.
The RUWE value provides an indication of the quality of the astrometric solution, which affects both the positions and PMs of sources.
Higher RUWE values suggest larger systematic errors in the astrometric measurements, potentially leading to less reliable PM estimates.
The median RUWE index for the selected $\sim28\,000$ LMC stars is $1.026$ with $90\%$ of the targets having RUWE less than $1.18$.
However, the reader will notice a systematic increase of the RUWE index in the central most crowded bar region, which indicates possible elevated systematic uncertainties in the astrometric solution there.
About $4\%$ of the LMC member stars have $\rm RUWE>1.4$, located predominantly in the bar.

   \begin{figure}
   \centering
   \includegraphics[width=\hsize]{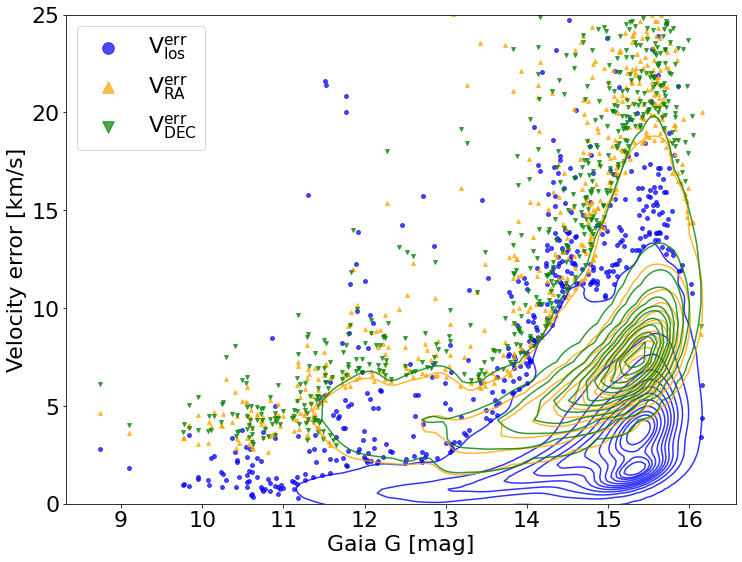}
      \caption{Velocity error distribution as a function of Gaia G magnitude along the LOS and in the plane of the sky. PMs are converted to linear velocities assuming a LMC distance of $49.9$\,kpc. The densest regions are represented only with contours due to the significant overlap between the points.}
         \label{fig:vel_err}
   \end{figure}

Throughout this work we assume a distance modulus to the LMC of $18.49$\,mag \citep{degrijs+2014, crandall+ratra2015}, derived from a large compilation of different tracers, which corresponds to a distance of $49.9$\,kpc.
This allows us to convert the Gaia DR3 PMs and their uncertainties to linear velocities and make a fair comparison between the velocity errors across the three spatial dimensions.
Fig. \ref{fig:vel_err} shows the individual velocity uncertainties along the LOS and in the plane of sky as a function of Gaia G magnitude.
The median RV error in our LMC sample is $3.2$\,\kms with $90\%$ of the stars having RV errors less than $6.9$\,\kms.
The PM uncertainties along the RA and DEC directions trace each other very closely and are noticeably larger than the RV uncertainties with a median $\rm \mu_{RA}$ error $5.1$\,\kms ($0.021$\,\masyr), $\rm \mu_{DEC}$ error $5.3$\,\kms ($0.022$\,\masyr) with $90$-th percentiles, respectively $9.0$\,\kms ($0.038$\,\masyr) and $9.6$\,\kms ($0.041$\,\masyr).

\section{Axisymmetric Jeans Models}\label{sec:jeans}
\subsection{Model setup}

In \citet{kacharov+cioni2023} we presented a series of discrete axisymmetric dynamical models of the LMC with varying gravitational potentials, based on the \citet{jeans1915} hydrodynamical equations, using $\sim1\times10^6$ PM measurements in the same VMC field (the complete sample down to Gaia $\rm G = 18$\,mag shown in Fig. \ref{fig:data1}, but with contaminants included).
We found that these models provide realistic estimates of the enclosed total mass within the extent of the data and describe well the rotation and velocity dispersion of the LMC disc.
However they do not capture well the kinematic properties of the central region of the galaxy, dominated by the LMC bar - an intrinsically non-axisymmetric feature.
In addition, when using solely PM data, it is not feasible to fully constrain the orientation of the line of nodes and the pointing of the angular momentum vector, which leads to uncertainties in the overall shape and morphology of the galaxy.

In this article we present a similar setup of discrete Jeans axisymmetric models of the LMC, but using only stars, for which we have 3D Gaia kinematics.
The inclusion of RV measurements allows us to fully constrain the morphology and orientation of the LMC disc, purely from its kinematical properties, albeit still bound to the assumption for axisymmetry, which is intrinsic to the solution of the Jeans equations \citep{cappellari2008}.

The Jeans equations, derived from the collisionless Boltzmann equation, link the gravitational potential of a self-gravitating system to its stellar density distribution and stellar velocity moments.
In order to solve them we need to make assumptions for the gravitational potential and the density distribution and parametrise them.

For the stellar density we assume that the galaxy surface brightness follows the projected stellar density and that it is exponential.
We adopt the LMC exponential surface brightness profile from \citet{gallart+2004} with a scale radius $\rm r_e = 98.7$\,arcmin and scale it to a total luminosity $\rm L_{LMC} = 1.31\times10^9$\,\Lsun~from the NASA/IPAC Extragalactic Database (NED).
While we do keep the above figures fixed in the dynamical model fit, we vary the projected flattening ($\rm q_{proj}$), defined as the projected minor to major axis ratio, the orientation angle ($\psi$), which gives the direction of LMC's projected minor axis on the sky measured from north to east), and the inclination angle ($\theta$), which is defined to be $0^{\circ}$ if seen face-on and $90^{\circ}$ if seen edge-on. They are free parameters in our setup. The intrinsic flattening can then be calculated as $\rm q_{intr} = \sqrt{\rm q_{proj}^2-\cos\theta^2} / \sin\theta$ in the axisymmetric case.

For the gravitational potential we have two cases.
\begin{enumerate}
\item We assume that the total mass of the LMC is contained within a spherical DM halo that follows a \citet*[][NFW]{nfw1997} mass density profile.
That means that mass contributions from the stellar and gaseous component are also included in the DM halo and not treated separately.
The NFW radial density profile (Eq.~\ref{eq:nfw}) is characterised by two parameters - a characteristic density ($\rm\rho_s$) and a scale radius ($\rm r_s$), which are free parameters in our Jeans dynamical model.
\begin{equation}\label{eq:nfw}
\rho(r) = \frac{\rho_s}{\left(\frac{r}{r_s}\right) \left(1 + \frac{r}{r_s}\right)^2}.
\end{equation}
\item We have a mixed potential, which consists of a NFW DM halo and a visible mass components.
The latter follows the exponential surface brightness distribution and an assumed stellar mass to light ratio ($\rm M/L_*$).
We explore models where $\rm M_*/L$ is a free parameter in the model or fixed to  $\rm M_*/L=1.5$\,\MLsun~for a total luminous mass of $2\times10^9$\,\Msun.
\end{enumerate}

We represent both the mass density and surface brightness distributions as Multi-Gaussian Expansions \citep[MGE;][]{emsellem+1994}, which makes them convenient to project and de-project - operations necessary to solve the Jeans equations. 

We use the {\sc python} version of the Jeans Anisotropic Modelling (JAM) code by \citet{cappellari2008, cappellari2012} and adopt a Bayesian framework with a discrete likelihood function, initially introduced by \citet{watkins+2013}, but see also \citet{zhu+2016,kamann+2020,kacharov+2022}.
In the discrete likelihood framework, the JAM code predicts the first ($\overline{v_z}$, $\overline{v_r}$, $\overline{v_{\phi}}$) and second ($\overline{v_z^2}$, $\overline{v_r^2}$, $\overline{v_{\phi}^2}$) velocity moments at the position of each star in the sample in the three cardinal directions - LOS ($z$), radial ($r$), and tangential ($\phi$).
We compute the probability of each star's three observed velocity components to be drawn from a 3D Gaussian with the first moments as the mean values in the three cardinal directions of the model and corresponding variances $\sigma^2 = \overline{v^2} - \overline{v}^2$, also in all three directions and call it the dynamical likelihood ($P_{dyn}^{\rm LMC}$).
We assume that there is no cross-term correlation between the two PMs and the RV.
This setup requires two additional model parameters, which are set free in our fit - the velocity anisotropy ($\beta_z = 1 - \frac{\overline{v_z^2}}{\overline{v_R^2}}$) and a systemic rotation parameter, defined as $\kappa=\frac{\overline{v_{\phi}}}{(\overline{v_{\phi}^2} - \overline{v_R^2})^{1/2}}$.
When $\kappa = \pm1$ and $\beta_z = 0$, the system reduces to an isotropic rotator and when $\kappa=0$, there is no net angular momentum.
Negative $\beta_z$ values indicate tangential anisotropy, while positive $\beta_z$ values indicate radial anisotropy \citep{cappellari2008}.
Although both the anisotropy and the rotation parameter can vary with radius, here we assume that they are radially constant.

We also include in our Jeans models an additional foreground / background population component with a flat surface density, designed to capture the foreground stellar and background galaxy contaminants in our Gaia DR3 data. 
It is controlled by a single free parameter ($\epsilon$), as a fraction of the central surface brightness of the LMC MGE, as in \citet{watkins+2013, zhu+2016, kacharov+2022}.
Hence we can write the spatial probability distribution functions of the LMC and contaminant population components as:
\begin{equation}
P_{spa}^{\rm LMC} = S / (S + \epsilon C)
\end{equation}
and
\begin{equation}
P_{spa}^{\rm cont} = 1 - P_{spa}^{\rm LMC} ~,
\end{equation}
where $S$ is the MGE surface brightness profile at the position of each star and $C$ - the central surface brightness of the LMC MGE.
We also define $P_{dyn}^{\rm cont}$ analogically to $P_{dyn}^{\rm LMC}$, but with $0$ mean motion and very large variance.

The posterior probability of a star ($i$) to belong to the LMC or to the foreground component of the model is then given by the joint probability of the above specified two probability distributions and we maximise the $\log$ posterior function of the entire sample:
\begin{equation}
\ln P = \sum_i \ln (P_{spa,i}^{\rm LMC} P_{dyn,i}^{\rm LMC} + P_{spa,i}^{\rm cont} P_{dyn,i}^{\rm cont}).
\end{equation}
Thus, when the best-fit model is obtained we can easily compute the probability for each star to be a genuine LMC member or to belong to the foreground / background contamination.

Finally, we have 5 additional trivial model parameters, which describe the kinematic centre of the LMC ($\alpha_0, \delta_0$) and its mean space motion ($v_{los}^0, \mu_{\alpha}^0, \mu_{\delta}^0$), which brings the total number of free parameters in our Jeans model to 13.

We use the {\sc emcee} affine invariant Markov Chain Monte Carlo (MCMC) algorithm \citep{goodman+weare2010} in {\sc python} \citep{foreman-mackey+2013}.
We run our models on a CPU cluster engaging 96 cores and use 192 walkers and 3000 steps in the MCMC, which we confirmed to be enough for the fit to converge.

We convert the stellar coordinates and PMs to an orthographic projection \citep[$\rm \alpha, \delta, \mu_{\alpha}, \mu_{\delta} \rightarrow X, Y, \mu_X, \mu_Y$, see Eq. $1-3$ in][]{gaiaedr3}, correct the three velocity vectors for perspective effects, as in \citet[][Eq. 6]{vandeven+2006}, and rotate the coordinate system ($\rm X, Y, \mu_X, \mu_Y \rightarrow X_p, Y_p, \mu_{X_p}, \mu_{Y_p}$) every time we call the likelihood function to account for the incremental random changes in the kinematic centre, mean space motion, and disc orientation during the fit.

\subsection{Model results: LMC orientation}

The best-fit parameters from our 3D Jeans model are summarised in Table~\ref{tab:jeans}.
We also include there the results from our 2D Jeans model, based solely on Gaia PM data and described in \citet[][this model includes the grey points in Fig.~\ref{fig:data1} as kinematic tracers]{kacharov+cioni2023}.
The way the 2D and 3D Jeans models are constructed and fit are essentially the same, which makes them directly comparable.
The only difference (besides not including RV data in the 2D model) is that we kept the centre fixed to the best-fit coordinates from \citet{gaiaedr3} in the 2D model, while here we fit for it, using the added knowledge of the LOS velocities.
The 2D model is also fit to significantly more and fainter stars with good PM data down to Gaia G $18$\,mag, compared to the 3D model (see Fig.~\ref{fig:data1}).
This may make the two models sensitive to different stellar populations in the LMC; the 2D model being more sensitive to the kinematics of old stars, represented by the numerous, but faint red giant branch (RGB) population.

\begin{table*}
\caption{Jeans results for the following model setups:  3D model for all stars with NFW gravitational potential (fiducial model);  3D model for all stars with mixed potential (NFW + scaled stellar density);  3D model for the young stellar population with NFW potential; 3D model for the old stellar population with NFW potential; 2D model for all stars with NFW potential.}             
\label{tab:jeans}      
\centering          
\begin{tabular}{c c c c c c}
\hline\hline       
   &  \multicolumn{4}{c}{3D Model} & 2D Model \\
   & \multicolumn{2}{c}{All} & Young & Old & All \\
\hline
   &  \multicolumn{5}{c}{Adopted quantities}  \\
\hline
m$-$M\tablefootmark{a} [mag] & $18.49$ & $18.49$ & $18.49$  & $18.49$  & $18.49$ \\
distance\tablefootmark{a} [kpc] & $49.9$ & $49.9$ & $49.9$  & $49.9$  & $49.9$ \\
$\rm L_{tot}$\tablefootmark{b} [$10^9\,$\Lsun] & $1.31$ & $1.31$ & $1.31$ & $1.31$ & $1.31$ \\
$\rm M_*/L$\,[\MLsun] & -- & 1.5 & -- & -- & -- \\
$\rm r_{e}$\tablefootmark{c} [arcmin] & $98.7$ & $98.7$ & $98.7$  & $98.7$ & $98.7$\\
$N_{stars}$ & $71\,794$ & $71\,794$ & $54\,734$ & $17\,060$ & $968\,613$ \\
\hline
   &  \multicolumn{4}{c}{Fitted parameters}  \\
\hline
$\alpha_0$ [deg] & $80.29\pm0.04$ & $80.26\pm0.04$ & $79.69\pm0.05$ & $80.99\pm0.06$  & $81.07$\tablefootmark{d} \\
$\delta_0$ [deg] & $-69.25\pm0.02$ & $-69.26\pm0.02$ & $-69.16\pm0.02$ & $-69.23\pm0.02$  & $-69.41$\tablefootmark{d} \\
$\mu_{\alpha}^0$ [mas\,yr$^{-1}$] & $1.88\pm0.01$ & $1.88\pm0.01$ & $1.87\pm0.01$ & $1.88\pm0.01$  & $1.87\pm0.01$\\
$\mu_{\delta}^0$ [mas\,yr$^{-1}$] & $0.32\pm0.01$ & $0.32\pm0.01$ & $0.29\pm0.01$ & $0.35\pm0.01$  & $0.36\pm0.01$\\
$v_{los}^0$ [km\,s$^{-1}$] & $264.83\pm0.16$ & $264.70\pm0.15$ & $266.97\pm0.25$ & $264.42\pm0.23$  & -- \\
$\theta$ [deg] & $25.5\pm0.2$ & $25.2\pm0.2$ & $26.9\pm0.2$ & $25.2\pm0.3$  & $34.0\pm0.1$ \\
$\psi$\tablefootmark{e} [deg] & $124.0\pm0.4$ & $124.4\pm0.5$ & $124.4\pm0.4$ & $122.7\pm0.5$  & $104.2\pm0.2$ \\
$\rm q_{proj}$ & $0.91\pm0.01$ & $0.91\pm0.01$ & $0.89\pm0.01$ & $0.91\pm0.01$  & $0.84\pm0.01$ \\
$\log r_s$ [pc] & $4.02\pm0.02$& $4.30\pm0.03$ & $4.06\pm0.02$ & $3.65\pm0.01$  & $3.66\pm0.01$ \\
$\log\rho_s$ [\Msun\,pc$^{-3}$] & $-1.99\pm0.03$ & $-2.47\pm0.04$ & $-2.05\pm0.02$ & $-1.89\pm0.03$  & $-1.40\pm0.01$ \\
$\kappa$ & $1.01\pm0.01$ & $1.00\pm0.01$ & $1.00\pm0.01$ & $1.02\pm0.01$  & $1.01\pm0.01$ \\
$\beta_z$ & $0.70\pm0.01$ & $0.64\pm0.01$ & $0.89\pm0.01$ & $0.69\pm0.01$  & $0.72\pm0.01$ \\
\hline
   &  \multicolumn{4}{c}{Derived quantities}  \\
\hline
$N_{mem} (P_{mem}>0.8)$ & $28\,000$ & $27\,951$ & $11\,043$ & $16\,621$ & $761\,850$ \\
$\rm q_{intr}$ & $0.23\pm0.01$ & $0.20\pm0.01$ & $0.15\pm0.01$ & $0.26\pm0.01$  & $0.25\pm0.01$ \\
$\rm M_{encl}\tablefootmark{f}$ [$10^{10}$\,\Msun] & $1.36\pm0.01$ & $1.34\pm0.01$ & $1.38\pm0.03$  & $1.30\pm0.02$ & $1.41\pm0.01$ \\
$\rm M_{200}$ [$10^{11}$\,\Msun] & $1.81\pm0.10$ & $4.34\pm0.50$ & $2.06\pm0.25$  & $1.45\pm0.09$ &  $0.75\pm0.01$ \\
$\rm R_{200}$ [kpc] & $75\pm2$ & $156\pm6$ & $79\pm4$  & $68\pm2$ &  $51\pm0.4$ \\
$\rm V_{circ}^{max}$ [\kms] & $115\pm2$ & $126\pm4$ & $118\pm4$  & $109\pm2$ &  $98\pm0.2$ \\
$\rm (M/L)_{centre}$ [\Msun/\Lsun] & $2.17\pm0.03$ & $2.35\pm0.03$ & $2.11\pm0.05$  & $2.33\pm0.05$ & $3.68\pm0.02$ \\
$\rm (M/L)_{encl}\tablefootmark{f}$ [\Msun/\Lsun] & $11.5\pm0.1$ & $10.7\pm0.1$ & $11.7\pm0.2$  & $11.0\pm0.1$ & $11.8\pm0.03$ \\
$\rm (M/L)_{total}$ [\Msun/\Lsun] & $140\pm10$ & $330\pm40$ & $160\pm20$  & $110\pm10$ & $59\pm1$ \\
\hline                  
\end{tabular}
\tablefoot{
\tablefoottext{a}{Distance modulus and corresponding distance from \citet{degrijs+2014, crandall+ratra2015}.}
\tablefoottext{b}{Total luminosity from the NED database.}
\tablefoottext{c}{Radius of the exponential surface brightness profile from \citet{gallart+2004}.}
\tablefoottext{d}{Centre coordinates were kept fix to the estimate by \citet{gaiaedr3}.}
\tablefoottext{e}{Line of nodes orientation angle measured from north to east.}
\tablefoottext{f}{Enclosed mass within the radial extent of the selected kinematic tracers ($6.2$\,kpc)}.
}
\end{table*}

The main difference between the 2D and 3D model results concerns the orientation of the LMC disc in space, its angular momentum vector, and the corresponding viewing angles from Earth.
The difference in the angle $\psi$, which describes the orientation of the rotation axis in the plane of the sky is $20^{\circ}$ ($124^{\circ}$ for the 3D model vs. $104^{\circ}$ for the 2D model).
The difference in the best-fit inclination angle $\theta$ is $8.5^{\circ}\pm0.2^{\circ}$: $\theta = 34.0^{\circ}\pm0.1^{\circ}$ for the 2D model, very close to the \citet{gaiaedr3} estimate of $\theta = 33.3^{\circ}$, vs. $\theta = 25.5^{\circ}\pm0.1^{\circ}$ for the 3D model.
Note, however, that the \citet{gaiaedr3} study makes an assumption for a flat LMC disc, which makes their inclination angle an upper limit.
They find a line of nodes orientation angle $\psi\sim130^{\circ}$, closer to our 3D model result.
The discrepancy between the 2D and 3D results also leads to a significant difference in the projected flattening of the galaxy - $q_{proj} = 0.91\pm0.01$ for the 3D model vs. $q_{proj} = 0.84\pm0.01$ for the 2D model.
Clearly, adding $\rm V_{LOS}$ data to the Jeans model fit of the LMC disc significantly changes how we perceive the galaxy from a purely kinematic perspective.
These results can be compared to photometric constraints, which we do in Sect.~\ref{sec:phot}.
Note that the resulting intrinsic flattening of the LMC disc is essentially estimated to be the same according to the 2D and 3D fits $q_{intr}\sim0.24$.
Previous literature studies put the orientation of the line of nodes at $120^{\circ} - 155^{\circ}$ and the inclination angle in the range of $25^{\circ}-40^{\circ}$ \citep[][and references there in]{vandermarel+kallivayalil2014, gaiaedr3}.
Our 3D kinematics analysis is consistent with these constraints, albeit on the lower end.
On the other hand the 2D kinematic analysis yields a line of nodes orientation angle closer to the photometric major axis and an inclination angle close to the middle of the distribution, but clearly incompatible with the LOS velocity field.
While the 3D velocities provide the most robust kinematic constraint of the galaxy orientation, PMs alone can also provide information about the line of nodes orientation and inclination assuming rotation of a thin axisymmetric disc.
The formalism for this is given in \citet{gaiadr2b}.
In the case of the LMC, it appears that the line of nodes orientation constrained from PMs alone and the axis of maximum $\rm V_{LOS}$ gradient are offset by $20^{\circ}$.
This discrepancy reflect the complex and perturbed nature of the LMC disc.
Furthermore, the mismatch between the orientation of LMC's morphological major axis and the axis of maximum $\rm V_{LOS}$ gradient is a known issue, which prompted \citet{vandermarel+cioni2001, vandermarel+kallivayalil2014} to argue for an elongated non-axisymmetric disc in the LMC.

   \begin{figure*}
   \centering
            {
            \includegraphics[scale=.25]{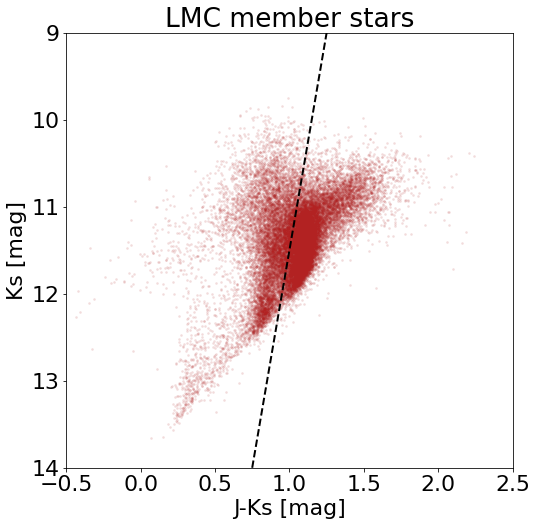}
            \includegraphics[scale=.178]{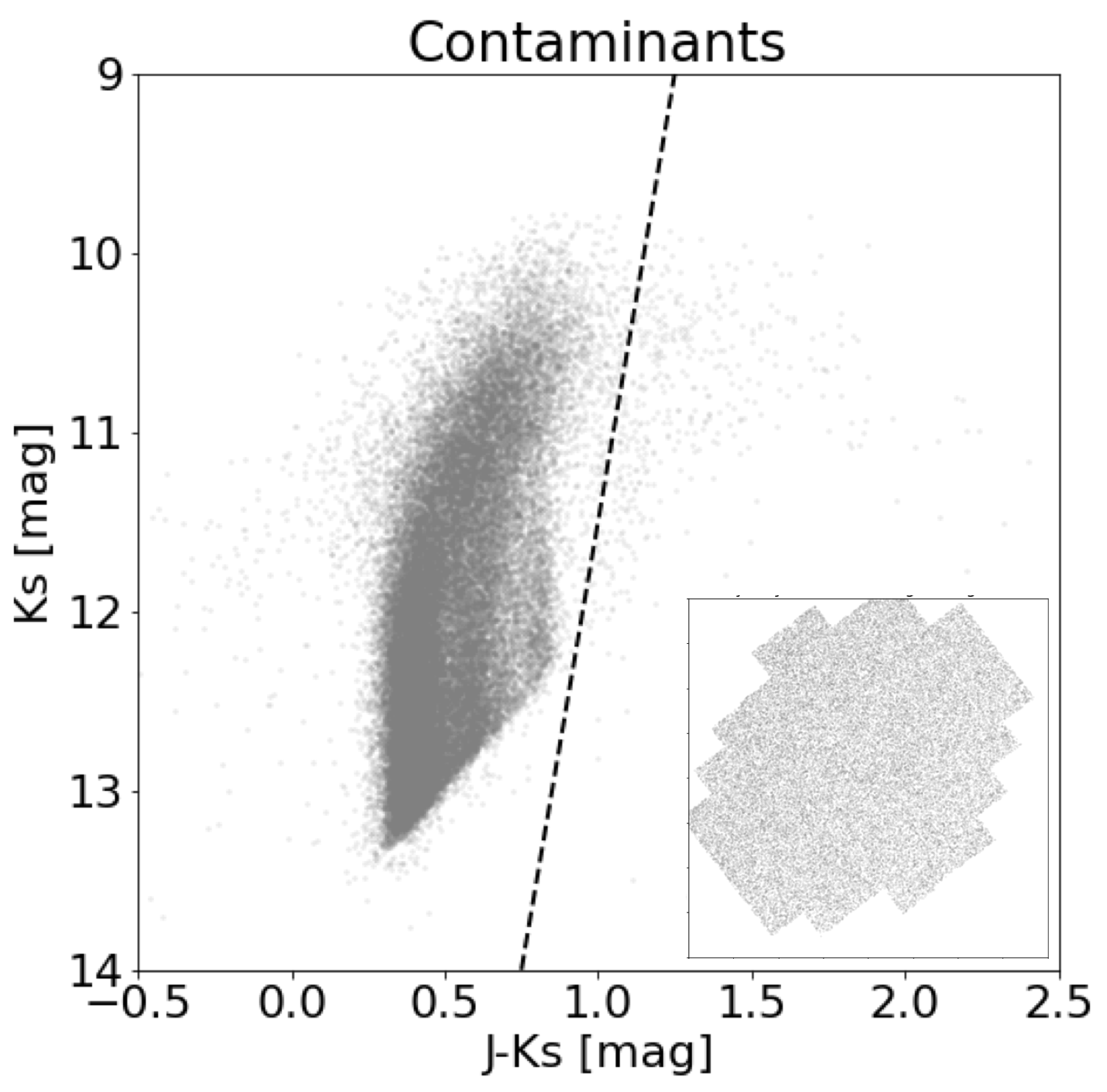}
            }
      \caption{{\em Left panel:} An infra-red CMD from the VMC survey of the high-probability LMC member stars according to the 3D Jeans model fit. {\em Right panel:} An infra-red CMD of the high-probability contaminant sources (foreground Milky Way stars and background galaxies) according to the 3D Jeans model fit. The inlet shows their spatial distribution in the VMC footprint. The dashed line is the same in both panels. It is used to separate the old and young LMC stellar populations.}
         \label{fig:jeans_membership}
   \end{figure*}

The left panel of Fig.~\ref{fig:jeans_membership} shows an infra-red CMD from the VMC survey of all stars in our 3D velocity sample that have a high probability ($P_{mem}>0.8$) of being genuine LMC member stars, according to our 3D axisymmetric Jeans dynamical model fit.
Similarly, the right panel of Fig.~\ref{fig:jeans_membership} shows the infra-red CMD of the high-probability contaminant sources (foreground Milky Way stars and background galaxies).
They have a very flat spatial distribution, as seen from the inlet of the figure, showing that the discrete Jeans model is very successful at separating genuine LMC member stars from foreground / background contamination.
We find $28\,000$ stars with LMC membership probability $>0.8$.
In comparison, applying the \citet{gaiaedr3} membership selection criteria, based on parallax and PM cuts, we are left with $27\,653$ LMC stars, which is an excellent agreement.

In order to investigate whether the difference in the inferred orientation of the LMC, based on the 3D and 2D kinematics, could be alleviated by using a more homogeneous stellar population, we fitted the same Jeans model to 3D kinematic data from only the old stellar population and only the young stellar population, separately.
We used the same photometric cuts as in \citet{elyoussoufi+2019, cioni+2019} in the infra-red VMC CMD to separate the stars into old and young populations (see Fig.~\ref{fig:jeans_membership}).
The young population includes main sequence stars, as well as blue, yellow, and red supergiants, while the old population consists of AGB and upper RGB stars.
The results from these two additional models are also summarised in Table~\ref{tab:jeans}.
They are essentially in line with the model that uses the 3D kinematic data of all stars.

Thus, the significant discrepancy of the LMC viewing angles inferred from the 3D and 2D Jeans axisymmetric models (the former are also in tension with photometric constraints) are not due to stellar population differences, but speak of a non-negligible deviation of the LMC disc from axisymmetry \citep[see also][]{vandermarel+cioni2001, vandermarel+kallivayalil2014}.

   \begin{figure}
   \resizebox{\hsize}{!}
            {
            \includegraphics[width=\hsize]{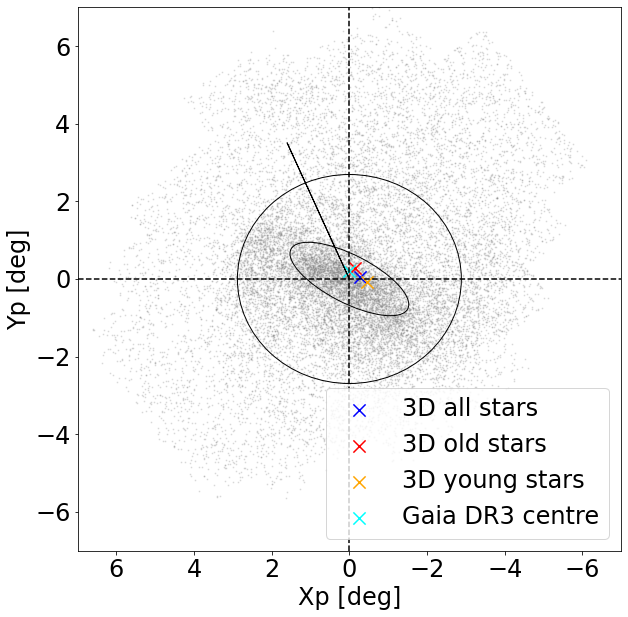}
            }
      \caption{The figure shows the distribution of the high probability LMC member stars according to our 3D Jeans model (grey dots), rotated at angle $\psi=124^{\circ}$, so that the best-fit projected rotation axis is vertical and the major axis horizontal. The best-fit kinematic LMC centres are indicated with crosses, based on all stars with 3D velocities (blue cross), old stars (red cross), young stars (yellow cross), and the centre based on 2D kinematics from \citet[][cyan cross]{gaiaedr3}. The large ellipse indicates the best-fit projected shape of the LMC disc, based on the axisymmetric Jeans model. The small ellipse indicates the shape of the bar, based on a photometric surface density fit (see Sect.~\ref{sec:phot}). The dashed lines indicate the photometric centre, based on the surface density fit. The solid black line shows the bulk PM of the LMC.}
         \label{fig:jeans_morphology}
   \end{figure}

Fig.~\ref{fig:jeans_morphology} shows the spatial distribution of the high-likelihood genuine LMC members with 3D kinematics in an orthogonal coordinates system, where the projected rotation axis is vertical and the line of nodes horizontal.
We show the best-fit ellipse that describes the projection of the assumed axisymmetric LMC disc, together with its bulk PM direction according to the 3D Jeans model.
We show the locations of the inferred kinematic centres from the three Jeans models (based on 3D kinematics of all stars, and the old, and young populations separately), as well as the 2D kinematic centre published in \citet{gaiaedr3}, with respect to the photometric centre inferred from the Gaia DR3 density map of all sources (see Sect.~\ref{sec:phot}).
We find that the distance between the 3D kinematic centre inferred from all stars and the photometric centre is $16.6$\,arcmin.
The 2D kinematic centre is closest to the photometric centre with an angular separation of $11.7$\,arcmin.

   \begin{figure*}
   \centering
            {
            \includegraphics[width=6cm]{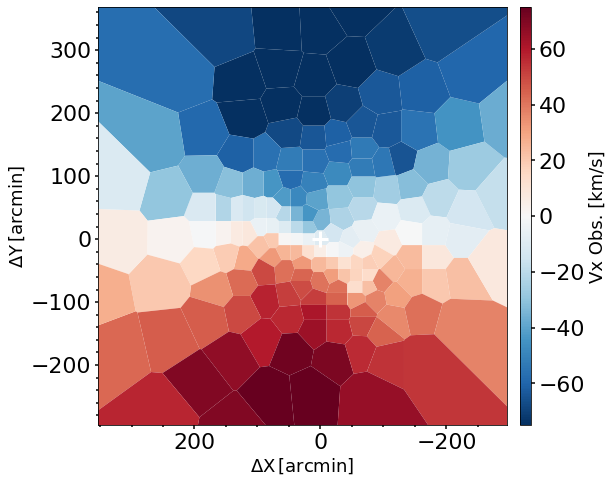}
            \includegraphics[width=6cm]{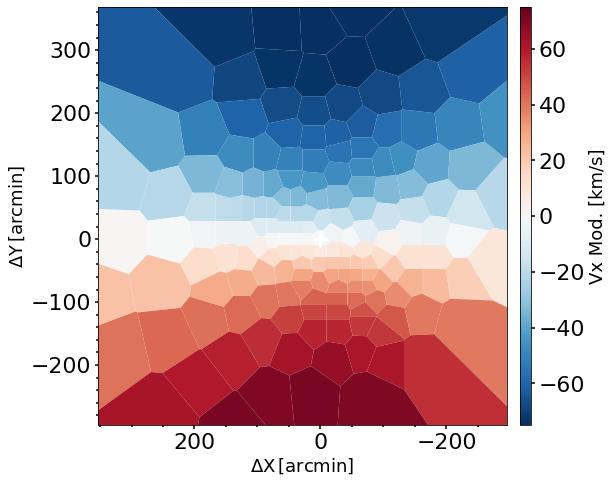}
            \includegraphics[width=6cm]{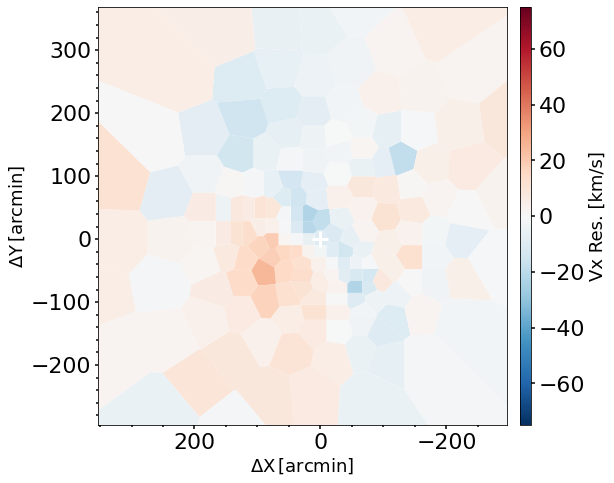}
            \includegraphics[width=6cm]{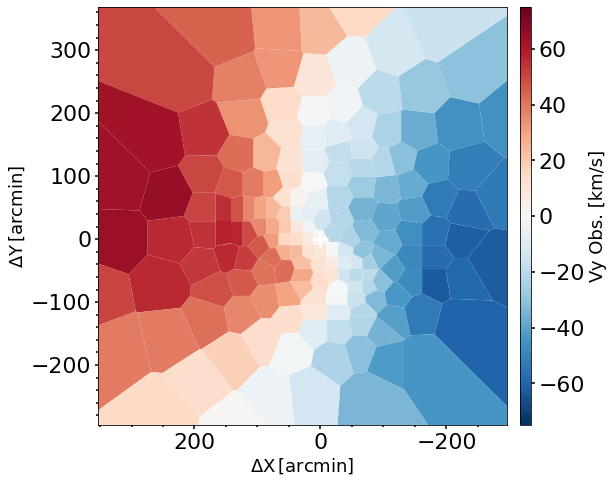}
            \includegraphics[width=6cm]{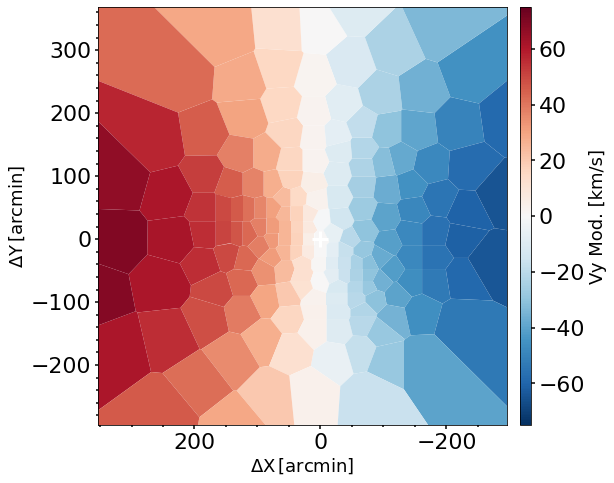}
            \includegraphics[width=6cm]{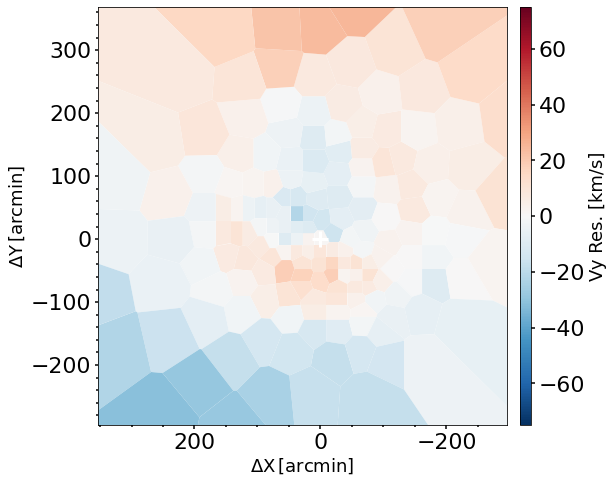}
            \includegraphics[width=6cm]{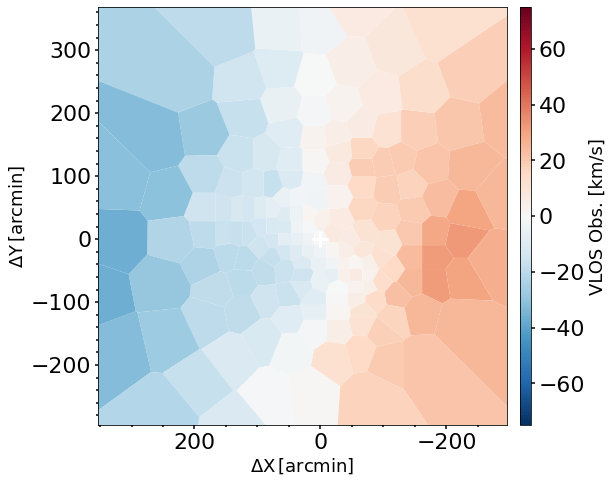}
            \includegraphics[width=6cm]{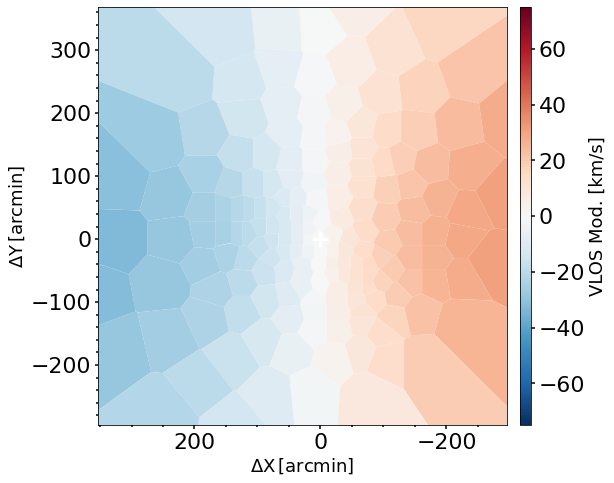}
            \includegraphics[width=6cm]{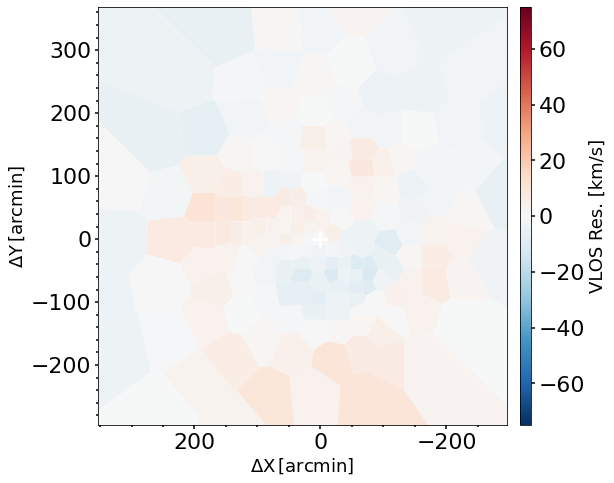}
            }
      \caption{Voronoi maps showing the 3D Jeans model fit to the LMC rotation field from the Gaia DR3 kinematic data. The first column represents the data in the X, Y, and LOS directions, respectively. Each Voronoi bin is colour-coded according to the median velocity of all stars that fall into it. The second column represents the Jeans model results and the third column the model residuals. PMs are converted into linear velocities. The velocity scale is kept unchanged between all maps for consistency and easier comparison.}
         \label{fig:jeans_rot}
   \end{figure*}
   
      \begin{figure*}
   \centering
            {
            \includegraphics[width=6cm]{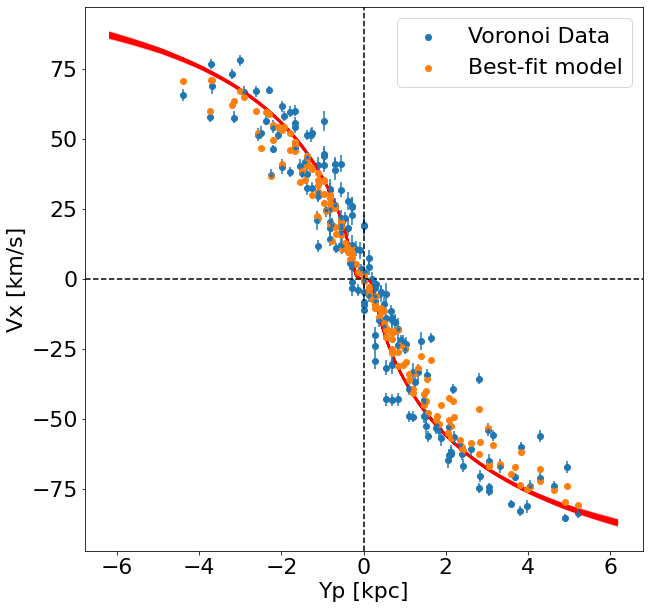}
            \includegraphics[width=6cm]{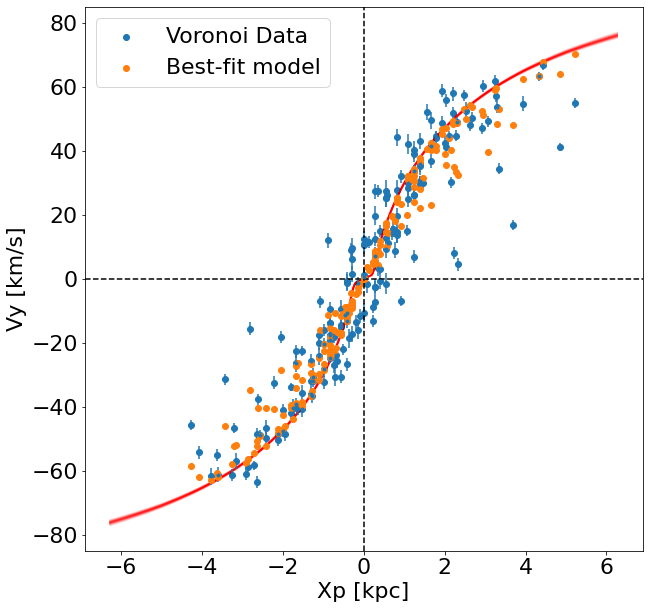}
            \includegraphics[width=6cm]{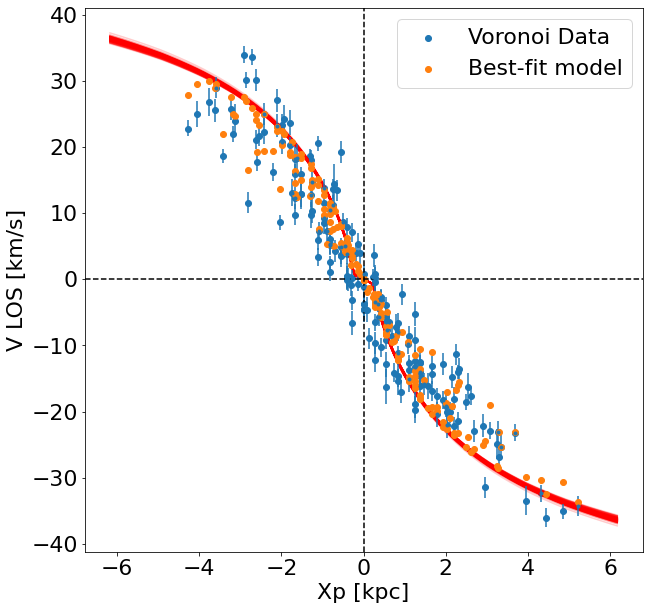}
            }
      \caption{Projected rotation curve of the LMC in the three cardinal directions. The blue points show the maximum likelihood mean velocities in each of the Voronoi bins defined in Fig.~\ref{fig:jeans_rot}. The yellow points show the corresponding mean velocities of the best-fit 3D Jeans model in the same Voronoi bins. The read lines show 150 random draws from the model posterior distribution along the horizontal axis (vertical axis in the case of $\rm V_x$).}
         \label{fig:rotation_profile}
   \end{figure*}
   
      \begin{figure*}
   \centering
            {
            \includegraphics[width=6cm]{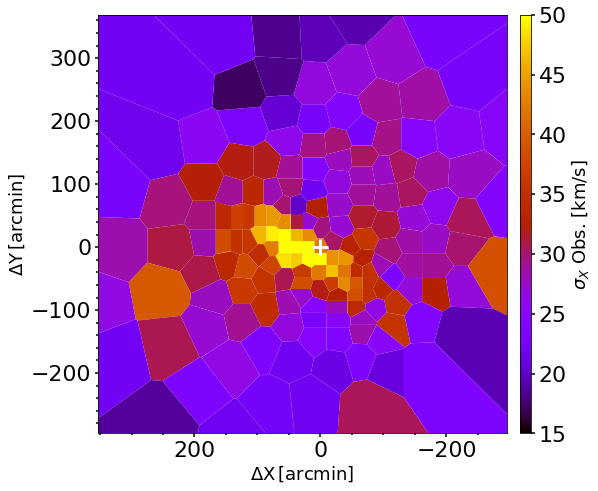}
            \includegraphics[width=6cm]{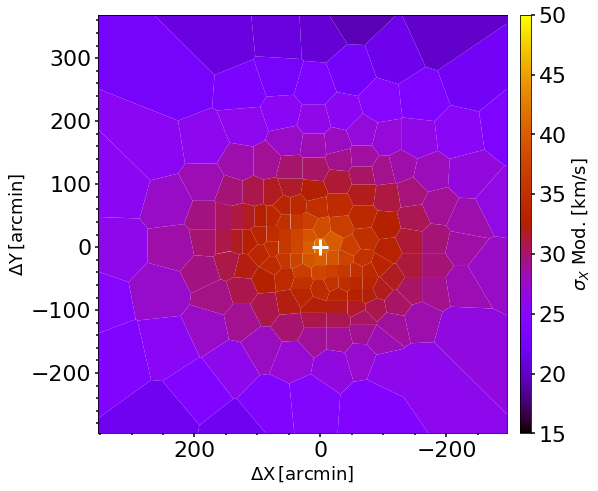}
            \includegraphics[width=6cm]{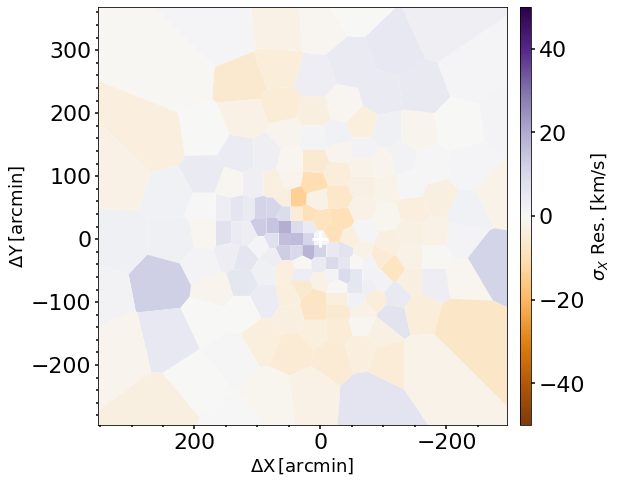}
            \includegraphics[width=6cm]{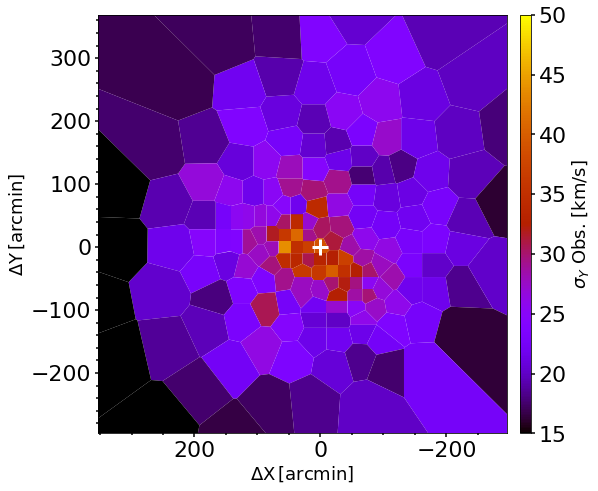}
            \includegraphics[width=6cm]{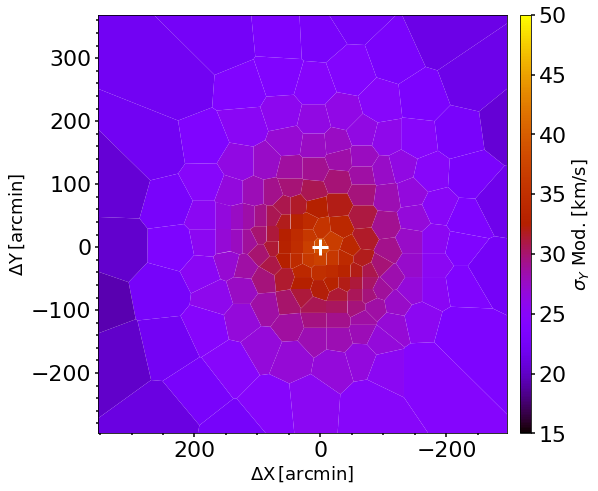}
            \includegraphics[width=6cm]{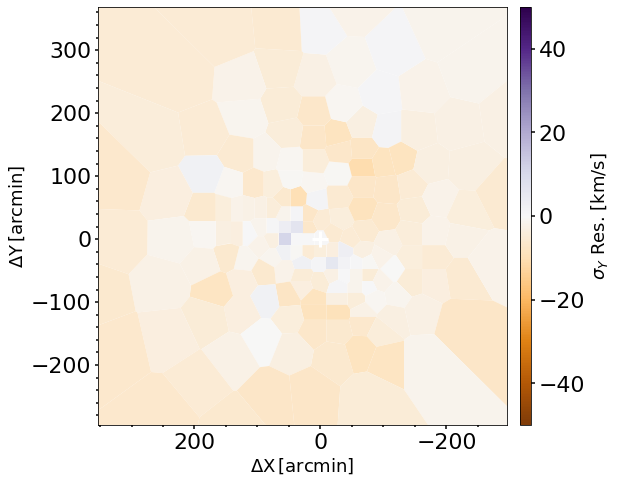}
            \includegraphics[width=6cm]{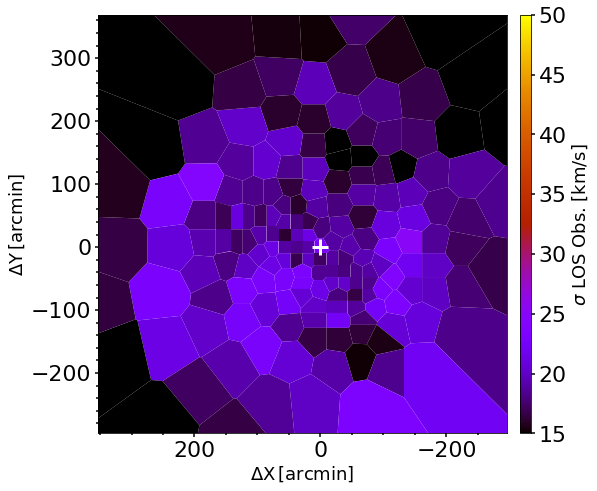}
            \includegraphics[width=6cm]{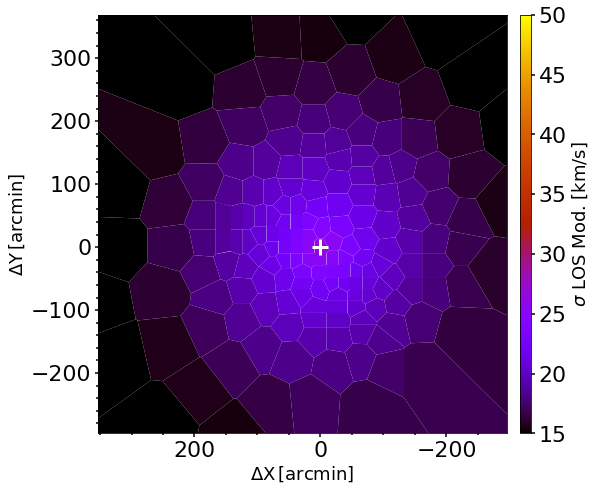}
            \includegraphics[width=6cm]{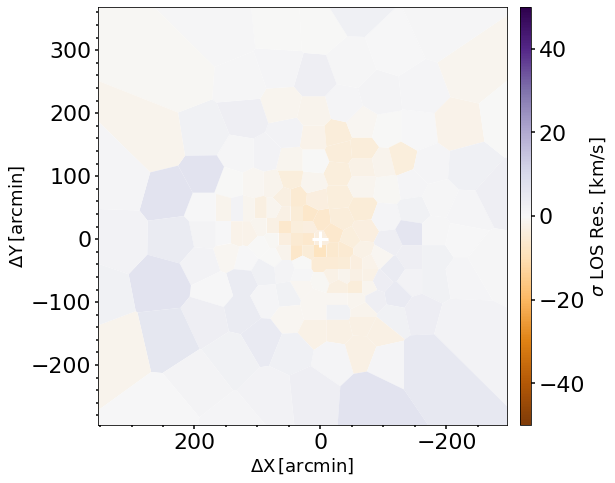}
            }
      \caption{Voronoi maps showing the 3D Jeans model fit to the LMC velocity dispersion field from the Gaia DR3 kinematic data. The first column represents the data in the X, Y, and LOS directions, respectively. Each Voronoi bin is colour-coded according to the maximum likelihood velocity dispersion of all stars that fall into it, taking into account their individual velocity uncertainties. The second column represents the Jeans model results and the third column the model residuals. PMs are converted into linear velocities. The velocity scale is kept constant among all maps for consistency and easier comparison.}
         \label{fig:jeans_disp}
   \end{figure*}

      \begin{figure*}
   \centering
            {
            \includegraphics[width=6cm]{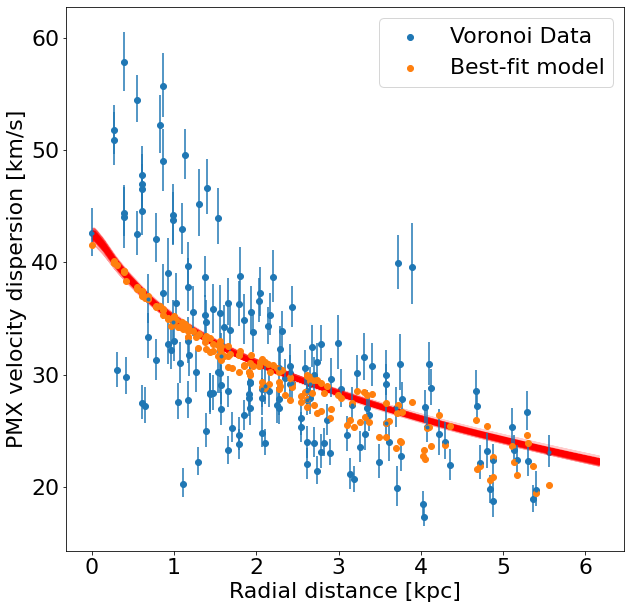}
            \includegraphics[width=6cm]{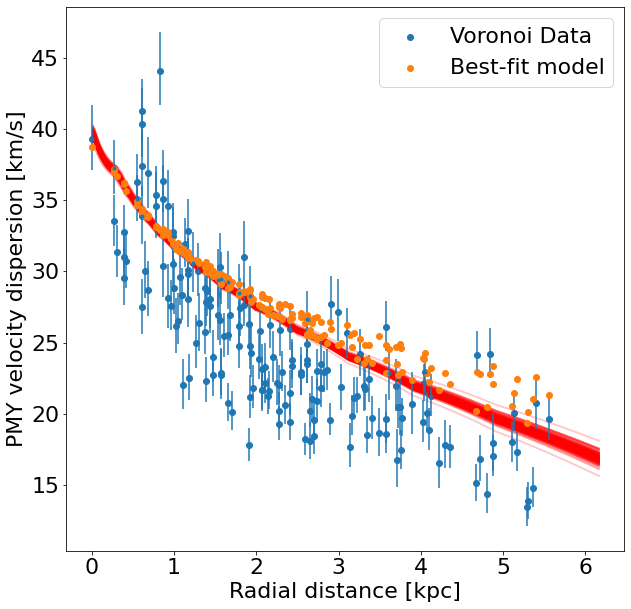}
            \includegraphics[width=6cm]{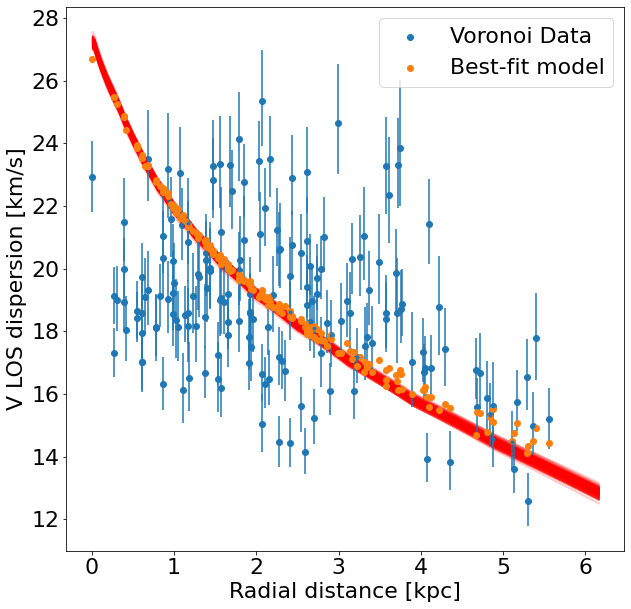}
            }
      \caption{Projected velocity dispersion profile of the LMC in the three cardinal directions. The blue points show the maximum likelihood velocity dispersion in each of the Voronoi bins defined in Fig.~\ref{fig:jeans_disp}. The yellow points show the corresponding velocity dispersions of the best-fit 3D Jeans model in the same Voronoi bins. The read lines show 150 random draws from the model posterior distribution along the horizontal axis.}
         \label{fig:dispersion_profile}
   \end{figure*}

\subsection{Model results: LMC kinematics}

While the Jeans model fits were performed on the discrete individual velocity measurements, we show the model performance using binned Voronoi maps of the Gaia DR3 kinematic data.
We divided our data in Voronoi bins, each containing 150 stars with 3D kinematic information and converted the PMs to linear velocities using the same adopted distance to the LMC throughout this article.
We also subtracted the systemic PM of the LMC, as fitted by our model.
For each bin we inferred the mean velocity, velocity dispersion, and their uncertainties in the three cardinal directions using a maximum likelihood approach that takes into account the individual velocity errors.

Fig.~\ref{fig:jeans_rot} shows the observed projected 3D rotation field of the LMC from the Gaia DR3 data, together with the results from our best-fit Jeans model, and the model residuals.
We kept the velocity scale consistent in all maps for easier comparison of the results.
There are several things worth noting from these kinematic maps.
The Jeans model fits very well the projected rotation of the LMC disc in all three directions, as shown by the residual maps, which are very flat and do not show much systematic structure.
In the central region, however, the data shows a noticeable twist in the rotation axis, aligned with the LMC bar, which is not caught by our axisymmetric model, by definition.
Most of the rotation happens in the plane of the sky, as evident from the large amplitude difference between the LOS and the $\rm V_X, V_Y$ velocity fields, confirming the smaller inclination angle that we find in this analysis, compared to other studies, based only on PM data, e.g. \citet{gaiaedr3}.
We plot the projected rotation curve of the LMC in Fig.~\ref{fig:rotation_profile}.
The rotation amplitude in the plane of the sky reaches $\sim75$\,\kms~at $6$\,kpc from the centre in the plane of the sky, while along the LOS direction it is only $\sim35$\,\kms at the same distance.
We sample the rotation curve from the Jeans model posterior distribution (shown with red lines in Fig.~\ref{fig:rotation_profile}) to find that the Gaia DR3 data and its uncertainties provide very tight constraints to the Jeans solution with almost no room for variation within the model limits.
 
Analogically, Fig.~\ref{fig:jeans_disp} shows the Jeans model match to the projected velocity dispersion map of the LMC in the three cardinal directions and Fig.~\ref{fig:dispersion_profile} shows the projected velocity dispersion profiles as a function of radius.
Again we have kept the velocity dispersion range consistent in all panels of Fig.~\ref{fig:jeans_disp} for better clarity.
The observational maps are quite anisotropic and show different velocity dispersion in the three directions with the $\rm X_p$ having the highest dispersion in the central region and the LOS direction - the lowest.
The Jeans model tries to accommodate this with a relatively high anisotropy parameter $\beta_z \propto 1 - \frac{\sigma_z^2}{\sigma_R^2} = 0.7\pm0.01$.
The highest velocity dispersion in the plane of the sky is observed along the LMC bar, which is a non-axisymmetric feature and cannot be precisely reproduced by our Jeans model.
The Gaia DR3 data shows that the velocity dispersion can reach above $50$\,\kms~in the plane of the sky, while only reaching $\sim24$\,\kms~in the LOS direction.
Our best-fit Jeans model predicts central velocity dispersions $\rm \sigma_X = 42$\,\kms, $\sigma_Y = 40$\,\kms, and $\sigma_{LOS} = 28$\,\kms.

\subsection{Model results: LMC potential}

      \begin{figure*}
   \centering
            {
            \includegraphics[width=6cm]{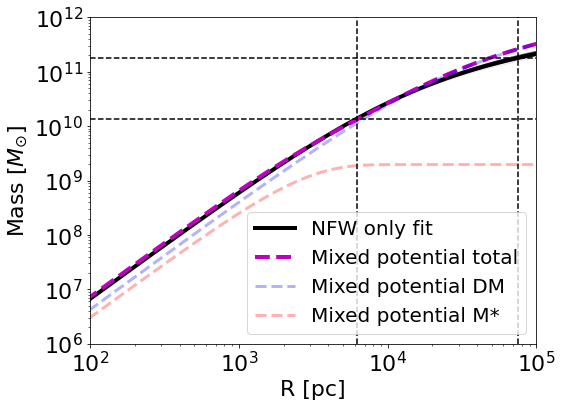}
            \includegraphics[width=6cm]{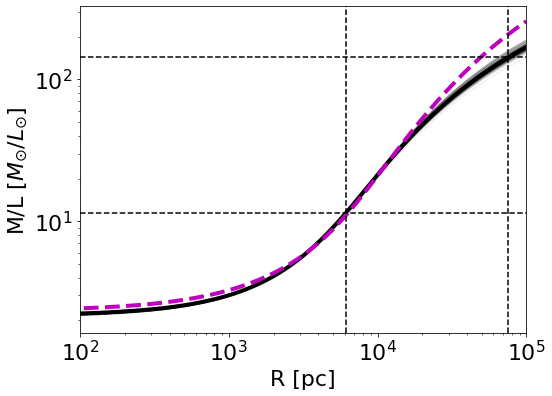}
            \includegraphics[width=6cm]{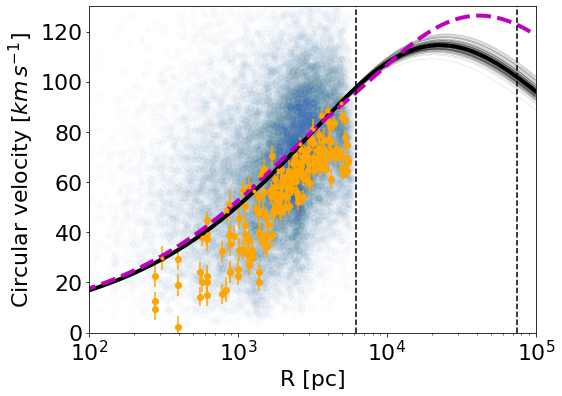}
            }
      \caption{Cumulative mass ({\em left panel}), mass-to-light ratio ({\em middle panel}), and rotation curve ({\em right panel}) of the LMC according to best-fit NFW only (black) and mixed (magenta) gravitational potentials of the axisymmetric Jeans model. The vertical dashed lines denote the extent of the kinematic tracers and the viral radius ($\rm R_{200}$) and the horizontal lines denote the enclosed mass and mass-to-light ratio within these radii. In the rotation curve plot, we overlay the total mean velocity of each Voronoi bin defined in Fig.~\ref{fig:jeans_rot} as orange dots, as well as the total velocity of all LMC member stars as a transparent blue cloud.}
         \label{fig:jeans_mass}
   \end{figure*}

Finally, we discuss the gravitational potential of the LMC, as constrained by the best-fit NFW only gravitational potential, characterised by $\rm \rho_s = 1\pm0.05\times10^{-2}\,M_{\odot}\,pc^{-3}$ and $r_s = 10.5\pm0.4$\,kpc and the mixed potential (visible $+$ DM mass), characterised by $\rm \rho_s = 3.4\pm0.3\times10^{-3}\,M_{\odot}\,pc^{-3}$, $r_s = 20.0\pm1.5$\,kpc, and a fixed $\rm M_*/L = 1.5$\,\MLsun.
Note that if we let the mass-to-light ratio to be a free parameter in our MCMC fit, it tends to go to $0$, so the overall best solution is the DM only model.
Fig.~\ref{fig:jeans_mass} shows the derived cumulative mass, mass-to-light (M/L) ratio, and the circular velocity curve according to our Jeans dynamical models with both types of gravitational potential.
The most distant kinematic tracers in our data selection are at a projected distance of $6.2$\,kpc from the centre of the LMC and both potentials have virtually undistinguishable mass properties within this radial extent.
We find an enclosed mass of $\rm M_{encl} = 1.36\pm0.01\times10^{10}\,M_{\odot}$ for the DM only model vs. $\rm M_{encl} = 1.34\pm0.01\times10^{10}\,M_{\odot}$ for the model with a mixed potential (see Table~\ref{tab:jeans}).

The difference between both choices of gravitational potentials becomes, however, noticeable if we extrapolate them beyond the extent of the kinematic tracers.
The DM only model has a virial mass $\rm M_{200} = 1.81\pm0.01\times10^{11}\,M_{\odot}$ at a virial radius $\rm r_{200} = 75\pm2$\,kpc.
This virial mass estimate is well in line with other independent LMC mass estimates from literature \citep[see][]{vasiliev2023}, albeit on the higher end, and matches precisely the estimate by \citet{watkins+2024}.
The latter study and our models assume the same NFW shape of the gravitational potential, but use kinematic tracers at different radial scales.
While we focus on the inner region stellar kinematics of the LMC disc and bar, \citet{watkins+2024} use halo GCs out to $13.2$\,kpc.
The fact that both studies find the same estimate for the LMC virial mass is astounding and shows that the NFW DM density approximation is reasonable at these radial scales.
But these total LMC mass estimates should not be taken at face value and must be treated with caution.
The Jeans model with a mixed potential yields a significantly higher virial mass of $\rm M_{200} = 4.3\pm0.5\times10^{11}\,M_{\odot}$ at a virial radius $\rm r_{200} = 156\pm6$\,kpc, clearly showing the dangers of extrapolating these results beyond the extent of the kinematic tracers.
To accommodate for the visible mass component that follows the adopted exponential distribution of the LMC surface brightness, the NFW halo in the mixed potential has a lower central density, but is significantly more extended than in the DM only case.
Both virial radius estimates are significantly larger than the LMC - Milky Way distance.
Our mass estimates are mostly driven by the strong assumption that the DM density of the LMC follows a strict NFW profile, which is most certainly not true that far out.

On the other hand, the virial mass derived from our DM only 2D Jeans model is about twice lower than from the DM only 3D Jeans model \citep[see Table~\ref{tab:jeans} and ][]{kacharov+cioni2023}, putting it on the lower end of LMC independent mass estimates from the literature.
The enclosed mass within the maximum radial extent of our kinematic tracers is much more reliable and the numbers are in a reasonable agreement between all five Jeans models that are outlined in Table~\ref{tab:jeans}.

The unrealistically small statistical uncertainties are due to the good constraints that the kinematic tracers provide to the scale density and scale radius of the NFW profile, but they are also very model dependent and the systematic errors from the choice of the potential and surface brightness parametrisation will dominate.
One could also see in Table~\ref{tab:jeans}, that changing the number and type of kinematic tracers will change the derived parameters more than their formal $1\,\sigma$ statistical uncertainties.

In the middle panel of Fig.~\ref{fig:jeans_mass} we plot the total M/L ratio as a function of radius for both types of gravitational potentials.
The total M/L ratio is computed by dividing the cumulative total mass by the cumulative brightness.
As mentioned in the previous section, we use an exponential surface brightness profile with a fixed total luminosity of $1.3\times10^9$\,\Lsun.
We have a central $\rm M/L = 2.17\pm0.03$\,\Msun/\Lsun~for the DM only fit, which is typical for an old stellar population, but there is no distinction between stellar and DM contribution in this model.
The mixed potential has a central $\rm M/L = 2.35\pm0.03$\,\Msun/\Lsun, which shows that the DM dominates the potential under the model assumption of a stellar $\rm M_*/L = 1.5$\,\Msun/\Lsun.
The M/L ratio gradually rises reaching $\sim11$\,\Msun/\Lsun~at $6.2$\,kpc - the maximum radial extent of our kinematic tracers.
This radius contains $93\%$ of the total LMC luminosity according to our exponential parametrisation.
At the virial radius the M/L ratio is $140\pm10$\,\Msun/\Lsun~according to our fiducial DM only parametrisation, emphasising that the LMC resides in a very massive DM halo.

In the right panel of Fig.~\ref{fig:jeans_mass} we plot the circular velocity curve of the LMC, as computed from the best-fit NFW only profile, which reaches a maximum of $115\pm2$\,\kms~and $90$\,\kms~at $5$\,kpc in excellent agreement with the result from \citet{vasiliev2018}.
The resulting rotation curve from the mixed potential follows very closely the fiducial DM only one within the extent of our kinematic tracers, but extrapolated outwards, it reaches a maximum of $126\pm4$\,\kms~with the caveats discussed above.
We compare the rotation curve to the total mean velocities of each of the Voronoi bins defined in Fig.~\ref{fig:jeans_rot}.
They all follow tightly the circular velocity curve and lie on, or under it if the Voronoi bin is far from the line of nodes (maximum velocity gradient).
This assures us that the NFW parametrisation of the gravitational potential is adequate at least within the radial extent of the kinematic data.

Overall the 3D Jeans model fits provide a good description of LMC disc's kinematics and the galaxy's fundamental dynamical parameters, but to do better in the central region we need to work with triaxial models, which can describe the kinematics of the LMC bar.

\section{Probing the axisymmetry of the LMC from 3D kinematics}

   \begin{figure*}
   \centering
            {
            \includegraphics[width=5.66cm]{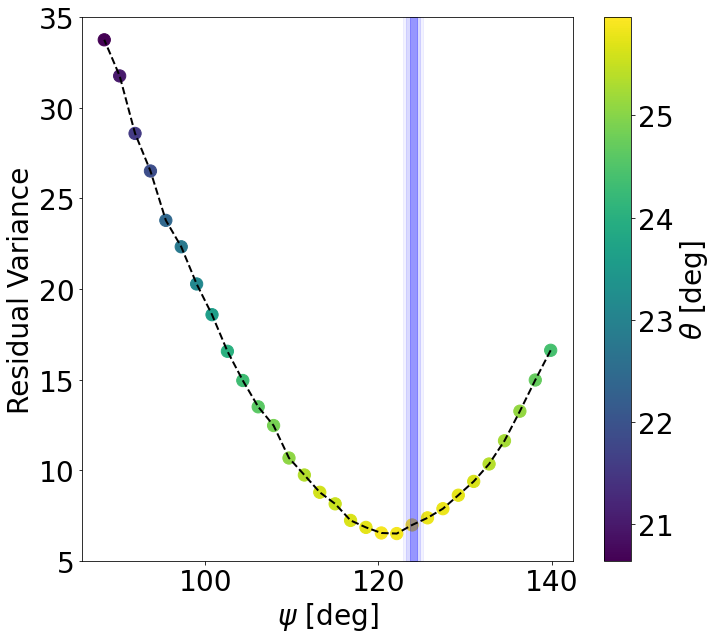}
            \includegraphics[width=5.66cm]{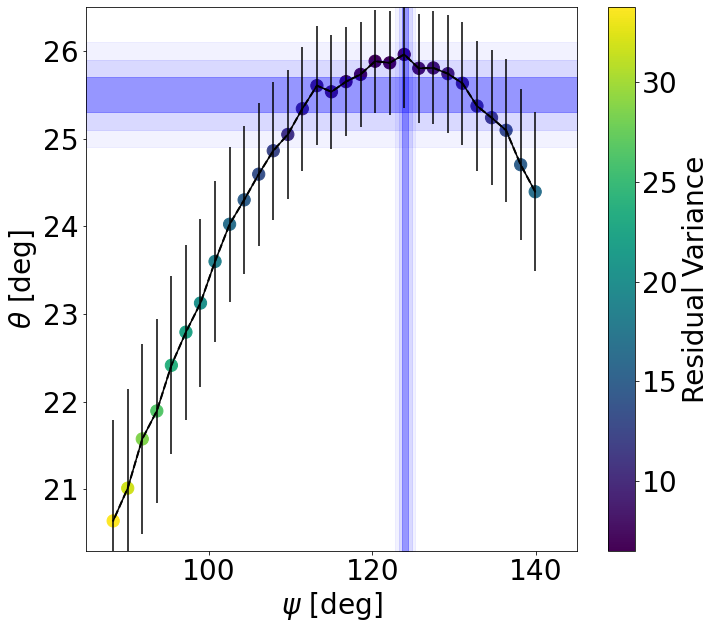}
            \includegraphics[width=6.05cm]{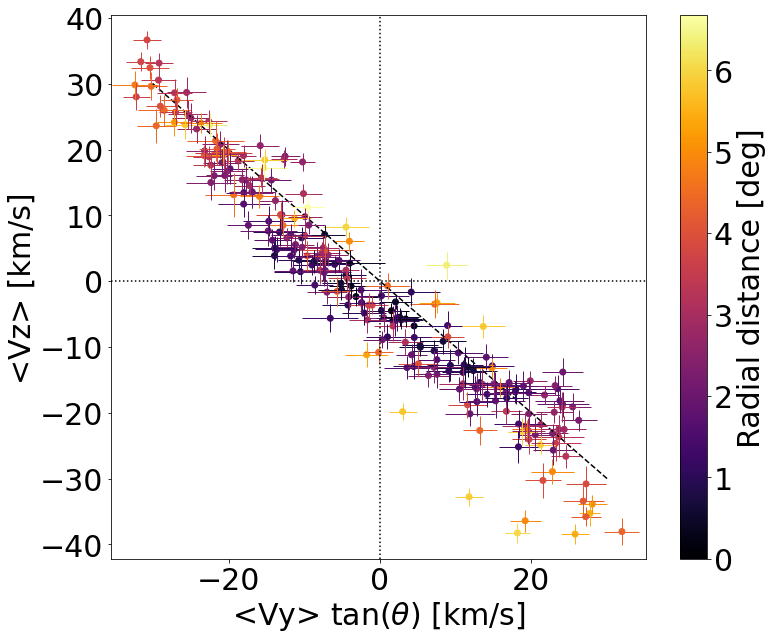}
            \includegraphics[width=5.66cm]{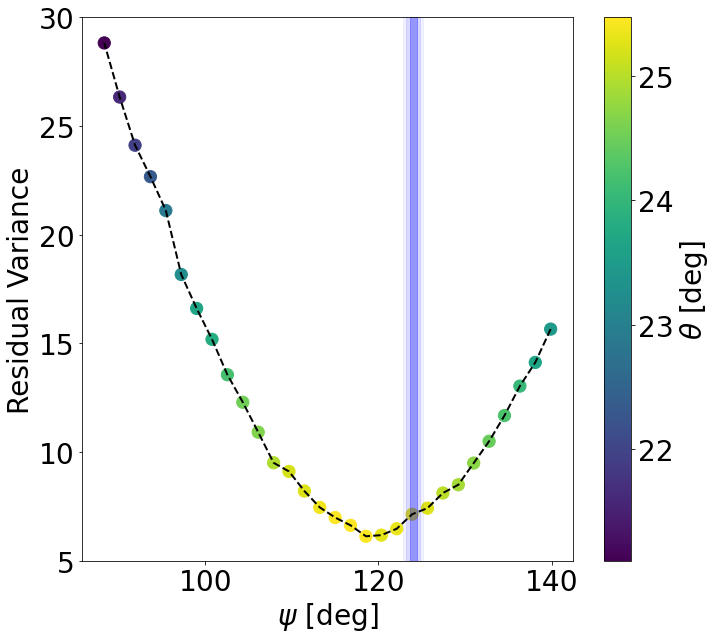}
            \includegraphics[width=5.66cm]{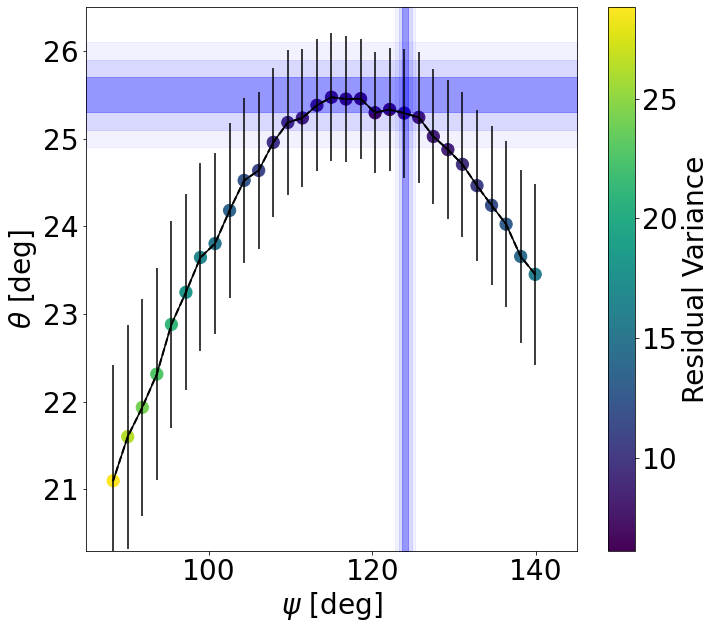}
            \includegraphics[width=6.05cm]{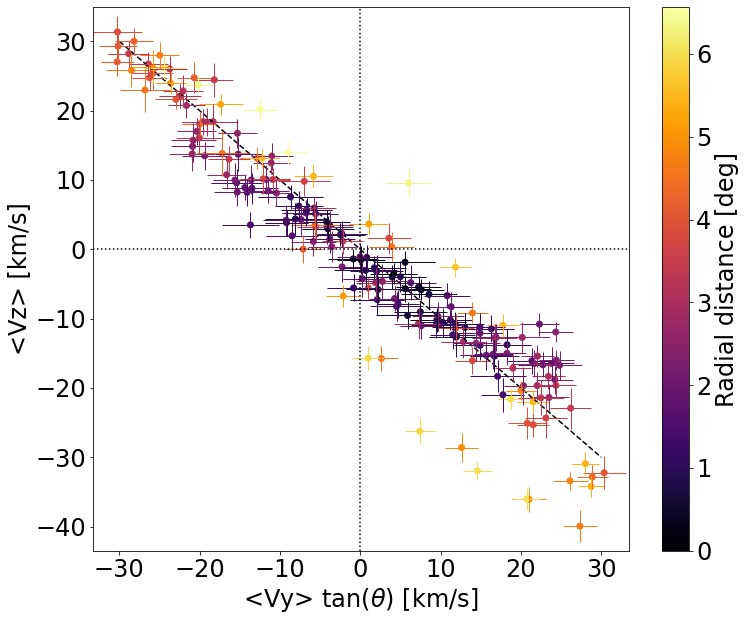}
            \includegraphics[width=5.66cm]{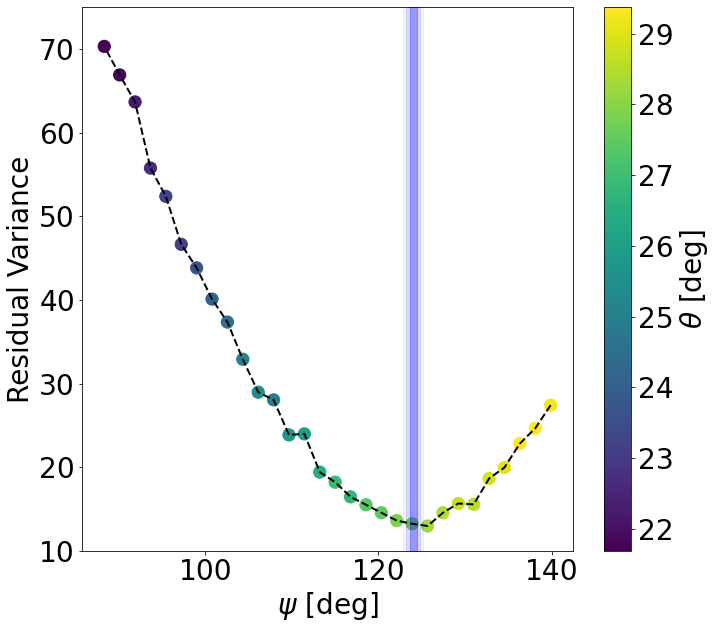}
            \includegraphics[width=5.66cm]{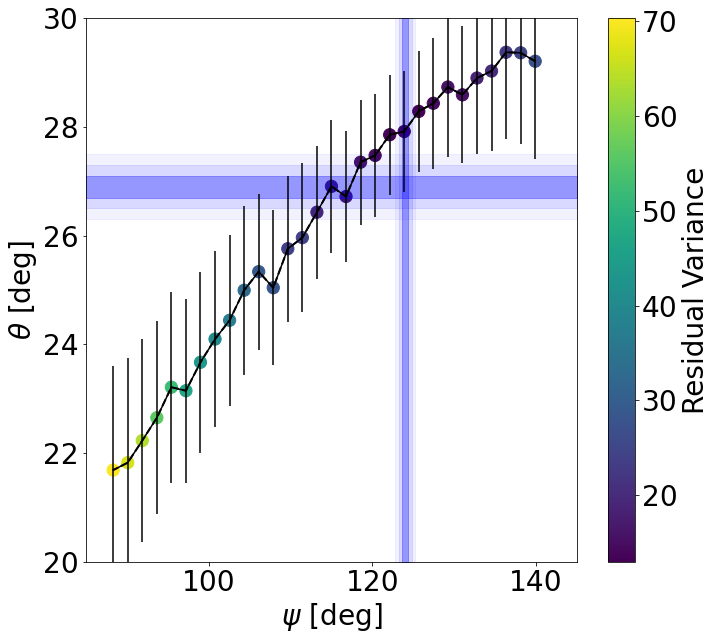}
            \includegraphics[width=6.05cm]{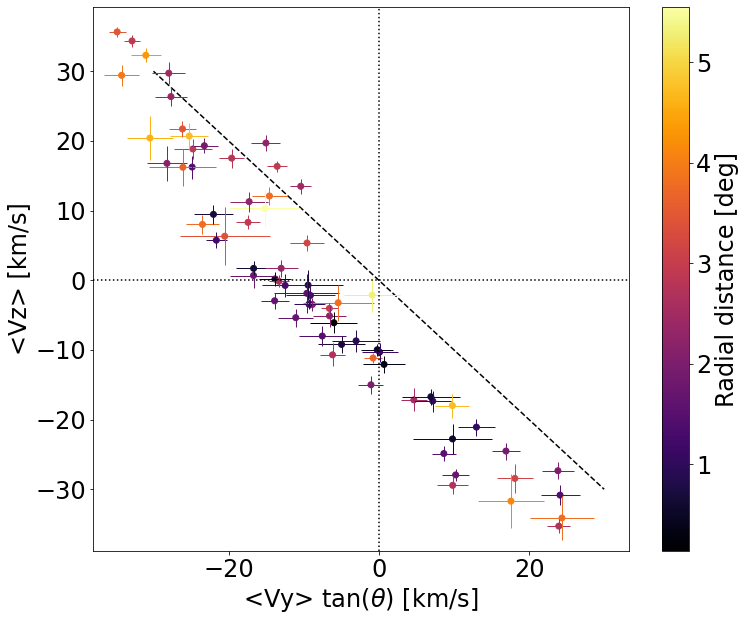}            
            }
      \caption{{\em Left panels:} residual variance of the mean $\rm V_{LOS}$ vs. the mean $\rm V_Y$ correlation for different line of nodes orientations, colour coded by the inclination angle for all stars ({\em top}), old stars ({\em middle}), and young stars ({\em bottom}) rows, respectively. {\em Middle panels:} line of nodes orientation vs. inclination angle, colour coded by the residual variance for the old stars and separated into old and young populations. The shaded regions in these panels correspond to the best fit values from the 3D Jeans model and their uncertainties. {\em Right panels:} mean $\rm V_{LOS}$ vs. mean $\rm V_Y\tan(\theta)$ for the LMC orientation with minimum residual variance for each respective population. These plots are colour-coded by distance from the centre of each Voronoi bin and the dashed lines indicate a perfect anti-correlation.}
         \label{fig:vz_vy_cor}
   \end{figure*}   

We can probe the overall axisymmetry of the LMC by examining the correlation between the mean velocities in the LOS and Y directions.
In an axisymmetric system they are related via the equality $\rm \langle V_{LOS}\rangle = -\langle V_Y\rangle\,\tan\theta$ \citep{vandeven+2006}.
We measure these quantities in Voronoi bins, similar to the ones shown in Fig.~\ref{fig:jeans_rot} and Fig.~\ref{fig:rotation_profile}, which are for one particular orientation - the one derived from the 3D Jeans dynamical model.
We estimate the residual variance between these two velocity components by fitting a linear function, assuming different line of node orientations (Fig.~\ref{fig:vz_vy_cor}).
The slope of the line corresponds to the inclination angle of the galaxy.
We do this exercise for all LMC member stars, as well as for the old and young LMC populations separately, according to the age separation defined in Fig.~\ref{fig:jeans_membership}.
We find a best fit line of nodes orientation angle $\psi = 122^{\circ}$ with a corresponding inclination angle $\theta=26^{\circ}$, when considering all LMC member stars.
The old population yields $\psi = 120^{\circ}$ and $\theta=25.5^{\circ}$, while for the young stars we find $\psi = 126^{\circ}$ and $\theta=28^{\circ}$.
These estimates are practically identical to the results obtained from the 3D Jeans model.

The resulting correlation between the mean $\rm V_{LOS}$ and the mean $\rm V_Y$ at the estimated line of nodes angle for the three cases that minimises the scatter is also shown in Fig.~\ref{fig:vz_vy_cor}.
We colour-coded the separate velocity Voronoi bins by their distance from the galaxy's centre to check whether the inner bins, most influenced by the kinematic effects of the bar, are outliers in this plot.
We do notice that the most central regions (dominated by the bar) are systematically offset from the line for all stars, while this effect is almost entirely mitigated, when considering only the old population.
On the other hand, the young population shows the largest offset from the perfect correlation line, which indicates a more significant deviation from axisymmetry.
This is not surprising, given the non-uniform and patchy spatial distribution of these young stars, which have likely not reached an equilibrium state.

In addition, the most distant Voronoi bins from the LMC centre in our footprint, located on the approaching side of the disc (and on the side of the SMC) also show an increased scatter in the old stellar population, likely an effect from the interaction between the three galaxies.

Overall, we do not see a distinct effect induced from the non-axisymmetry of the LMC bar in these plots.
The bar is prominent in both the old and young stellar populations.
The kinematic effect of the bar can mostly be seen when looking directly at the Voronoi maps in Fig.~\ref{fig:jeans_rot}, as a twist in the velocity gradient direction in the bar region.
This effect, however, appears to follow the same pattern in both the $\rm \langle V_{LOS}\rangle$ and $\langle V_Y\rangle$ Voronoi maps, keeping the correlation between them.



\section{De-projecting the LMC surface density distribution assuming a triaxial bar}\label{sec:phot}

The goal of this study is to at least partially abandon the common assumption for axisymmetry when it comes to the dynamical modelling of the LMC and include a triaxial bar, along with an axisymmetric disc.
To achieve this, we need a new parametrisation of the LMC surface brightness distribution, which in the case of the Jeans models, was assumed to be purely exponential.
Here we provide a new two-component photometric decomposition of the Gaia DR3 stellar density map of the LMC\footnote{The Gaia DR3 stellar density map of the LMC was obtained as fits image from \url{
https://alasky.cds.unistra.fr/ancillary/GaiaEDR3/density-map/}. Copyright: ESA/Gaia mission/DPAC (\url{https://dx.doi.org/10.5270/esa-1ugzkg7}).} \citep{gaiaedr3_main} using {\sc galfit}\footnote{\url{https://users.obs.carnegiescience.edu/peng/work/galfit/galfit.html}} \citep{peng+2010}, which we use to de-project the galaxy and calculate its viewing angles.

\subsection{Photometric decomposition of the LMC}\label{sec:ecomposition}

\begin{figure*}
   \centering
            {
            \includegraphics[width=4.5cm]{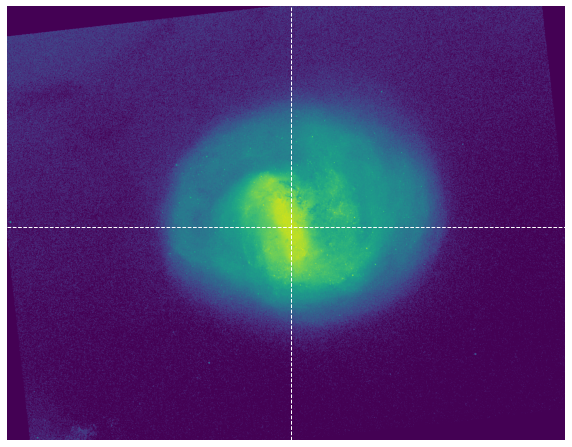}
            \includegraphics[width=4.5cm]{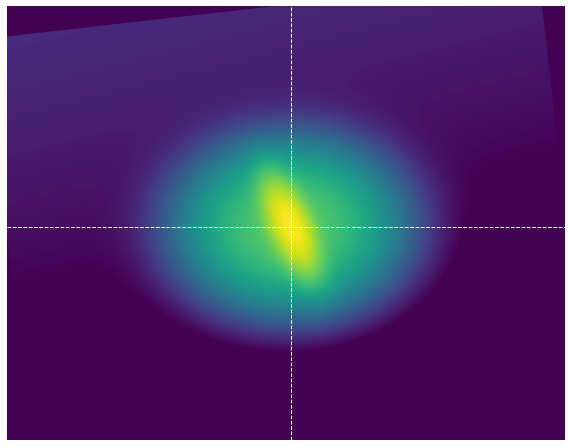}
            \includegraphics[width=4.5cm]{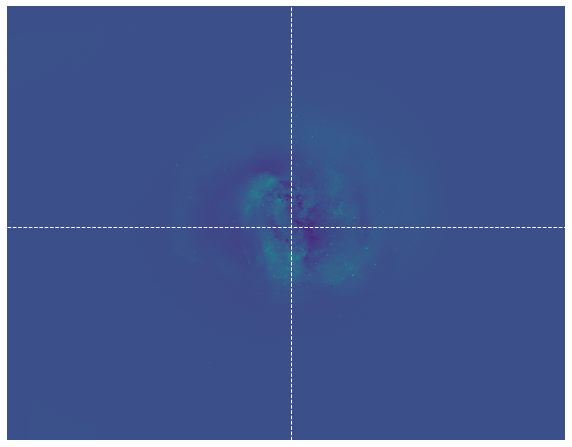}
            \includegraphics[width=4.1cm]{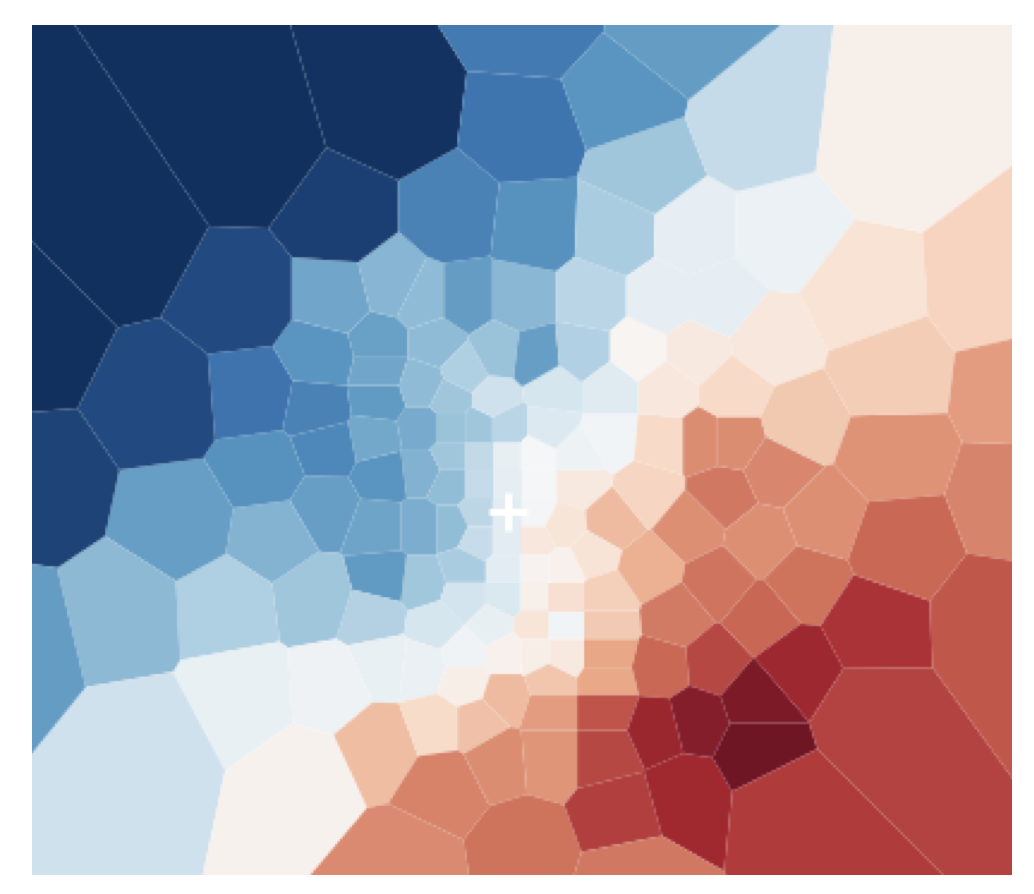}
            \includegraphics[width=4.5cm]{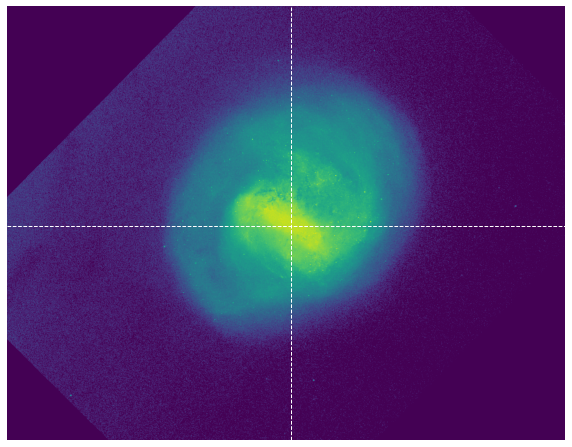}
            \includegraphics[width=4.5cm]{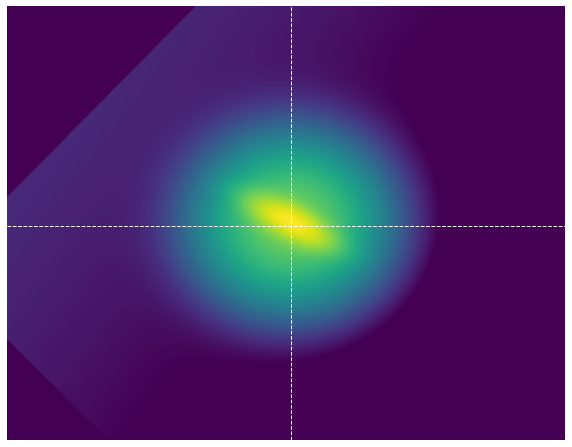}
            \includegraphics[width=4.5cm]{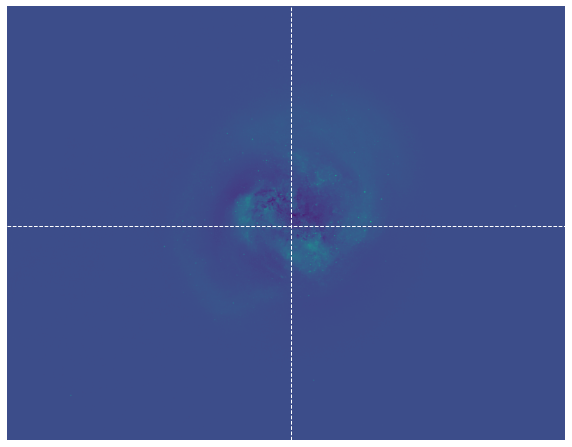}
            \includegraphics[width=4.1cm]{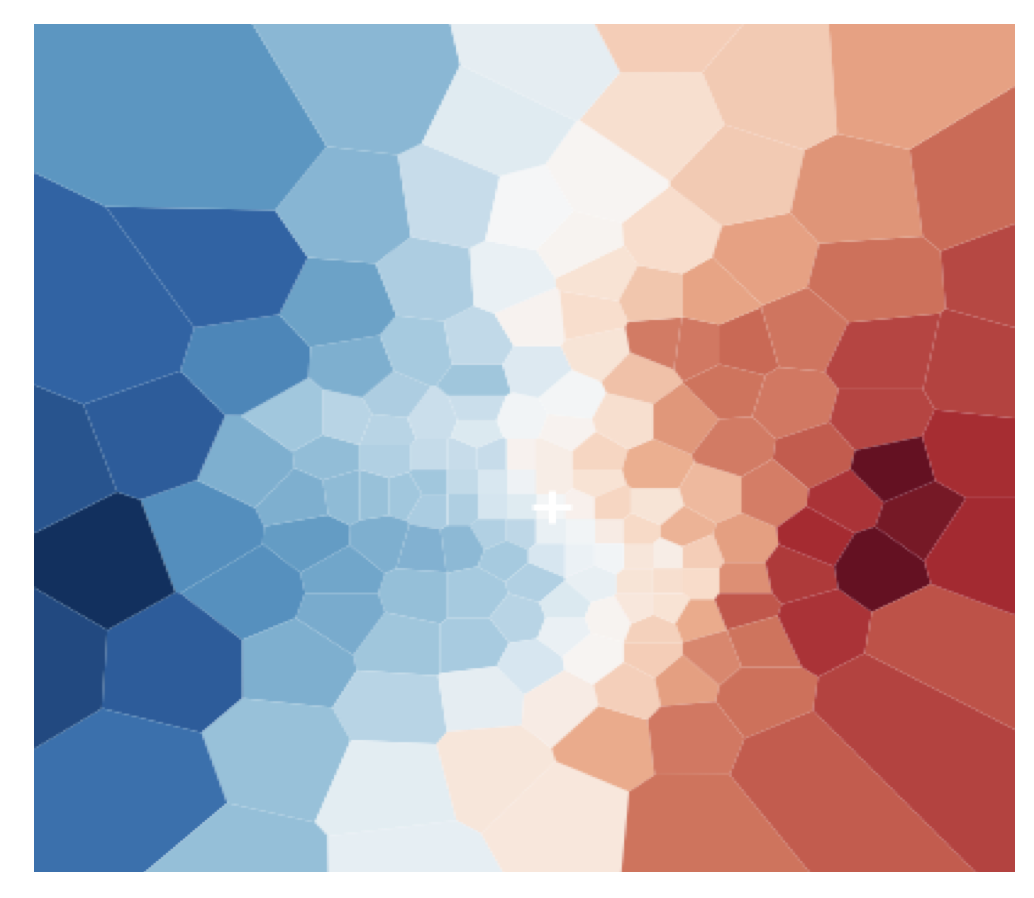}
            }
      \caption{{\it From left to right:} The Gaia DR3 density map of the LMC, the best-fit {\sc galfit} model, the {\sc galfit} model residuals, and the Gaia DR3 LOS Voronoi map in the same orientation. The white dashed lines denote the best-fit major and minor photometric axes that cross at the best-fit photometric centre according to the {\sc galfit} solution. The {\em top row} is the best-fit solution, when all {\sc galfit} parameters are left free; the {\em bottom row} shows the {\sc galfit} solution when the orientation of the large photometric axis and projected flattening are set to the orientation of the line of nodes and projected flattening according to the 3D Jeans model and using the adopted S\'ersic indices of the disc and bar $\rm n_{d} = \rm n_{b} = 0.5$.}
         \label{fig:galfit1}
   \end{figure*}

The Gaia DR3 density map of the LMC is shown in the left panels of Fig.~\ref{fig:galfit1}.
We have obtained it as a fits image with pixel resolution $1271 \times 991$\,px and pixel scale $81.84$\,arcsec\,px$^{-1}$.
The fits file contains a WCS coordinate solution, which allows us to convert pixels to equatorial coordinates.
We convolve this image with a Gaussian PSF kernel of $5$\,px and complement it with a corresponding Poissonian error image (each pixel has the square root of the value in the original image) for the {\sc galfit} analysis.

Then we proceed with the {\sc galfit} fit and morphological decomposition of the LMC.
We look for a two component solution, employing two S\'ersic profiles - one describing the LMC disc and the other one - the LMC bar.
We also allow for a background component with a spatial gradient.
The main free parameters of the fit are summarised in Table~\ref{tab:galfit}.
Each of the two S\'ersic profiles is characterised by five parameters - integral magnitude ($M$), scale radius ($R$), S\'ersic index ($n$), axis ratio ($q$), and position angle ($\psi$).
In addition we fit for the centre of the LMC, requiring that the bar and the disc share the same centre.
The latter requirement might be viewed as controversial, since many studies have reported an off-centred bar for the LMC \citep[e.g.][]{besla+2012}, but it is necessary for running our triaxial dynamical models.
Our study was especially motivated by the close coincidence of the PM kinematic centre of the LMC as inferred by \citet{gaiaedr3} with the centre of the bar.

{\sc galfit} uses the minimum $\chi^2$ approach to determine the best fit.
We give the minimum reduced $\chi^2_R = \chi^2 / N_{\rm px}$ value in the bottom of Table~\ref{tab:galfit}.
Since the variance of the $\chi^2$ distribution would be $2N_{\rm px}$, the reduced $\chi^2_R$ has a standard deviation of $\sqrt{2}$.
We did several fits with varying different additional constraints, discussed below.
The results are shown in Fig.~\ref{fig:galfit1} with the best-fit centre indicated with white lines.

\begin{table}
\caption{{\sc galfit} model results.}             
\label{tab:galfit}      
\centering          
\begin{tabular}{c c c c}
\hline\hline       
   & Free fit &  \multicolumn{2}{c}{Fixed $\psi_d$} \\
\hline
$\alpha_0$ [deg] & 80.437 & 80.576 & 80.560 \\
$\delta_0$ [deg] & $-$69.487 & $-$69.503 & $-$69.502  \\
$M_d - M_b$ [mag] & $-$1.00 & $-$1.42 & $-$1.37 \\
\hline
\multicolumn{4}{c}{Disc} \\
\hline
$R_d$ [deg] & 3.41 & 2.89 & 2.91 \\
$n_d$ & 0.43 & 0.50 & 0.50\tablefootmark{*}  \\
$q_d$ & 0.78 & 0.93 & 0.91\tablefootmark{*}   \\
$\psi_d$\tablefootmark{**} [deg] & 85.5 & 124.0\tablefootmark{*} & 124.0\tablefootmark{*} \\
\hline
\multicolumn{4}{c}{Bar} \\
\hline
$R_b$ [deg] & 1.83 & 1.74 & 1.68  \\
$n_b$ & 0.47 & 0.34 & 0.50\tablefootmark{*}  \\
$q_b$ & 0.40 & 0.35 & 0.37  \\
$\psi_b$\tablefootmark{***} [deg] & 66.9 & 27.7 & 27.4 \\
\hline
$\chi^2_R$  & 84.2 & 93.4 & 94.2  \\       
\hline     
\end{tabular}
\tablefoot{
\tablefoottext{*}{Fixed parameter.}
\tablefoottext{**}{Orientation of the LMC disc measured from north to east.}
\tablefoottext{***}{Angle between the major axes of the disc and bar.}
}
\end{table}

The result of the general photometric decomposition, where all S\'{e}rsic parameters for both components were free during the fit is shown in the top row of Fig.~\ref{fig:galfit1}.
The residual image shows that the disc and bar of the LMC are modelled reasonably well.
The most noticeable components in the residuals are the LMC spiral arms and dust lanes, which we did not attempt to include in the fit.
The on sky projected orientation and flattening of the LMC disc according to this photometric fit is close to what we obtain from the Jeans 2D kinematic fit with $|\Delta\psi_d| = 18.7^{\circ}$ and $|\Delta q_d| = 0.06$.
And similarly to the 2D Jeans kinematic fit, it does not agree with the direction of the LOS velocity gradient.
The $\rm V_{LOS}$ rotation axis (as obtained from the 3D Jeans kinematic solution) appears to be inclined by $38.5^{\circ}$ with respect to the photometric major axis.
The LOS velocity map is also shown in the top row of Fig.~\ref{fig:galfit1} in the same orientation.

In the second {\sc galfit} solution we decided to fix the orientation angle $\psi_d = 124^{\circ}$ to correspond to the orientation angle of the projected rotation axis, according to our best fit 3D Jeans dynamical model.
All other {\sc galfit} parameters were kept free as before.
The projected flattening of the LMC disc that we find from this photometric decomposition is very close to what we find from the 3D kinematic model ($|\Delta q_d| = 0.02$).
However, as one would expect, this photometric fit is significantly worse than the first one, having a minimum $\chi^2$ $6.5$ standard deviations higher than the original fit.
This is a compromise that we have to make in order to find an axisymmetric disc orientation, compatible with the galaxy's line of nodes orientation.

Furthermore, we notice that the best-fit S\'ersic index values for both the disc and the bar components are equal or slightly below $0.5$.
A S\'ersic index of $0.5$ corresponds to a Gaussian distribution.
Usually, galaxies have S\'ersic indices between $1$ and $4$, with values $1-2$ being typical for discs and bars, and $n>2$ more typical for spheroidal components \citep{kormendy+kennicutt2004}.
But the fact that we find so low $n$ values for the LMC may not be too surprising if the stellar component of the LMC is severely tidally limited by its proximity to the Milky Way.
In addition, the surface density in the central region of the LMC remains approximately constant in an extended region, which is described better by a Gaussian distribution rather than an exponent.
Hence, the third {\sc galfit} solution that we present fixes $n_d = n_b = 0.5$, as well as $q_d = 0.91$, in addition to the fixed $\psi_d$ angle (bottom row of Fig.~\ref{fig:galfit1}).
We found that adopting $q_d = 0.91$ from the dynamical solution of the 3D Jeans model, provides better constraints on the parameters of our Schwarzschild models, compared to using the photometric fit of $q_d = 0.93$.
The {\sc galfit} fits with free and fixed S\'ersic indices are practically identical and within $1\,\sigma$ of the $\chi^2$ distribution, but we need to impose this additional constraint, because we need to be able to represent the derived LMC surface brightness distribution with a MGE.
In this case the MGE consists of two Gaussian components - one for the disc and one for the bar (Table~\ref{tab:mge1}).
It is scaled to the same total LMC luminosity of $1.31\times10^9$\,\Lsun~used in our Jeans models (Table~\ref{tab:jeans}).

\begin{table}
\caption{Adopted LMC MGE for the Schwarzschild orbit superposition modelling using Gaussian disc and bar.}             
\label{tab:mge1}      
\centering          
\begin{tabular}{c c c c c}
\hline\hline       
   & $L$ & $\sigma$ & $q$ & $\psi$ \\
   & [\Lsun\,pc$^{-2}$] & [arcsec] & & [deg] \\
\hline
disc & 55.54  & 7416.50 & 0.91 & 0.0 \\
bar & 117.02 & 4271.76 & 0.37 & 27.4 \\
\hline     
\end{tabular}
\end{table}

\subsection{LMC de-projection and viewing angles}\label{sec:angles}

   \begin{figure}
   \resizebox{\hsize}{!}
            {
            \includegraphics[width=\hsize]{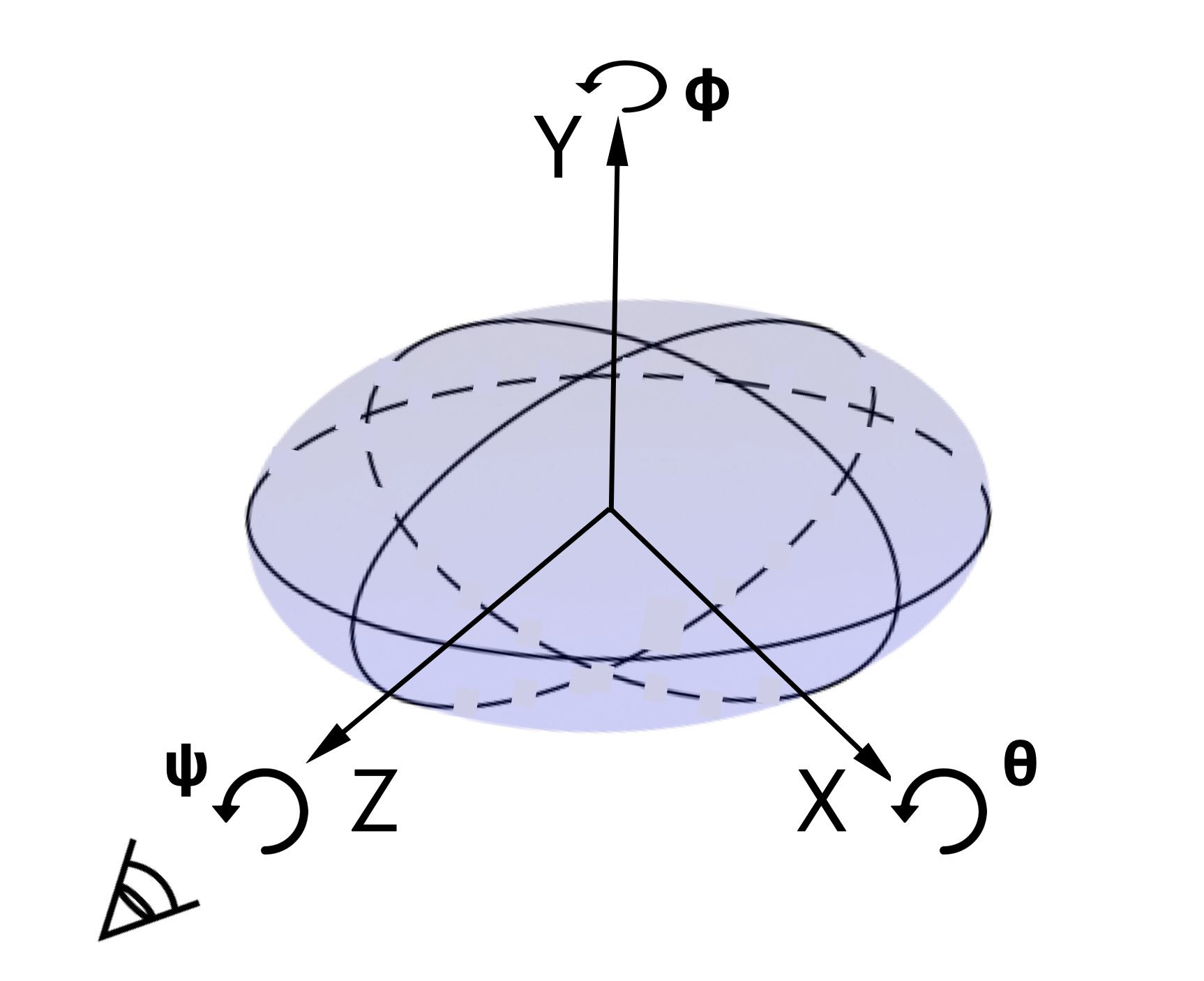}
            }
      \caption{Definition of the viewing angles of a triaxial ellipsoid. The LOS direction is along the Z axis.}
         \label{fig:ellipsoid}
   \end{figure}

Once we have defined the disc - bar morphological decomposition of the LMC on the plane of the sky, we can proceed with the 3D de-projection of the galaxy - essentially determining the intrinsic shapes of the disc and bar, as well as the viewing angles of the projected image.

The intrinsic shape of the LMC triaxial bar is determined by two axis ratios ($q_b^0 = c_b^0/a_b^0$ and $p_b^0 = b_b^0/a_b^0$), where $a_b^0, b_b^0, c_b^0$ are the intrinsic sizes of the long, intermediate, and short bar axes, respectively.
The intrinsic shape of the axisymmetric disc is determined by a single axis ratio ($q_d^0 = b_d^0 / a_d^0$), where $a_d^0, b_d^0$ are the intrinsic sizes of the long and short disc axes.
There are also two unknown viewing angles - the inclination ($\theta$) and the tip angle ($\phi$).
The latter is only relevant for triaxial ellipsoids.
The third angle ($\psi$) can be fixed, such that the projected short axis of the disc is vertical.
See Fig.~\ref{fig:ellipsoid} for the definition of the viewing angles.

In the general case, the de-projection of a triaxial ellipsoid is a degenerate problem, but we can break the degeneracy with the aid of the co-existing axisymmetric disc.
The problem becomes solvable when we assume that the short axis of the bar is aligned with the short axis of the disc \citep{tahmasebzadeh+2021}.

We use a MCMC sampling approach to infer the intrinsic shapes of the LMC disc and bar from our morphological decomposition of the galaxy described in the previous section.
This works as follows - we propose intrinsic shapes of the two morphological components and project them with a set of proposed viewing angles.
The likelihood function maximises the similarity of the resulting projected morphology to the modelled image from {\sc galfit}.
We aim to match the projected flattening of both the disc, and bar, as well as the angle between the projected major axes of the two components.

The projection equation is from \citet{binney1985}:
\begin{equation}
\begin{pmatrix}
X_p \\
Y_p \\
0
\end{pmatrix}
=
\begin{pmatrix}
-\sin\phi & \cos\phi & 0 \\
-\cos\theta\cos\phi & -\cos\theta\sin\phi & \sin\phi \\
\sin\theta\cos\phi & \sin\theta\sin\phi & \cos\theta
\end{pmatrix}
\cdot
\begin{pmatrix}
X \\
Y \\
Z
\end{pmatrix}
.
\end{equation}

         \begin{figure*}
   \centering
            {
            \includegraphics[width=5.7cm]{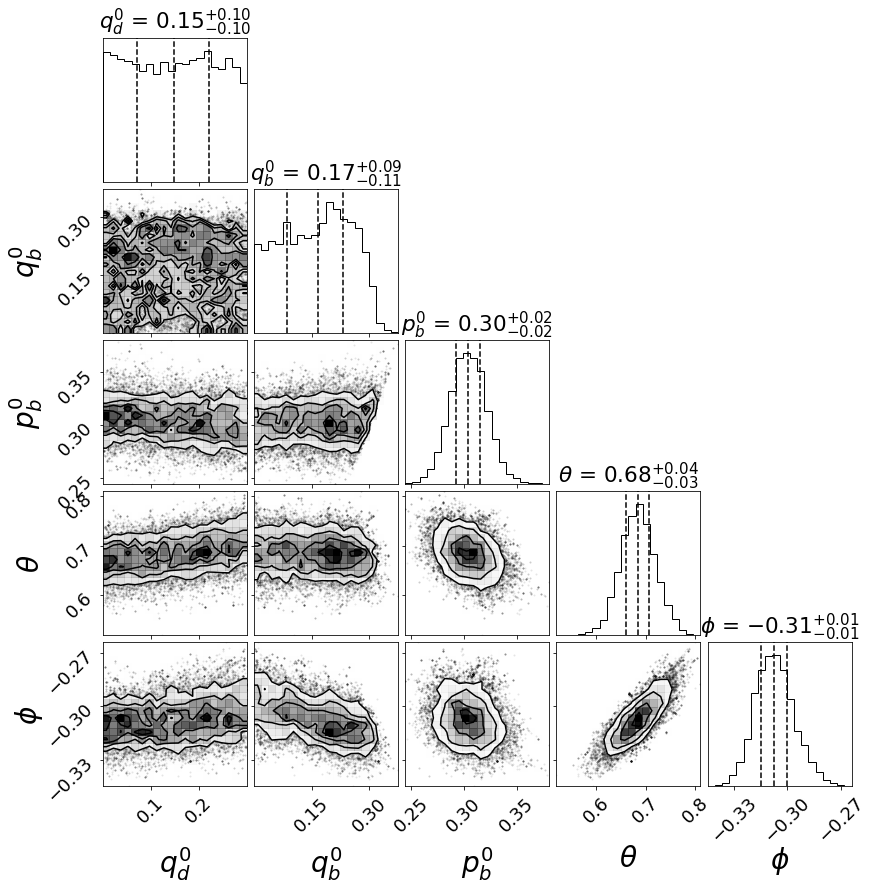}
            \includegraphics[width=6.8cm]{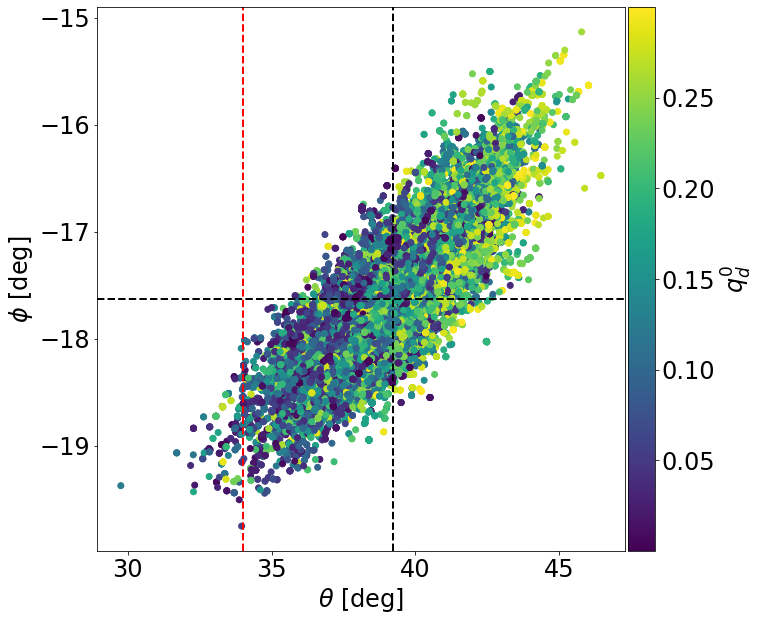}
            \includegraphics[width=5.7cm]{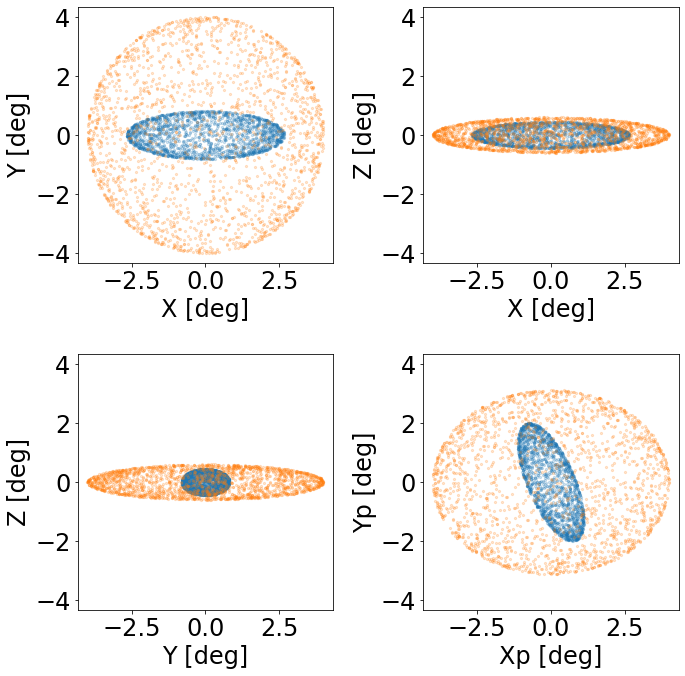}
             \includegraphics[width=5.7cm]{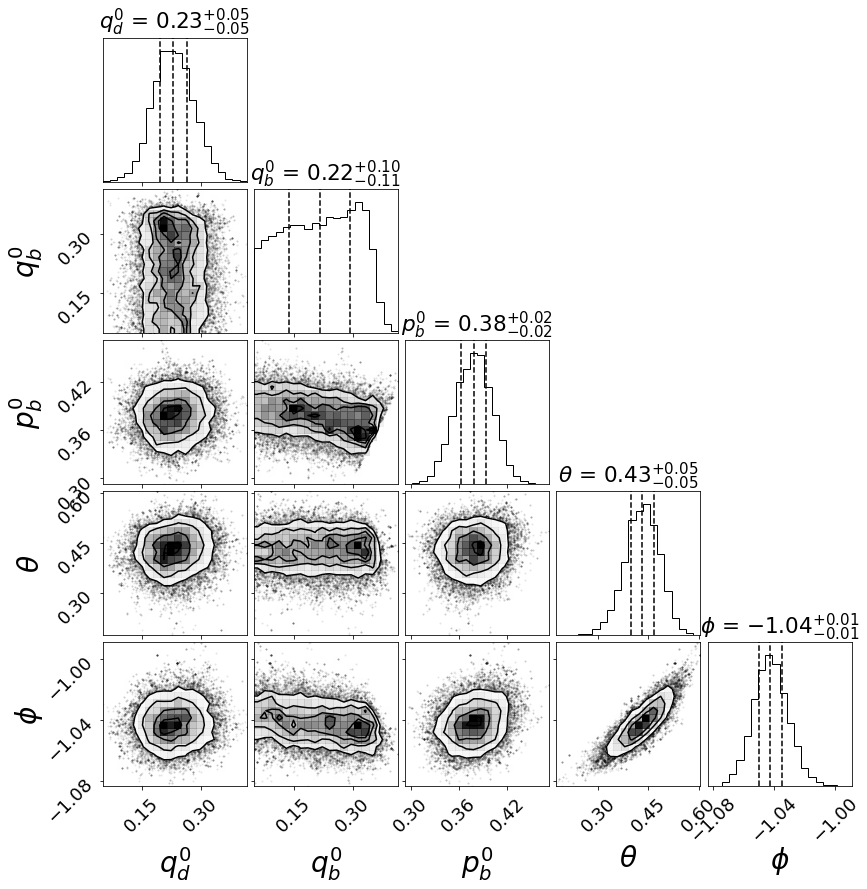}
            \includegraphics[width=6.8cm]{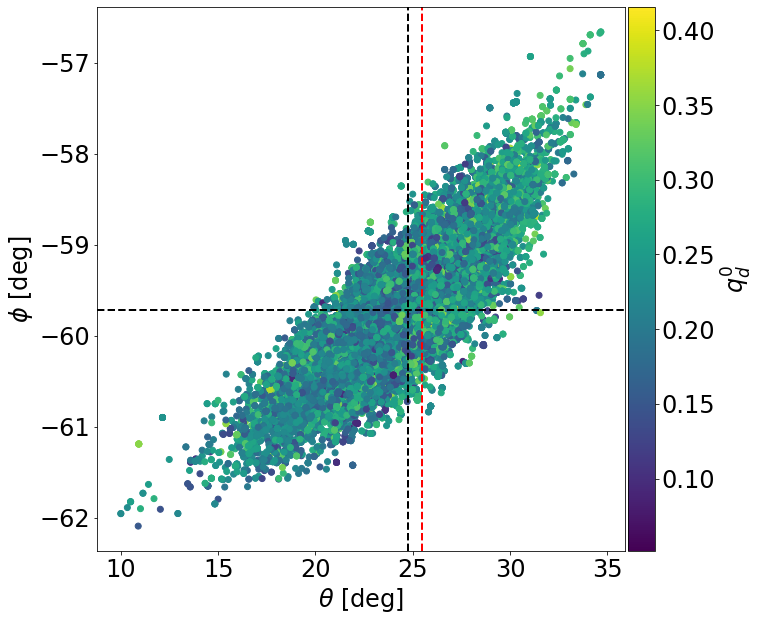}
            \includegraphics[width=5.7cm]{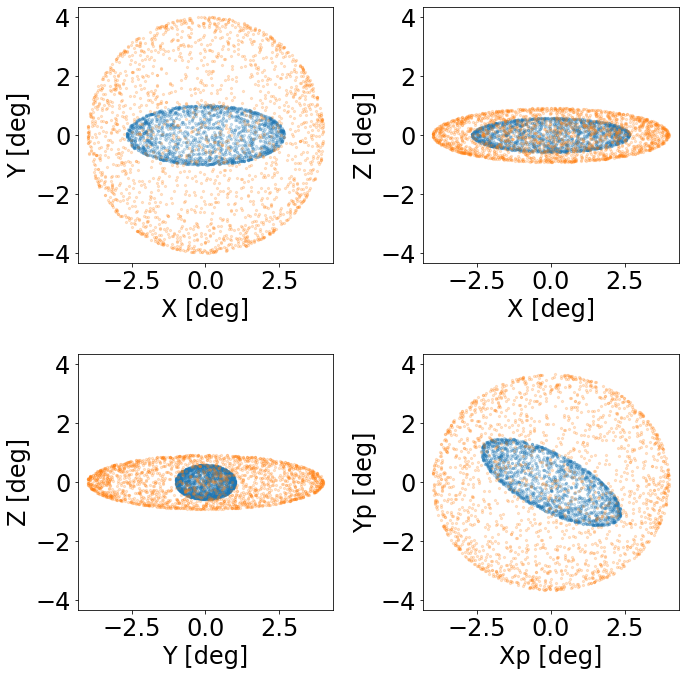}
            }
      \caption{{\it Left panels:} De-projection results - corner plots showing the MCMC distributions of the viable intrinsic axis ratios of the axisymmetric disc ($q_d^0$), the triaxial bar ($q_b^0, p_b^0$), and the viewing angles $\theta$ and $\phi$ in rad, that result in the observed morphological decompositions of the LMC from the two cases shown in Fig.~\ref{fig:galfit1}. {\it Middle panels:} zoom-in to the ($\theta, \phi$) planes, colour-coded by the disc intrinsic flattening. The dashed black lines show the median angles from the de-projection, while the red dashed line shows the best-fit inclination angle ($\theta$) from the 2D Jeans dynamical model in the top panel and the 3D Jeans model in the bottom panel. {\it Right panels:} Embedded ellipsoids of the de-projected  disc and bar with axis ratios, as inferred from the de-projection of the LMC morphological decomposition from Fig.~\ref{fig:galfit1}, viewed from the three cardinal directions, and from viewing angles $\theta$ and $\phi$.}
         \label{fig:deprojection1}
   \end{figure*}
   
The results of the de-projection routine are summarised in Fig.~\ref{fig:deprojection1} for the two LMC decompositions shown in Fig.~\ref{fig:galfit1} in corresponding rows.

First we turn our attention to the top row of Fig.~\ref{fig:deprojection1}, corresponding to the general {\sc galfit} solution with a mismatch between the projected long axis of the disc and the line of nodes orientation.
The corner plot shows that the viewing angles are well constrained and we find an inclination angle $\theta = 39.0^{\circ}\pm2.3^{\circ}$ and a tip angle $\phi = -17.8^{\circ}\pm0.6^{\circ}$.
The inclination is slightly larger than what we inferred from the 2D Jeans dynamical model for the disc ($\theta=34^{\circ}\pm0.1^{\circ}$), but in this de-projection routine we used a slightly lower projected axis ratio $q_d = 0.78$ from {\sc galfit}, compared to $q_d = 0.84$, as inferred from the 2D Jeans model (see Tables \ref{tab:jeans} and \ref{tab:galfit}).
The lower angle inferred from the dynamical modelling of $\theta=34^{\circ}\pm0.1^{\circ}$ is still within the allowed range of angles for the de-projection (see the middle panel of the top row in Fig.~\ref{fig:deprojection1}).
The intermediate-to-major axis ratio for the bar is also well constrained at $p_b^0 = 0.30\pm0.02$, while we are able to only put an upper limit to the minor-to-major axis ratio for the bar at $q_b^0\lesssim0.3$.
The intrinsic flattening of the disc, however, is degenerate in the targeted projection, so we need to adopt the additional prior constraint that $q_d^0 < 0.3$.
The top right panel of Fig.~\ref{fig:deprojection1} shows points sampled on the surfaces of two ellipsoids, representing the LMC disc and bar, using the median axis ratios, as inferred from our de-projection routine and how they project on the sky at the median viewing angles - a 2D geometry essentially the same as the one inferred from the {\sc galfit} LMC decomposition shown in Fig.~\ref{fig:galfit1}.

Now we look at the de-projection results shown in the bottom rows of Fig.~\ref{fig:deprojection1}, corresponding to the {\sc galfit} solution, where the major axis of the disc and the line of nodes orientation are aligned.
This is the de-projection that we adopt for our Schwarzschild models described in the next section.
The viewing angles that we find for this targeted projection are $\theta = 25^{\circ}\pm3.0^{\circ}$ and $\phi = -61.3^{\circ}\pm1.1^{\circ}$.
The inclination angle is in line with the disc inclination that we infer from the 3D Jeans dynamical model ($\theta = 25.5^{\circ}\pm0.2^{\circ}$; Table~\ref{tab:jeans}) within the uncertainties.
We remind the reader that the targeted projected flattening of the disc in this case is $q_d = 0.91$ - the same as inferred from the 3D Jeans model.
We also manage to get a good constraint of the intrinsic flattening of the LMC disc ($q_d^0 = 0.23\pm0.05$) from this targeted projected geometry without the need for prior assumptions.
The best-fit axis ratios for the bar are $p_b^0 = 0.38\pm0.02$ and again an upper limit for $q_b^0\lesssim0.35$.
The bottom right panel of Fig.~\ref{fig:deprojection1} illustrates the inferred 3D geometry of the LMC under our simplifying assumptions and the resulting projected morphology from the fitted median viewing angles, which is in excellent agreement with the {\sc galfit} model shown in the bottom row of Fig.~\ref{fig:galfit1}.

\section{Schwarzschild models and the bar pattern speed}\label{sec:schwarzschild}

The orbit-superposition technique \citep{schwarzschild1979} is a powerful tool to study galaxy dynamics \citep{rix+1997, gebhardt+2003, cappellari+2007}, not limited to the assumption of axisymmetry \citep{vandenbosch+2008, neureiter+2021}.
It relies on the idea of superposing a large number of individual stellar orbits to simulate the overall mass distribution and kinematics of galaxies.
The method starts with an assumed gravitational potential.
Then, a collection of orbits is numerically integrated in this potential.
Each orbit spans a certain region of phase space and represents many stars.
Individual orbits are weighted and superimposed to match the observed surface brightness and kinematic data, among other constraints.
The method has been further developed to include the stellar age and metallicity, enabling chemo-dynamical decomposition of galaxy structures \citep{zhu+2022, ding+2023, jin+2024}.

Here we use a modified version of the Schwarzschild code originally developed by \citet{vandenbosch+2008} and later published as {\sc dynamite}\footnote{\url{https://dynamics.univie.ac.at/dynamite_docs/index.html}} \citep{jethwa+2020, thater+2022} to model the LMC as a two-component stellar system, consisting of an axisymmetric disc and a triaxial bar, following an approach outlined in \citet{tahmasebzadeh+2022}.
The most important modification is that the orbit integration is done in a rotating frame of reference, in which the barred galaxy potential is stationary.
The frame rotation corresponds to the bar pattern speed ($\Omega$) - a free parameter in our Schwarzschild models.
As a consequence, the retrograde orbits cannot be derived by reversing the sign of the prograde orbit family and need to be integrated separately.
In addition, the 8-fold orbit symmetry in non-rotating reference frames is reduced to 4-fold symmetrisation.

\subsection{Data preparation}

Currently {\sc dynamite} takes as input data Voronoi tessellation maps of the velocity field, velocity dispersion, and higher velocity moments with the their corresponding uncertainty maps in the LOS direction.
The kinematic data set is complemented with a MGE of the projected surface luminosity density distribution and viewing angles.
We used the projected luminosity density distribution MGE derived with {\sc galfit} from the Gaia DR3 density map of the LMC and shown in Table~\ref{tab:mge1}.
The corresponding intrinsic shapes and viewing angles are as defined in Sect.~\ref{sec:angles} (bottom panel of Fig.~\ref{fig:deprojection1}).
We allow for $1\%$ relative uncertainty in the projected and intrinsic luminosity distributions.

Here we focus on the preparation of the LOS kinematic data for the Schwarzschild modelling.
The input Voronoi maps are based on the ones shown in the bottom left panels of Fig.~\ref{fig:jeans_rot} and Fig.~\ref{fig:jeans_disp} to illustrate the performance of the discrete 3D Jeans model in recovering the observed LOS velocity field and LOS velocity dispersion of the LMC, respectively.
To construct the Voronoi maps, we consider only stars with high probability of being genuine LMC members according to our 3D Jeans model.
To be on the conservative site, we impose a minimum LOS velocity error of $1$\,\kms~for individual stars (if the quoted Gaia DR3 LOS velocity uncertainty is less than $1$\,\kms, we replace it with $1$\,\kms) and remove stars with Gaia DR3 LOS velocity errors larger than $15$\,\kms.
Some Gaia DR3 sources have unrealistically low LOSV uncertainty entries. The minimum LOSV error in our sample is only $0.11$\,\kms. In total $\sim1900$ sources (out of $\sim 28\,000$) have Gaia DR3 errors below $1$\,\kms. Imposing a minimum LOSV uncertainty allows a bit more flexibility in our Schwarzschild models and prevents them from getting stuck in local $\chi^2$ minima driven by spurious kinematic features.
The resulting median LOS velocity error of the individual stars is $3$\,\kms.
Each individual Voronoi cell contains about 100 stars with Gaia DR3 LOS velocity measurements and its mean velocity, velocity dispersion, and respective uncertainties are computed using a maximum likelihood approach, taking the individual errors of the stellar velocity measurements into account.
We ignore higher velocity moments ($h_3, h_4, ...$) because these limited statistical samples are insufficient to estimate them reliably.

The resulting LOS dispersion map is, however, quite noisy and initial tests showed that this becomes an issue for the Schwarzschild models to find a global $\chi^2$ minimum.
We noticed that there is a significant difference in the velocity dispersion of the old and young LMC populations with the young stars showing considerably lower dispersion.
We opted to use only the old stars of the LMC, as defined in Fig.~\ref{fig:jeans_membership}, as kinematic tracers for both the LOS rotation pattern and velocity dispersion.
We also applied a Gaussian smoothing filter with a kernel size of $25$\,arcmin to the velocity dispersion map of the old stellar population to further decrease the remaining irregularities.
Such a smoothing was not applied to the rotation field, not to lose any subtle effects in the map caused by the bar kinematics.
The median uncertainty of the Voronoi bins is $2$\,\kms~in the velocity field and $1.5$\,\kms~in the velocity dispersion field.

Since in these models the bar is centred by construction, we adopted the kinematic centre determined by \citet{gaiaedr3} in our input Voronoi maps, as it is closest to the photometric centre of the bar.
 
\subsection{Gravitational potential}

The gravitational potential consists of visible (mass-follows-light) and DM (a classical NFW halo) components, defined by three parameters - stellar mass-to-light ratio $\rm M_*/L$, virial mass-to-stellar mass ratio ($\rm M_{200}/M_*$), and NFW halo concentration ($C=r_{200}/r_s$), where $r_{200}$ is the virial radius and $r_s$ - the NFW halo scale radius.
We add a very light black hole in the centre with a mass of $10$\,\Msun, which only purpose is to provide some additional stability to the orbit integration in the innermost region of the galaxy, but we expect that its overall effect on the dynamical properties of the modelled galaxy is negligible.

We examined a dense grid of possible potentials, varying $\rm \log M_{200}/M_*$ and $\rm \log C$ in step size of $0.05$, and $\rm M_*/L$ in step size of $0.2$\,\MLsun.
In addition the potential is stationary in a rotating reference frame defined by the bar pattern speed $\Omega$.
We tested a wide range of $\Omega$ ranging from $-10$\,\kmskpc \citep[counter-rotating bar, motivated by the possible solutions found by][]{jimenez-arranz+2024a}, up to $30$\,\kmskpc~with a step size of $1$\,\kmskpc.

\subsection{Orbit library and model setup}

Next we setup the orbit library used in our Schwarzschild dynamical models.
We use a $30\times15\times10$ three-dimensional main grid, defined by the three integrals of motion - orbit energy ($E$), $I_2$ (similar to the vertical angular momentum), and $I_3$.
The orbits are sampled logarithmically in radius from $r_{min} = 10^{1.5}$\,arcsec to $r_{max} = 10^{4.5}$\,arcsec ($7.6$\,pc to $7.6$\,kpc), covering the radial extent of our kinematic tracers.
We also define a retrograde orbit grid, using the exact same settings and in addition, we use an orbit dithering parameter ($d=3$), which defines a mini-grid around each of the main grid bins for a total of $2\times30\times15\times10 = 9\,000$ orbit bundles (each bundle consists of $d^3=27$ orbits) to integrate.
We impose a $10^{-6}$ relative energy conservation accuracy and integrate them in the selected gravitational potential for $200$ periods.
We then sample $50\,000$ points in the meridional plane for each orbit.

   \begin{figure}
   \resizebox{\hsize}{!}
            {
            \includegraphics[width=\hsize]{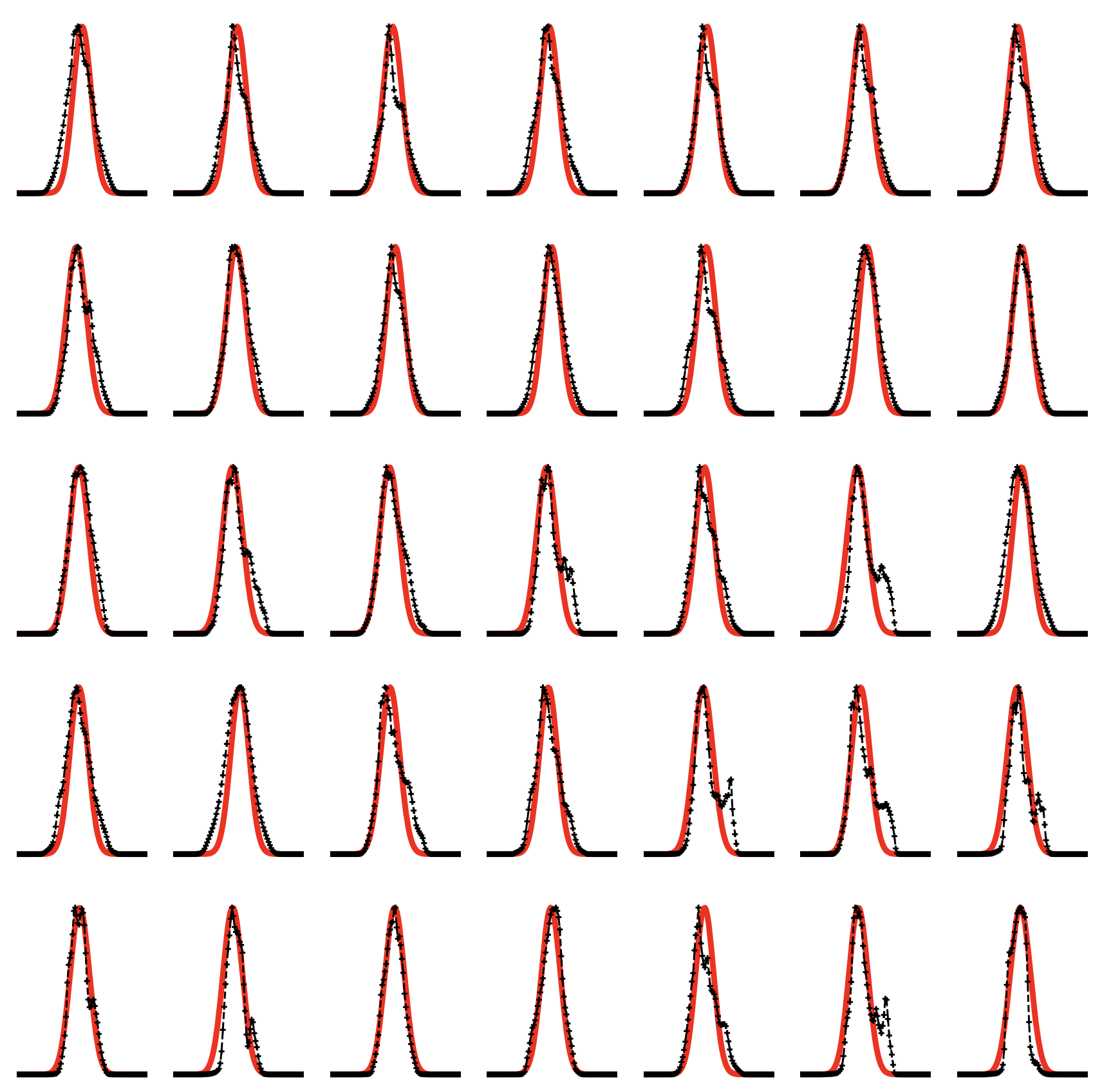}
            }
      \caption{Example LOSVD match between the input data (red) and the orbit superposition (black) for the 35 central Voronoi bins (out of 243). The input LOSVDs are represented by Gaussians, while the model LOSVDs are histograms with $200$ velocity bins $3$\,\kms~wide, constructed from all sampled points from the weighted orbit library, that fall in the same projected Voronoi bins.}
         \label{fig:losvd_match}
   \end{figure}

The Schwarzschild dynamical solution is a linear combination of all orbit bundles, from which we can compute the resulting projected model surface brightness and model LOSVD in the same Voronoi tessellation as the input data.
We find the orbit bundle weights, using a non-linear least squares methodology \citep[NNLS][]{lawson+hanson1974}.
The model LOSVD of each spatial bin is represented by a velocity histogram with $200$ velocity bins $3$\,\kms~wide, that are matched to the input Gaussian LOSVDs (Fig.~\ref{fig:losvd_match}) and a $\chi^2$ value is computed.

This procedure is repeated for different gravitational potentials and bar pattern speeds.

\subsection{Model results}

   \begin{figure*}
  \centering
            {
            \includegraphics[width=9.1cm]{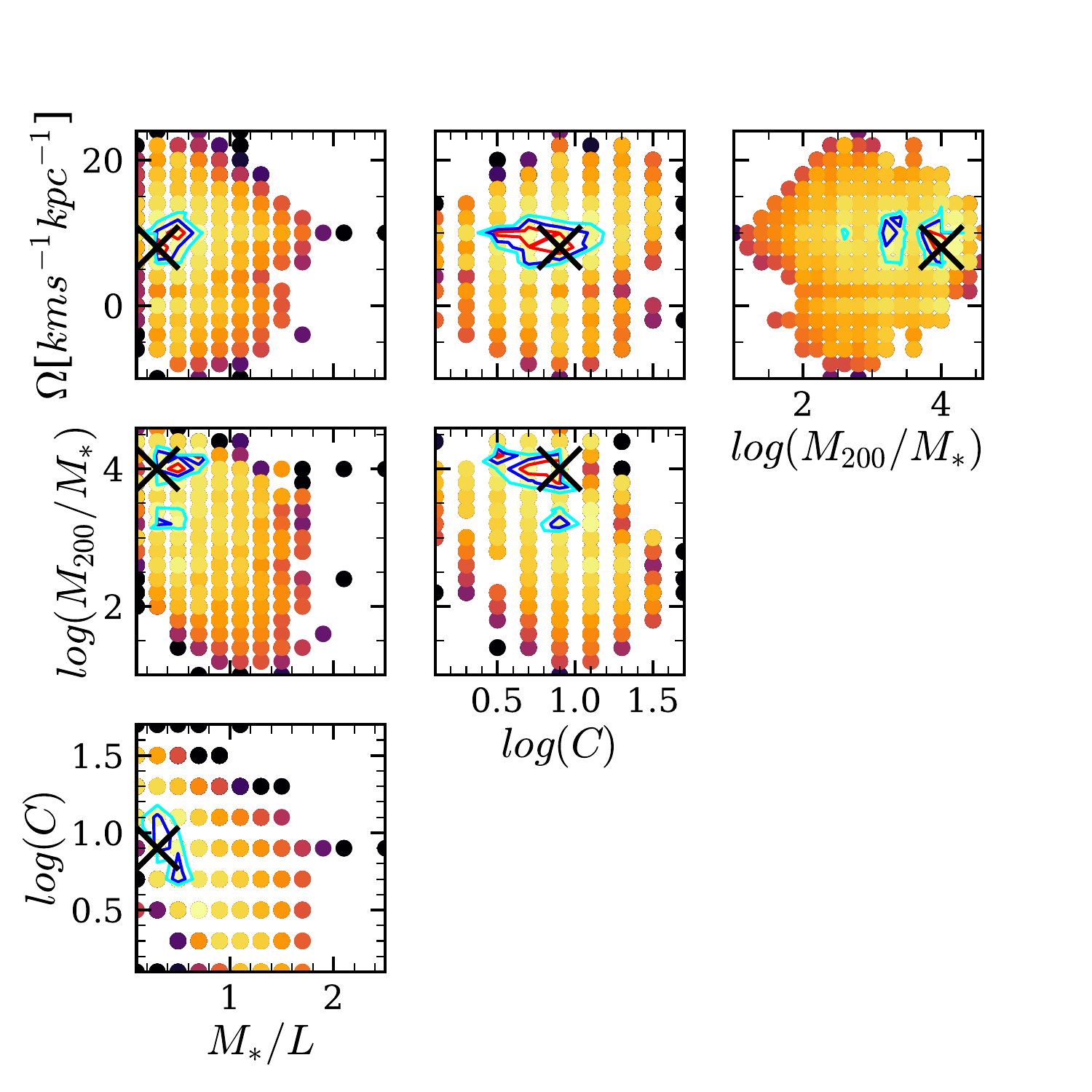}
            \includegraphics[width=9.1cm]{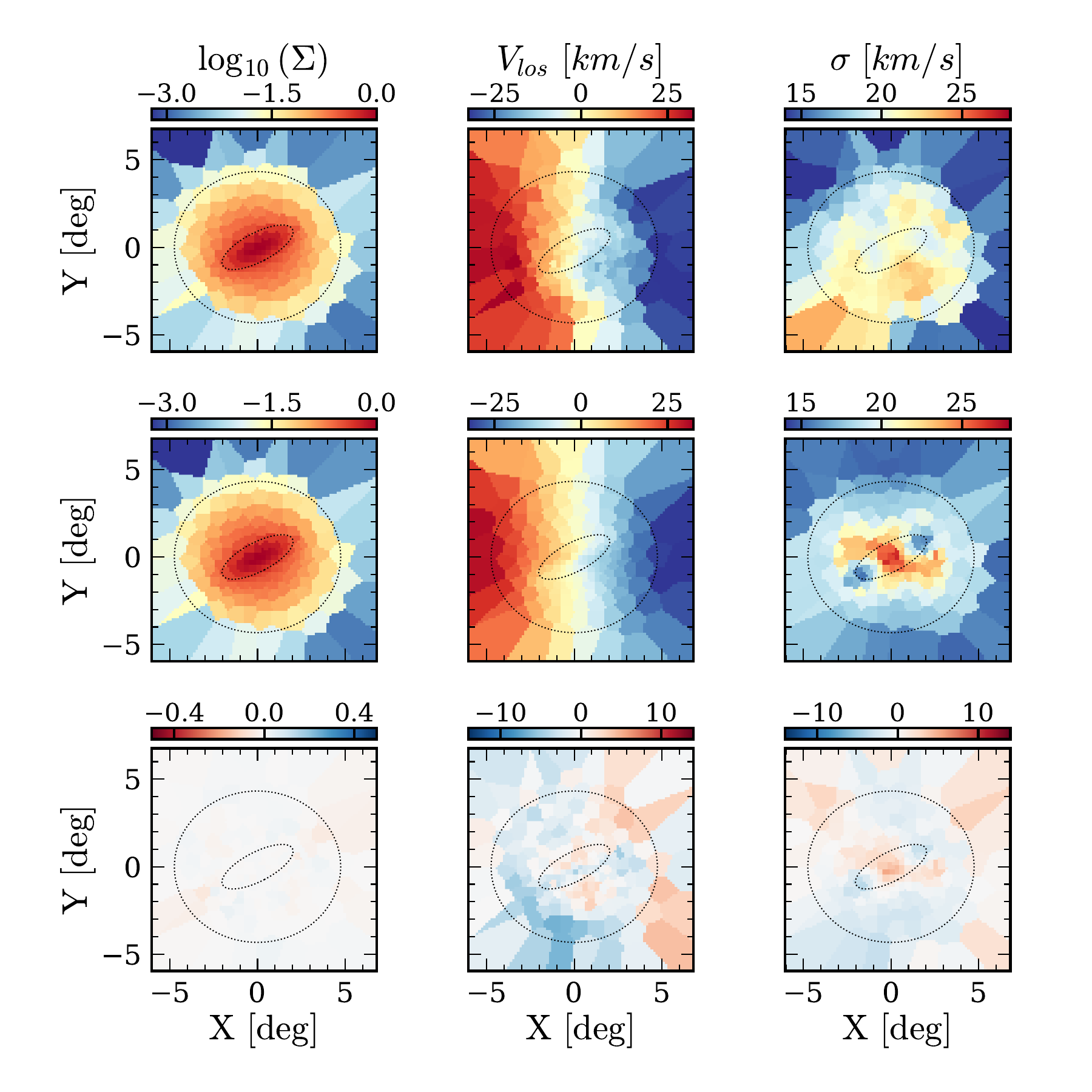}
            \includegraphics[width=6.2cm]{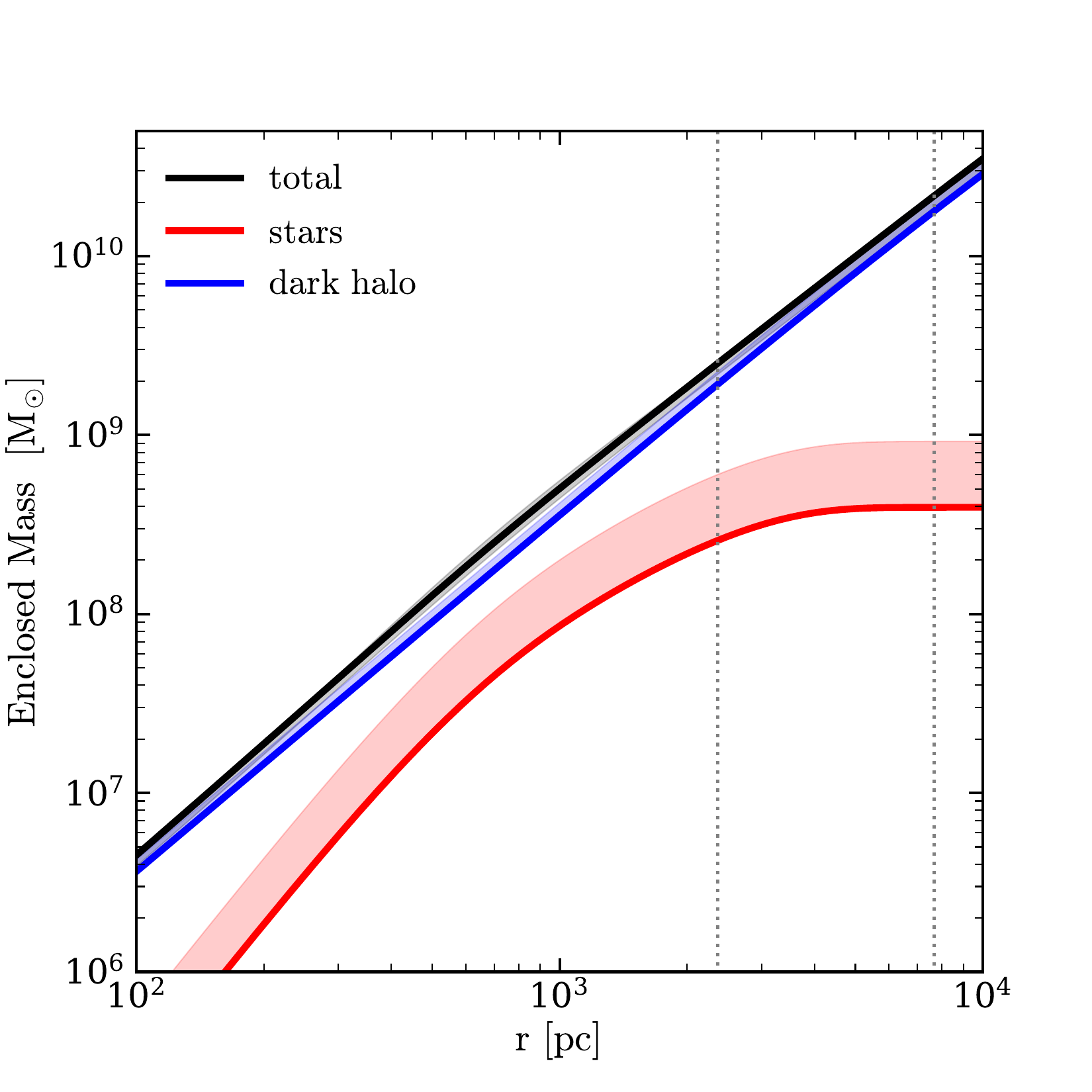}
            \includegraphics[width=6.2cm]{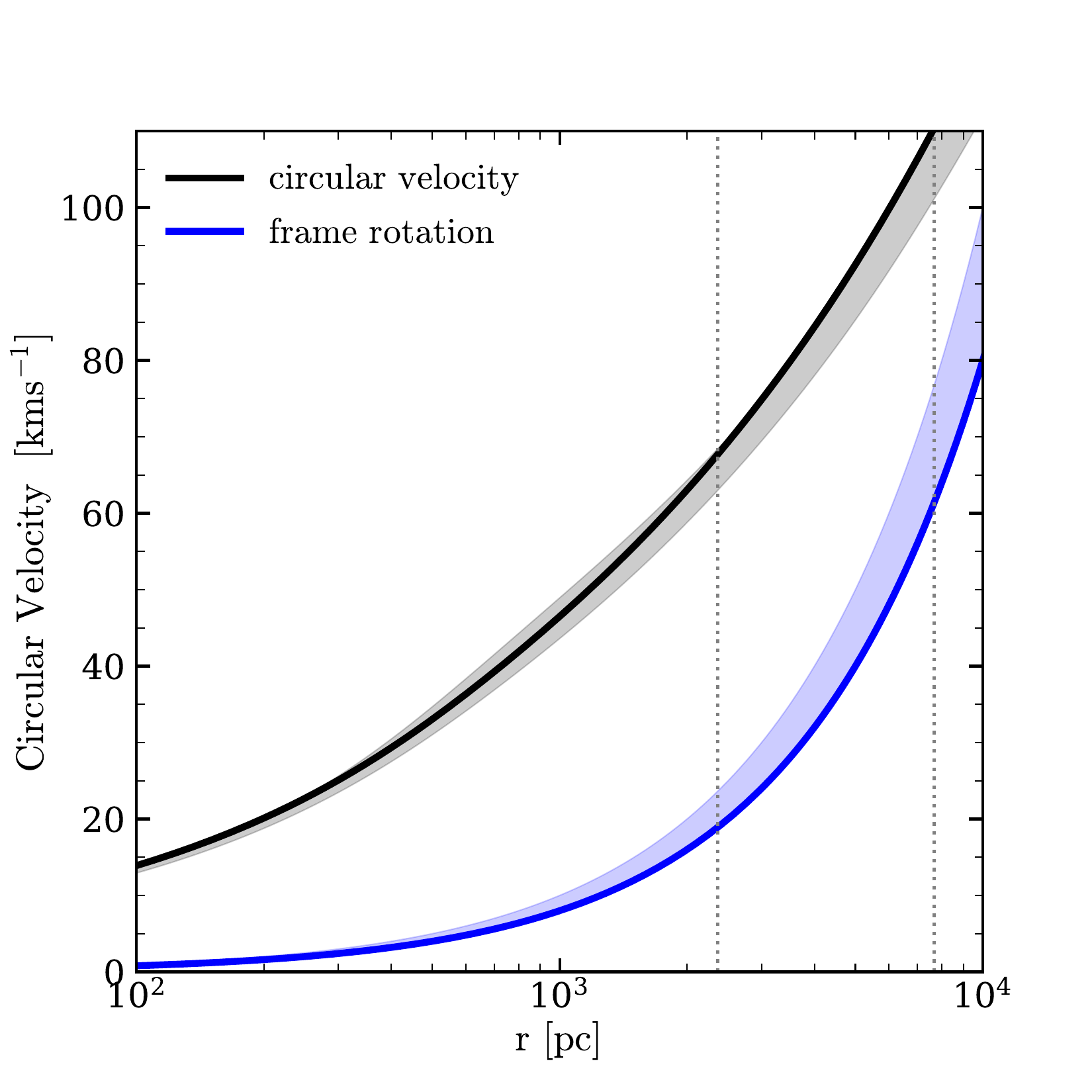}
            \includegraphics[width=5.5cm, trim={0 0 8cm 0}, clip]{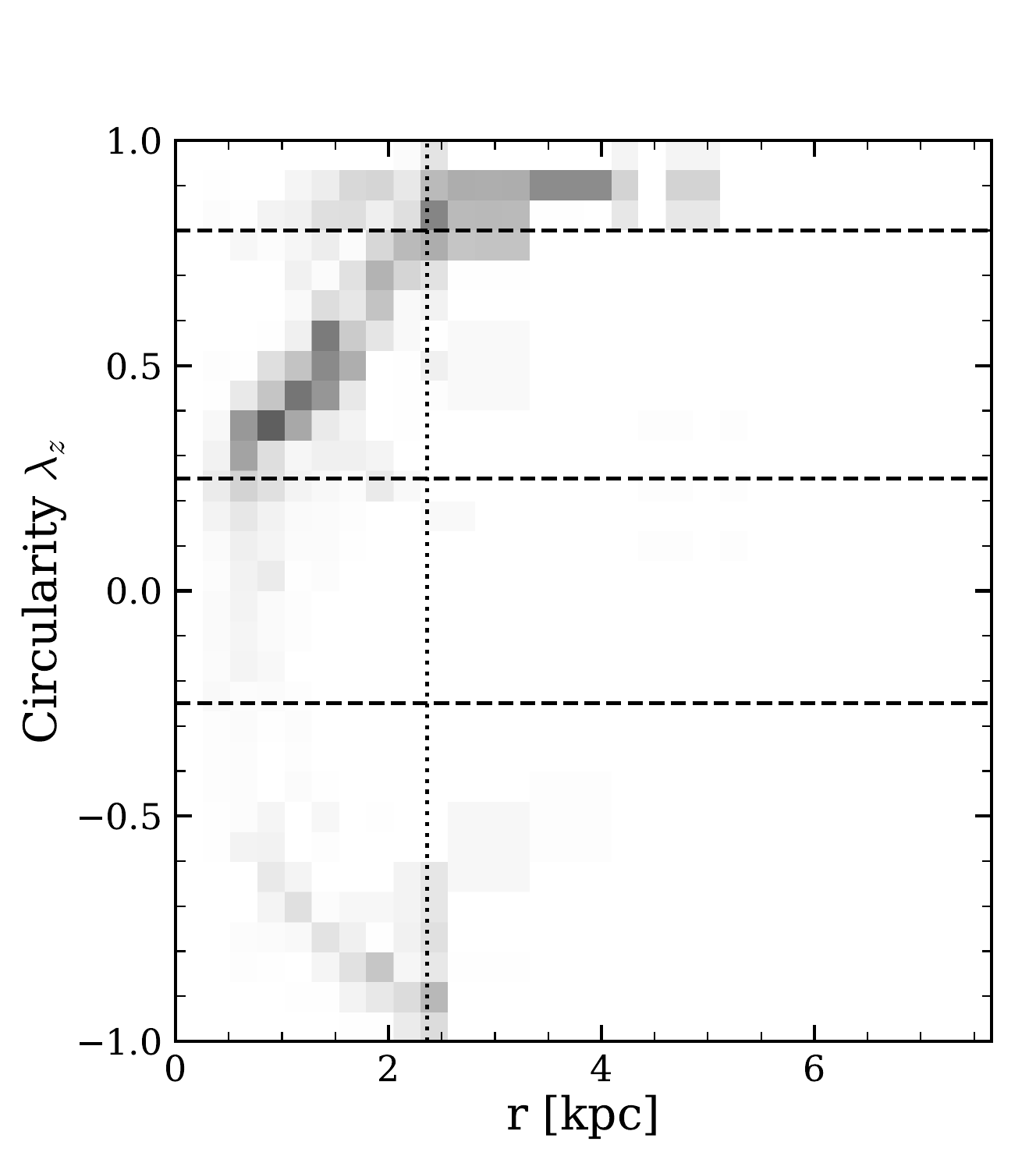}
            }
      \caption{{\em Top left panel:} the full grid of Schwarzschild models (potential parameters and bar pattern speed) that we computed, colour coded by $\chi^2$. The best-fit model is indicated with a cross and the one-, two-, and three-$\sigma$ contours are over-plotted with red, blue, and cyan, respectively. {\em Top right panel:} performance of the best Schwarzschild model with $\Omega=8$\,\kmskpc. {\em Top row:} model input - surface brightness MGE, LOS velocity, and smoothed LOS velocity dispersion. {\em Middle row:} model output - the same quantities as computed from the weighted orbits superposition. {\em Bottom row:} model residuals. {\em Bottom left panel:} enclosed cumulative mass profiles of the stellar ($\rm M_*/L = 0.3$\,\MLsun), DM, and total mass distributions according to the best-fit Schwarzschild model. {\em Bottom middle panel:} Circular velocity curve according to the same model, together with the frame rotation with $\Omega=8$\,\kmskpc. The shaded regions in both panels indicate the $1\,\sigma$ confidence intervals. {\em Bottom right panel:} Orbit weights as a function of their circularity and radius for the the best-fit model. The vertical line denotes the size of the bar and the horizontal lines differentiate different types of orbits - $\lambda_z\sim0$ are radial orbits, $\lambda_z\sim1$ are highly circular orbits, and $\lambda_z<0$ are retrograde.}
         \label{fig:bar_lmc_32}
   \end{figure*}

The upper left panel of Fig.~\ref{fig:bar_lmc_32} shows the full grid of Schwarzschild models that we computed varying the NFW parameters, the stellar $\rm M_*/L$ ratio, and the bar pattern speed.
We find a global $\chi^2$ minimum for $\Omega=8$\,\kmskpc, $\rm \log C = 0.9$, $\rm \log M_{200}/M_* = 4.0$, and $\rm M_*/L=0.3$\,\MLsun~(with an upper limit of $0.7$\,\MLsun~at the 3$\sigma$ level).
All model parameters appear to be well constrained by the explored $\chi^2$ grid.
We do expect a certain level of correlation between the mass parameters, given that our kinematic tracers are mostly concentrated in the inner region of the LMC and do not trace the full potential.
There are multiple combinations of halo virial mass, concentration, and $\rm M/L_*$ that can result in the similar inner mass density distributions.
While the gravitational potential parameters do have a physical meaning, we caution that here we treat them merely as pseudo-parameters that describe the potential shape within the extent of the kinematic tracers, as best as the chosen parametrisation allows.
$\rm M_*/L=0.3$\,\MLsun~is unrealistically low and the corresponding $\rm \log M_{200}/M_* = 4.0$ implies a virial mass $M_{200} \sim 4\times10^{12}$\,\Msun, which is also unphysical.
On the other hand, the measurement of the bar pattern speed is tightly constrained by the Schwarzschild model between $6$ and $13$\,\kmskpc~at the $3\,\sigma$ level.

If we restrict the explored Schwarzschild model grid to more realistic stellar $\rm M_*/L$ ratios ranging from $1$ to $1.5$\,\MLsun, motivated from predictions of composite stellar population models for LMC type galaxies \citep{vandermarel+2002}, the best fit DM halo parameters correspond to $\rm M_{200} = 1-2\times10^{12}$\,\Msun, which is still an order of magnitude higher, than independent virial mass estimates for the LMC \citep[$1-2\times10^{11}$\,\Msun][]{vasiliev2023}.
The result is a manifestation of the same extrapolation issue that we discussed with our Jeans models, however even more pronounced due to the steeper stellar mass distribution adopted here, motivated from the Gaia DR3 density image of the LMC (Fig.~\ref{fig:galfit1}), vs. the exponential profile adopted in the Jeans analysis.

The enclosed mass profile and the corresponding circular velocity curve of the LMC are shown in the bottom left and middle panels of Fig.~\ref{fig:bar_lmc_32}.
The circular velocity plots are also complemented with a curve showing the best fit rate of the bar frame rotation in each model grid.
The shaded regions correspond to the $1\sigma$ confidence intervals.

   \begin{figure*}
   \centering
            {
            \includegraphics[width=6.5cm]{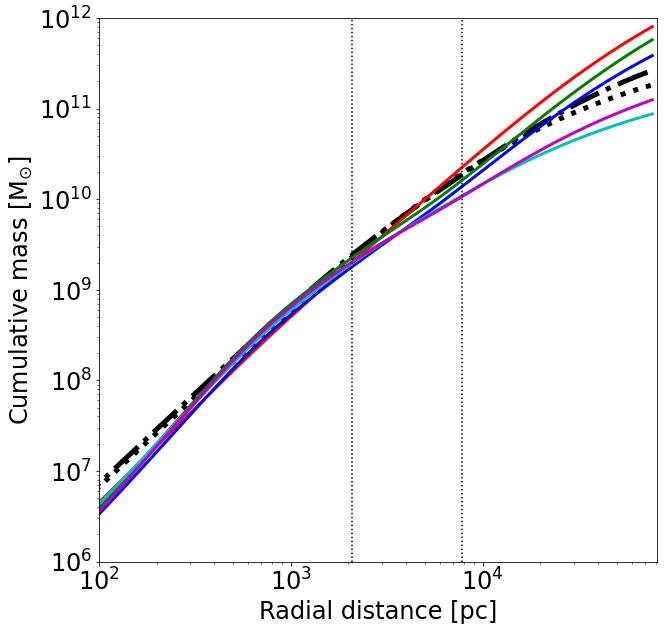}
            \includegraphics[width=11.6cm]{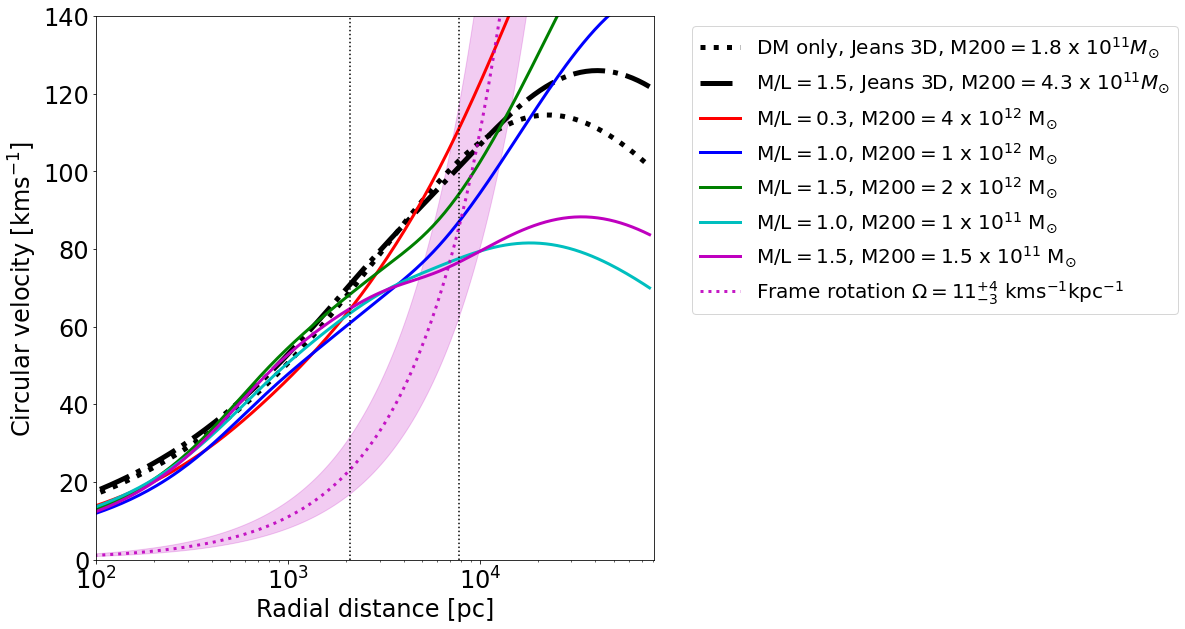}
            }
      \caption{{\em Left panel:} enclosed cumulative total mass profiles for select Schwarzschild dynamical models discussed in this work, compared to the 3D Jeans models. {\em Right panel:} Corresponding circular velocity curves. We over-plot the figure rotation rate for a bar pattern speed of $11$\,\kmskpc~(with a shaded region from $8$ to $15$\,\kmskpc). The vertical lines correspond to the size of the LMC bar and the extent of our kinematic data. Everything beyond the second vertical line is an extrapolation of the model.}
         \label{fig:schw_profile_all}
   \end{figure*}

In Fig.~\ref{fig:schw_profile_all} we plot together the cumulative total mass profiles for select Schwarzschild models from our model grid and the best fit Jeans 3D models (DM only and with mixed gravitational potential), as well as their corresponding circular velocity curves.
The selected Schwarzschild models include the one with the global $\chi^2$ minimum (Fig.~\ref{fig:bar_lmc_32}) and four additional ones with higher $\chi^2$ values but more realistic LMC physical parameters.
The first pair includes the best fit models at fixed stellar $\rm M_*/L = 1~\&~1.5$\,\MLsun, corresponding to virial masses $\rm M_{200} = 1~\&~2\times10^{12}$\,\Msun, respectively.
To select the second pair we additionally constrain the grid to only models with a fixed $\rm \log(M_{200}/M_*) = 1.9$, corresponding to a realistic LMC virial mass for the above stellar $\rm M_*/L$ ratios.
The bar pattern speed In all five models is constrained between $8$ and $15$\,\kmskpc.
The estimated enclosed mass profiles within the radial extent of our kinematic tracers ($6.2$\,kpc) agree within the uncertainties.
The tight constraints on the mass profile that all models impose assures us that we have properly explored the entire parameter space and that there is not further room to expand the parameter grids for the Schwarzschild modelling, unless we change the potential parametrisation entirely.
One could see that the over all best fit Schwarzschild model follows the 3D Jeans models most closely up to the radial extent of the kinematic tracers, but it does not have an extrapolative power at larger distances, as it becomes extremely steep. 
While all models agree well within a radial distance of $\sim3$\,kpc, we observe a significant divergence going outwards in the models with a realistic virial mass ($1-2\times10^{11}$\,\Msun), especially noticeable in the circular velocity plots, which can explain their worse $\chi^2$ results.

In the circular velocity plot we have also added the frame rotation rate for a bar pattern speed of $11^{+4}_{-3}$\,\kmskpc, in line with the $1\,\sigma$ confidence intervals of the five Schwarzschild models combined.
The point where the circular velocity curve and the frame rotation curve cross is the co-rotation radius, which according to the different dynamical models appears at roughly $R_{cor} \sim 10$\,kpc, depending on which rotation curve yields the most realistic results that far out in the LMC disc. 

The resulting kinematic maps based on this best performing model are shown in the top right panel of Fig.~\ref{fig:bar_lmc_32}.
The best fit model captures well the surface brightness and morphology of both the LMC disc and bar, which we modelled with Gaussian surface brightness distributions (Table.~\ref{tab:mge1}). 
The triaxial bar model is also successful in capturing the complexity of the LOS velocity field in the central region.
There is a hint of a decoupled bar rotation pattern in the input data, that creates asymmetries and twists in the LOS velocity field in the central region, which are also present in the dynamical model with $\Omega\sim10$\,\kmskpc.
All tested models, however, have trouble reproducing properly the input velocity dispersion in the inner region, predicting significantly higher values in the centre and a stronger gradient than the data.
This could be owing to either the complexity of the stellar populations in the LMC (we already noted that the young stars have significantly lower velocity dispersion than the old stars, but the latter are also not a uniform population, spanning a wide age range), deviations of the real surface brightness distribution in the central region from our simple MGE input, or velocity bias in the LOS measurements of the Gaia DR3 catalogue due to crowding.
Still qualitatively, we can find similarities between the input velocity dispersion field and the Schwarzschild model result.
The model forecasts dips in dispersion at both extremities of the bar.
These dips can also be discerned in our input velocity dispersion map, although they are somewhat less pronounced, likely due to the implemented smoothing filter.
They also appear slightly displaced from their predicted locations according to the model, likely a consequence of the actual LMC bar being off-centred.
Going outwards, the disc's velocity dispersion aligns well with the predictions made by the Schwarzschild model.

   \begin{figure*}
   \centering
            {
            \includegraphics[width=15cm]{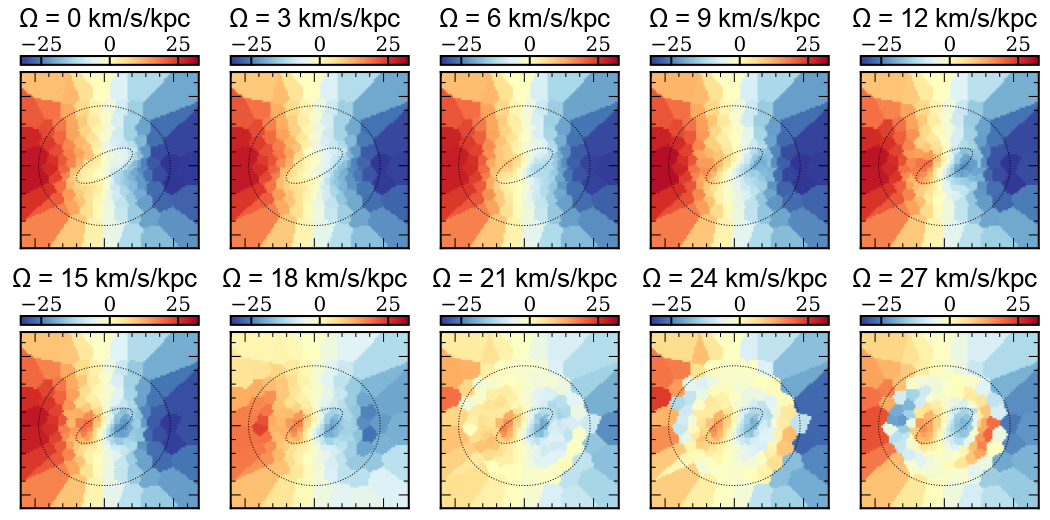}
            \includegraphics[width=15cm]{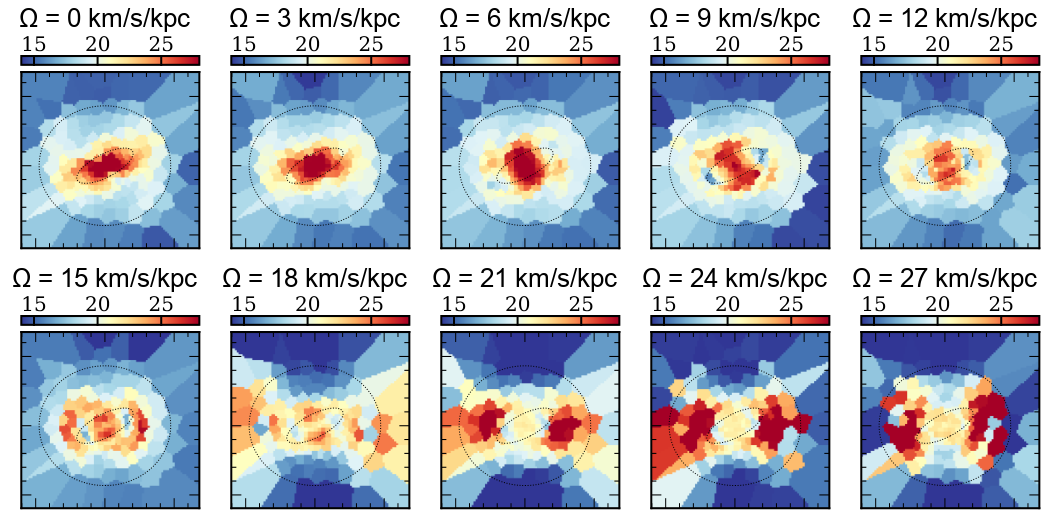}
            }
      \caption{The projected LOS velocity and velocity dispersion maps for the best-fit Schwarzschild potential with $\log C=0.7$,  $\rm \log M_{200}/M_*=2.9$, and $\rm M_*/L=1.0$\,\MLsun, but varying $\Omega$ from $0$ to $27$\,\kmskpc.}
         \label{fig:omega_kin}
   \end{figure*}

Fig.~\ref{fig:omega_kin} shows how the projected LOS velocity and velocity dispersion fields change by varying $\Omega$, when all other model parameters are kept fixed ($\rm M_{200}/M_*=2.9$, $\log C = 0.7$, and $\rm M_*/L = 1.0$\,\MLsun).
Note that the orbit weights are re-fitted for each case to the input data.
By observing these maps the reader can more easily understand the behaviour of the $\chi^2$ distribution with respect to the bar pattern speed.
Models with $\Omega=0$ to $\Omega\sim12$\,\kmskpc~are qualitatively most similar to the observed $\rm V_{LOS}$ field, however the telltale twist in the rotation axis that we observe in the data starts to become noticeable in the models only for $\Omega>6$\,\kmskpc.
Similarly, the model velocity dispersion maps are most similar to the input velocity dispersion for $\Omega$ in the range $9 - 15$\,\kmskpc.
By increasing $\Omega$ above $15$\,\kmskpc, a strong kinematic effect from the bar develops, which shows a distinctive counter-rotating pattern, not visible in our input data.

In the bottom right panel of Fig.~\ref{fig:bar_lmc_32} we show the orbit weights distribution for the best fit Schwarzschild model.
We characterise each orbit by its circularity ($\lambda_z \equiv J_z/J_{max}(E)$, where $J_z$ is the angular momentum of the orbit along the $z$-axis and $J_{max}(E)$ is the the maximum angular momentum of a circular orbit with the same binding energy $E$) and radius ($r$) and bin them in a grid, where the weight of each bin is the sum of the weights of all orbits that fall into that bin.
Not surprisingly, the disc dominated outer region of the LMC can be described with highly circular orbits ($\lambda_z\sim1$), while the bar motions require a wide range of different orbit families, including radial ($\lambda_z\sim0$) and retrograde orbits ($\lambda_z<0$).
The reader may notice that the modelled LOSVD in some central Voronoi bins (Fig.~\ref{fig:losvd_match}) is double peaked. This is owing to contribution from retrograde orbits. We did test model fits, where we gradually removed the most extreme retrograde orbits ($\lambda_z<-0.75$) from our library and confirmed that the double peaks disappear. However, we were not able to obtain a better $\chi^2$ solution than using the full orbit library. By cutting retrograde orbits, even if the few affected bins are better fitted, the overall central velocity dispersion of the rest of the modelled bins increases, making them less compatible with the observational data.

   \begin{figure}
   \resizebox{\hsize}{!}
            {
            \includegraphics{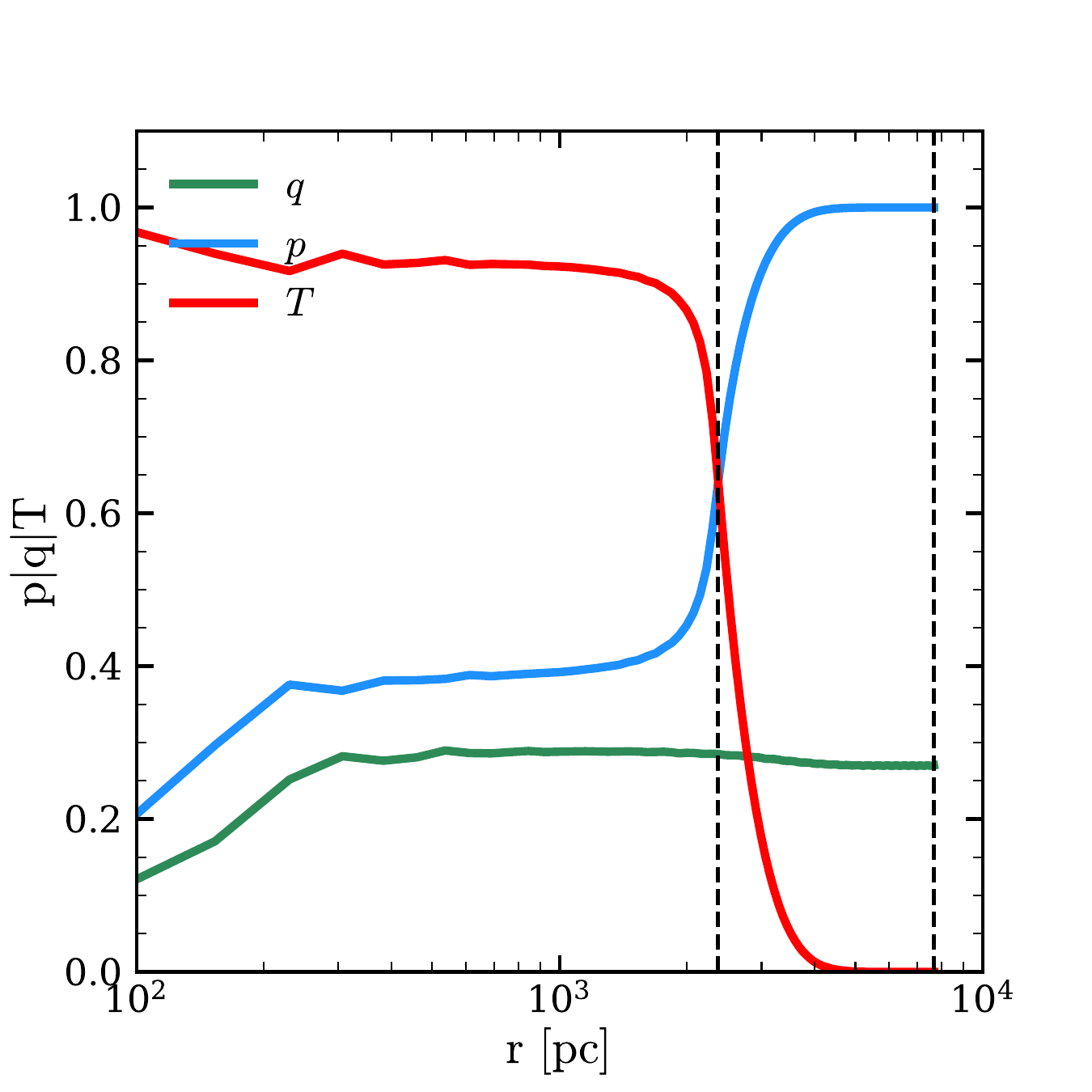}
            }
      \caption{Triaxiality profile of the LMC as reconstructed from the orbit superposition. The vertical lines in all panels indicate the size of the LMC bar and the radial extent of our kinematic tracers.}
         \label{fig:triaxiality}
   \end{figure}

Fig.~\ref{fig:triaxiality} shows the intrinsic axis ratios $p_0$, $q_0$, and a triaxiality parameter $T = (1-p_0^2)/(1-q_0^2)$ as a function of radius as recovered from the weighted orbit superposition.
We use the variations of these parameters to estimate the size of the LMC bar.
As expected, the $q_0$ ratio is relatively flat around $0.25$ in accordance with our input for the flattening of the LMC disc.
The $p_0$ ratio is settled around $0.4$ in the central region dominated by the bar and then rapidly increases to $1$ when the bar relevance fades and the axisymmetric disc starts to dominate.
Similarly the $T$ parameter is around $0.9$ (high triaxiality) in the bar dominated region and then falls to zero going outwards.
We consider the crossing of these two curves at $\sim2.2$\,kpc to indicate the intrinsic size of the LMC bar ($R_{bar}$).
This size also corresponds to $2\,\sigma$ of the Gaussian surface brightness density profile that represents the bar in our MGE model for the LMC and is similar to what one would measure by eye from the Gaia DR3 density map of the LMC.
These latter quantities are, however, projected, although probably not too far off the intrinsic value of the model due to the low inclination of the LMC that we have adopted here.

Considering $\Omega\sim11\pm4$\,\kmskpc~and a co-rotation radius $R_{cor}\sim10\pm4$\,kpc, we find a dimensionless bar rotation parameter $R_{cor} / R_{bar} \simeq 4.5\pm1.8$.
The main uncertainty here is the circular velocity of the LMC, since the co-rotation radius appears slightly outside of the extent of our kinematic tracers and the dynamical solution becomes entirely dependent on the parametrisation of the LMC gravitational potential (see Fig.~\ref{fig:schw_profile_all}).
Galaxy bars can be classified as fast if $1 < R_{cor} / R_{bar} < 1.4$ or slow if $R_{cor} / R_{bar} > 1.4$ \citep{debattista+sellwood2000}.
Hence, we argue that the LMC hosts a slow bar in agreement with the recent studies \citep{jimenez-arranz+2024a, jimenez-arranz+2024b}, based on independent methods.
In addition, galaxy bars that are heavily influenced by tidal interactions tend to be slower than secular bars with a dimensionless bar rotation parameter in the $2-3$ range \citep{gajda+2017} and possibly larger closer to pericentric passages \citep{gajda+2018}, which fits well with our measurement for the LMC bar pattern speed.
In such cases, the bar length and rotational velocity can change on a short time-scale, therefore, we can only report instantaneous (present-day) values that might be very different from genuine ones. The significant divergence in the LMC bar parameters existing in the literature \citep{jimenez-arranz+2024a} could be understood, as different techniques based on different tracers may capture different stages of the bar parameter fluctuations.

\section{Summary and conclusions}\label{sec:conclusion}

In this work we present a series of equilibrium dynamical models for the LMC utilising 3D kinematic data for $\sim 3\times10^4$ genuine member stars from Gaia DR3.
While the LMC is significantly out of equilibrium, as a result of complex tidal interactions with the Milky Way and a recent collision with the SMC \citep[e.g.][]{besla+2012}, that have undoubtedly disrupted the LMC's kinematic and structural properties, equilibrium dynamical models still remain a valuable tool for analysing the galaxy's kinematic properties.
By applying these models, we can quantify the degree of the LMC's deviation from equilibrium, offering crucial insights into its current dynamical state.

We utilise discrete 3D Jeans dynamical models to establish the basic dynamical properties of the LMC, its enclosed mass profile, and in particular its disc shape and orientation, under the assumption of axisymmetry.
We also use these models to separate high probability LMC member stars from the Milky Way foreground contaminants and unresolved background galaxies, based on their 3D kinematic properties and spatial distribution.
We show that our Jeans model based Bayesian membership selection is as good as the selection method outlined in \citet{gaiaedr3}, based on PM and parallax.

The fiducial Jeans model uses all member stars with 3D velocities from Gaia within the VMC footprint as kinematic tracers and assumes a spherical NFW gravitational potential.
In addition we compute models where only the kinematics of the old or young LMC stellar populations are used as tracers and find similar results, although noting that the young stellar population is significantly more clumpy and has noisier kinematic distribution than the old stars, which have a smoother density and velocity profiles.
We find an enclosed mass within $6.2$\,kpc of $\sim1.4\times10^{10}$\,\Msun~and a virial mass $\rm M_{200} \simeq 1.8\times10^{11}$\,\Msun, based on our fiducial model.
The latter result is in excellent agreement with independent estimates of the total LMC mass \citep{vasiliev2023}, although we take it we caution as we do not want to imply that a simple NFW potential is sufficient to describe the dynamical complexity of the LMC at all radii.
We also test a mixed gravitational potential that adds the scaled surface brightness profile by a fixed stellar $\rm M_*/L=1.5$\,\MLsun~to the NFW halo.
This model performs the same way as our fiducial model within the extent of the kinematic tracers, although the fitted NFW parameters define a more extended DM halo that extrapolates to an unrealistically high virial mass in order to accommodate the added visible mass.
If we let the $\rm M_*/L$ to be a free parameter in our model fit, it tends to go to $0$, thus reducing the model to the fiducial one.

Our 3D Jeans models constrain the orientation and shape of the LMC disc based purely on its kinematic properties, without any prior knowledge of the surface brightness distribution, besides the assumption of a radially exponential decline.
We find a projected flattening of $0.91$ at an inclination angle of $25.5^{\circ}$ and an orientation angle of $124^{\circ}$ in reasonable agreement with previous studies \citep[e.g.][]{vandermarel+kallivayalil2014, gaiaedr3}.
We also acknowledge the discrepancy between the best fit kinematic axis and the photometric major axis, which we estimate at an orientation angle of $\sim86^{\circ}$ from the Gaia DR3 density map of the LMC.
These discrepancies speak of the deviation of the LMC disc from axisymmetry, also noted in previous studies \citep{vandermarel+cioni2001, vandermarel+kallivayalil2014}.

Despite these caveats, our axisymmetric Jeans models describe the LMC disc velocity field quite well, capturing the projected velocity gradients due to the disc rotation and the velocity dispersion profiles in the three cardinal directions.
Additional tests, exploring the velocity gradient correlations in the $V_{LOS}$ and $V_Y$ directions \citep{vandeven+2006}, confirm the findings of the Jeans model and that at least purely kinematically the old LMC disc behaves as if it were axisymmetric.
The young stars, on the other hand, show a significant discrepancy from this assumption.

We use the Gaia DR3 density image and the {\sc galfit} software to decompose the LMC into co-centric axisymmetric disc and triaxial bar components and de-project them to obtain their intrinsic shapes, density distributions, and viewing angles.
The de-projection of a triaxial ellipsoid is a degenerate problem, however, in this case we can solve it if we assume that the short axis of the triaxial bar is aligned with the short axis of the axisymmetric disc \citep{tahmasebzadeh+2021}.
We also opted to keep the disc shape and orientation as derived kinematically from the Jeans model.
Although, we fitted the Gaia density image of the LMC with a mixture of S\'ersic density distribution to represent the two morphological components, we found S\'ersic indices $n_d=0.50$ and $n_b=0.34$ for the disc and bar, respectively.
Since a S\'ersic distribution with $n = 0.5$ reduces to the Gaussian distribution, we represent the surface brightness of the LMC as a MGE with two Gaussian components - one representing the disc and the other one the bar, scaled to a total luminosity of $1.31$\,\Lsun, as an input for our Schwarzschild models, together with the derived viewing angles.
As kinematic input we provide Voronoi binned maps of the LOS projected rotation and velocity dispersion fields of LMC's old stellar population from Gaia DR3.

We fit Schwarzschild orbit superposition dynamical models to this dataset using a modified version of the {\sc dynamite} code \citep{jethwa+2020, thater+2022}, which allows for orbit integration in a rotating reference frame.
The rotation rate that best fits the input data for a given gravitational potential corresponds to a stationary reference frame for the bar and thus gives us the bar pattern speed ($\Omega$).
Our gravitational potential consists of the scaled de-projected MGE density distribution plus a spherical NFW DM halo.
We explore a large grid of models, sampling the potential using three parameters - stellar mass to light ratio ($\rm M_*/L$), dark mass to stellar mass ratio ($\rm\log M_{200}/M_*$), and dark matter concentration ($\log C$), testing frame rotation rates from $-10$ to $30$\,\kmskpc.
The Schwarzschild models recover a very similar mass distribution to our Jeans models and fit better the LOSVD of the inner region of the LMC, in particular its bar.
We find a bar pattern speed $\Omega=11\pm4$\,\kmskpc, in line with other recent studies that use independent methodologies \citep{jimenez-arranz+2024a, jimenez-arranz+2024b}.
With a bar size of $\sim2.2$\,kpc and a co-rotation radius of $\sim10$\,kpc, we argue for a slow bar in the LMC, in line with expectations for bars born from tidal interactions \citep{gajda+2017, gajda+2018}.

This is the first study that attempts to include LMC's bar as a triaxial component in an equilibrium dynamical model, based on the Schwarzschild orbit superposition method, fit directly to the observed data.
Despite a significant success in deriving the bar pattern speed of the LMC and describing reasonably well the LOSVD in the inner region of the LMC, there are important caveats to the current models.
We plan next to expand our Schwarzschild models to work effectively with 3D velocity data.
The incorporation of the full velocity field will advance our understanding of the dynamics of the LMC, introducing a new layer of depth and complexity to existing studies.
Furthermore, our models are still hindered by the assumption of an axisymmetric disc, an assumption that may not hold in reality, as it is plausible that the disc of the LMC is not axisymmetric \citep{vandermarel+cioni2001}.
This assumption, therefore, risks oversimplifying the actual configurations of the stellar system and we plan to abandon it and present a fully triaxial Schwarzschild model of the LMC in the near future.
Another factor to consider is that the system might be entirely out of equilibrium and models assuming so may not provide a meaningful description of its complex stellar kinematics.
In light of this, it is worth noting that the LMC bar could be a transient feature born from a recent collision with the SMC \citep{besla+2012}.
Furthermore, it is plausible that the bar is off-centred \citep{devaucouleurs+freeman1972}, introducing potential asymmetries in the stellar system's morphology and dynamics that our models fail to account for.

There might be problems also in the Gaia data itself due to the high stellar density in LMC's innermost regions.
As we look towards the future, upcoming projects such as 4MOST "1001 Magellanic Fields" survey \citep{cioni+2019} and Gaia DR4 SIF data \citep{gaia2023_fpr} promise to significantly improve the kinematic information in the centre of the galaxy.
The 4MOST project, for instance, is set to deliver significantly more accurate LOS velocities in the crowded bar region of the LMC.
Likewise, Gaia DR4 SIF data will feature considerably more accurate proper motions in the bar, thus providing a wealth of improved 2D kinematic information.
This influx of precise new data has the potential to greatly enhance the fidelity of our dynamical models and their predictions.

\begin{acknowledgements}
We thank Alice Zocchi, Sabine Thater, Edward Lilley, and Prashin Jethwa for insightful discussions and help with the Schwarzschild modelling technique using {\sc dynamite}.
This work has made use of data from the European Space Agency (ESA) mission {\it Gaia} (\url{https://www.cosmos.esa.int/gaia}), processed by the {\it Gaia} Data Processing and Analysis Consortium (DPAC, \url{https://www.cosmos.esa.int/web/gaia/dpac/consortium}).
Funding for the DPAC has been provided by national institutions, in particular the institutions participating in the {\it Gaia} Multilateral Agreement.
This research has made use of NASA’s Astrophysics Data System.
This research has made use of the NASA/IPAC Extragalactic Database (NED), which is funded by the National Aeronautics and Space Administration and operated by the California Institute of Technology.
\end{acknowledgements}

\bibliographystyle{aa} 
\bibliography{lmc_dyn} 

\end{document}